\documentclass[twocolumn]{aastex61}

\usepackage{natbib}
\usepackage{multirow}
\usepackage{color}
\usepackage{bm}
\usepackage[normalem]{ulem}

\newcommand\aastex{AAS\TeX}

\received{\today}
\revised{}
\accepted{}
\shorttitle{\aastex\ sample article}
\shortauthors{Gaigalas et al.}
\begin{document}

\title{Extended calculations of energy levels and transition rates of Nd II-IV ions for application to neutron star mergers}

\correspondingauthor{Gediminas Gaigalas}
\email{Gediminas.Gaigalas@tfai.vu.lt}

\author{Gediminas Gaigalas}
\affil{Institute of Theoretical Physics and Astronomy, 
Vilnius University, Saul\.{e}tekio Ave. 3, Vilnius, Lithuania}

\author{Daiji Kato}
\affiliation{National Institute for Fusion Science, 
322-6 Oroshi-cho, Toki 509-5292, Japan}
\affiliation{Department of Advanced Energy Engineering,
 Kyushu University, Kasuga, Fukuoka 816-8580, Japan}

\author{Pavel Rynkun}
\affil{Institute of Theoretical Physics and Astronomy, 
Vilnius University, Saul\.{e}tekio Ave. 3, Vilnius, Lithuania}

\author{Laima Rad\v{z}i\={u}t\.{e}}
\affil{Institute of Theoretical Physics and Astronomy, 
Vilnius University, Saul\.{e}tekio Ave. 3, Vilnius, Lithuania}

\author{Masaomi Tanaka}
\affiliation{Astronomical Institute, Tohoku University, Aoba, Sendai 980-8578, Japan }



\begin{abstract}
  Coalescence of binary neutron star give rise to electromagnetic emission, kilonova,
  powered by radioactive decays of $r$-process nuclei.
  Observations of kilonova associated with GW170817 provided unique opportunity
  to study the heavy element synthesis in the Universe.
  However, atomic data of $r$-process elements to decipher the light curves and spectral features
  of kilonova are not fully constructed yet.
  In this paper, we perform extended atomic calculations of neodymium (Nd, $Z=60$)
  to study the impact of accuracies in atomic calculations to the astrophysical opacities.
  By employing multiconfiguration Dirac-Hartree-Fock and relativistic configuration interaction methods, we calculate 
  energy levels and transition data of electric dipole transitions
  for Nd II, Nd III, and Nd IV ions.
  Compared with previous calculations, our new results provide better agreement
  with the experimental data. The accuracy 
	of energy levels was achieved in the present work 10 \%, 3 \% and 11 \% for Nd II, Nd III and Nd IV, respectively, comparing with the NIST database.
  We confirm that the overall properties of the opacity
  are not significantly affected by the accuracies of the atomic calculations.
  The impact to the Planck mean opacity is up to a factor of 1.5,
    which affects the timescale of kilonova at most 20\%.
  However, we find that the wavelength dependent features in the opacity
  are affected by the accuracies of the calculations.
  We emphasize that accurate atomic calculations, in particular
  for low-lying energy levels, are important 
  to provide predictions of kilonova light curves and spectra.

%Employing multiconfiguration Dirac-Hartree-Fock method
% the excitation energies for the lowest 6 888 states of the $4f^4 \{6s,6p,5d\}$, 
%$4f^3 \{5d^2,5d 6p, 5d 6s,6s 6p\}$ configurations for Nd II, 1 488 states of the  
% $4f^4$, $4f^3 \{5d,6p,6s\}$, $4f^2 \{5d^2,5d 6s\}$ configurations for Nd III, and 
%241 states of the $4f^3$ and $4f^2 \{5d,6p,6s\}$ configurations for Nd IV. 
%Electric dipole (E1)  transition rates, line strengths, and oscillator strengths among
%these states were calculated for all of these ions.
\end{abstract}

%\keywords{energy spectra, transition data, opacity, neutron stars}
\keywords{radiative transfer --- opacity --- stars: neutron}



%%%%%%%%%%%%%%%%%%%%%%%%%%%%%%%%%%%%%%%%%%%%%%%%%%%%%%
% Introduction
%%%%%%%%%%%%%%%%%%%%%%%%%%%%%%%%%%%%%%%%%%%%%%%%%%%%%%
\section{Introduction} 
\label{sec:intro}

On 2017 August 18,
the first observation of gravitational waves (GWs)
from neutron star (NS) merger was achieved
(GW170817, \cite{abbott17}).
In addition to GWs, electromagnetic (EM) counterparts
across the wide wavelength range were also observed
\citep{abbott17MMA}.
In particular, intensive observations
of the optical and near-infrared (NIR) counterpart
(SSS17a, also known as DLT17ck or AT2017gfo)
have been performed 
and dense photometric and spectroscopic data were obtained
\citep{andreoni17,arcavi17,chornock17,coulter17,cowperthwaite17,diaz17,drout17,evans17,kasliwal17,kilpatrick17,lipunov17,muccully17,nicholl17,pian17,shappee17,siebert17,smartt17,soares-santos17,tanvir17,tominaga18,troja17,utsumi17,valenti17}.
SSS17a shows characteristic properties
that are quite different from those of supernovae.
The optical light curves decline rapidly while NIR light curves
evolve more slowly.
The spectra show feature-less, broad-line features
implying a high expansion velocity.
These properties are broadly consistent with theoretically
suggested kilonova or macronova emission from NS mergers
\citep{li98,kulkarni05,metzger10}.

Kilonova is EM emission powered by radioactive decay energy
of $r$-process nuclei that are newly synthesized in the NS mergers
\citep[see][for reviews]{rosswog15,tanaka16,fernandez16,metzger17}.
The timescale, luminosity, and color of the emission are
mainly determined by the mass and velocity of the ejecta
and opacities in the ejecta.
Among $r$-process elements, lanthanide elements have high
optical and NIR opacities \citep{kasen13,tanaka13}.
Therefore, if the ejecta include lanthanide elements,
the emission becomes red and faint.
On the other hand, if the ejecta is free from lanthanide elements,
the emission is blue and bright \citep{metzger14,kasen15,tanaka18}.

In fact, SSS17a shows both
blue and red components, which implies the presence of
multiple components with different lanthanide contents.
This fact suggests the production of a wide range of $r$-process elements
\citep{kasen17,tanaka17,rosswog17b}.
This is also consistent with the expectation from numerical relativity
simulations \citep[see e.g.,][]{shibata17,perego17}.
The ejecta mass to explain the luminosity of SSS17a
is about $0.03-0.06 M_{\odot}$.
Although it is still unclear if the $r$-process yields from NS mergers
are consistent with the solar ratios,
NS mergers may be the dominant site for the $r$-process elements in the Universe
\citep{rosswog17b,hotokezaka17}.

Although the observed properties can be explained by kilonova scenario,
physics included in current kilonova simulations is not yet perfect.
In particular, atomic data of $r$-process elements are not complete:
so far calculated data are available only for limited number of 
$r$-process elements
\citep{kasen13,fontes17,wollaeger17,kasen17,tanaka18}.
Even when the data are available, they are almost entirely based on
theoretical calculations, and derived energy levels often
deviates from experimental data by up to $\sim$ 30 \%
(note that experimental data are also insufficient).
It is not yet clear if these issues bring systematic impacts
to the opacities as well as properties of kilonova.

In this paper, we study impacts of the accuracies
in atomic calculations to the opacities by performing extensive,
accurate calculations. For this purpose, we choose
a lanthanide element, neodymium (Nd, $Z=60$),
which has also been studied by \citet{kasen13,fontes17,tanaka18}.
We focus on singly to triply ionized Nd,
for which accurate calculations are possible with
the multiconfiguration Dirac-Hartree-Fock method.
In Sections \ref{sec:methods} and \ref{sec:scheme},
we describe methods and strategies of our atomic calculations.
In Section \ref{sec:results}, we show and evaluate results of 
atomic calculations.
In Section \ref{sec:opacity}, we show the impact of the accuracy of
atomic calculations to the astrophysical opacities.
Finally we give summary in Section \ref{sec:summary}.

%%%%%%%%%%%%%%%%%%%%%%%%%%%%%%%%%%%%%%%%%%%%%%%%%%%%%%
% Methods
%%%%%%%%%%%%%%%%%%%%%%%%%%%%%%%%%%%%%%%%%%%%%%%%%%%%%%
\section{Methods} 
\label{sec:methods}

\subsection{Computational procedure}
The GRASP2K package \citep{graspV3} is based on the multiconfiguration Dirac-Hartree-Fock (MCDHF) 
and relativistic configuration interaction (RCI) methods taking
into account the transverse photon interaction (Breit interaction)
and quantum electrodynamic (QED) corrections \citep{grant, topical_rev}.

The MCDHF method used in the present work is
based on the Dirac-Coulomb Hamiltonian
\begin{eqnarray}
H_{DC} = \sum_{i=1}^N \left( c \; {\bm{ \alpha }}_i \cdot
                    {\bm{ p }}_i
         + (\beta_i -1)c^2 + V^N_i \right)
         + \sum_{i>j}^N \frac{1}{r_{ij}},
\end{eqnarray}
where $V^N$ is the monopole part of the electron-nucleus Coulomb interaction, $\bm{ \alpha }$ and $\beta$ are the $4 \times 4$ Dirac matrices, and $c$ is the speed of light in atomic units.
The atomic state functions (ASF) were obtained as linear
combinations of symmetry adapted configuration state functions (CSFs)
\begin{equation}
\label{ASF}
\Psi({\mathit \gamma} PJM)  = \sum_{j=1}^{NCSFs} c_{j} \Phi(\gamma_{j}PJM).
\end{equation}
Here $J$ and $M$ are the angular quantum numbers and $P$ is parity.
$\gamma_j$ denotes other appropriate labeling of the configuration state function $j$,
for example orbital occupancy and coupling scheme. Normally the label ${\mathit \gamma}$ of the atomic state
function is the same as the label of the dominant CSF.
The  CSFs are built from products of one-electron Dirac orbitals. 
Based on a weighted energy average of several states, the so called extended optimal level (EOL) scheme \citep{EOL},
both the radial parts of the Dirac orbitals and the 
expansion coefficients were optimized to self-consistency in the relativistic self-consistent field procedure.
Note that accurate calculations with the MCDHF method is much more difficult for neutral atoms than ions \citep{grant}, we focus on ionized Nd.

For these calculation, we used the spin-angular approach \citep{Gaigalas_1996,Gaigalas_1997}
which is based on the second quantization in coupled
tensorial form, on the angular momentum theory in
three spaces (orbital, spin, and quasispin) and on the reduced coefficients of fractional parentage. It allow us to study configurations with open \textit{f}-shells without any restrictions.

In subsequent RCI calculations the Breit interaction
\begin{eqnarray}
\label{eq:Breit}
         H_{\mbox{{\footnotesize Breit}}} =  - \sum_{i<j}^N \left[ \bm{\alpha}_{i} \cdot \bm{\alpha}_{j}\frac{ \cos(\omega_{ij} r_{ij}/c)}{r_{ij}}  \right.
			\nonumber \\
         + \left. (\bm{\alpha}_{i} \cdot \bm{\nabla}_i ) (\bm{\alpha}_{j} \cdot \bm{\nabla}_j )\frac{ \cos(\omega_{ij}r_{ij}/c) -1}{\omega_{ij}^2 r_{ij}/c^2} \right]
%         B_{ij} =  - \frac{1}{2 r_{ij}}
%         \Biggl[ \bm{\alpha}_{i} \cdot
%         \bm{\alpha}_{j}
%         +
%         \frac{ (\bm{\alpha}_{i} \cdot {\bm{ r_{ij} }})
%                (\bm{\alpha}_{j} \cdot {\bm{ r_{ij} }}) }
%         {r_{ij}^2} \Biggr]
\end{eqnarray}
was included in the Hamiltonian. The photon frequencies $\omega_{ij}$, used for calculating
the matrix elements of the transverse photon interaction,
were taken as the difference of the diagonal Lagrange multipliers
associated with the Dirac orbitals  \citep{Breit}. In the
RCI calculation the leading QED corrections, self-interaction and vacuum polarization, were also included.

In the present calculations, the ASFs were obtained as expansions
over $jj$-coupled CSFs. To provide the $LSJ$ labeling system,
the ASFs were transformed from a $jj$-coupled CSF basis into an $LSJ$-coupled CSF basis using the
method provided by \cite{jj2lsj_atoms}.

\subsection{Computation of transition parameters}
The evaluation of radiative transition data (transition probabilities, oscillator
strengths) between two states: $\gamma' P'J'M'$ and $\gamma PJM$, built on different and independently optimized orbital sets is non-trivial. 
The transition data can be expressed in terms of the transition moment, which is
defined as 
\begin{eqnarray}
\langle \,\Psi(\gamma PJ)\, \|  {\bf T} \| \,\Psi(\gamma' P'J')\, \rangle  &=&
\nonumber \\
 \sum_{j,k} c_jc'_k \; \langle \,\Phi(\gamma_j PJ)\, \|  {\bf T} \| \,\Phi(\gamma'_k P'J')\, \rangle,
\end{eqnarray}
where ${\bf T}$ is the transition operator. 
For electric dipole and quadrupole (E1 and E2) transitions there are two forms of the transition operator:
 the length (Babushkin) and velocity (Coulomb) forms, which for the exact solutions of the Dirac-equation give the same value of the transition moment \citep{gauge}.
The quantity $dT$,  characterizing the accuracy of the computed transition rates, is defined as
\begin{equation}
\label{accuracy}
dT = \frac{|A_{\rm l}-A_{\rm v}|}{\max(A_{\rm l},A_{\rm v})},
\end{equation}
where $A_{\rm l}$ and $A_{\rm v}$ are transition rates in length and velocity forms.

The calculation of the transition moment breaks down to the task of 
summing up reduced matrix elements between different CSFs.
The reduced matrix elements can be evaluated using standard techniques assuming
that both left and right hand CSFs are formed from the
same orthonormal set of spin-orbitals. This constraint is severe, since a high-quality and compact
wave function requires orbitals optimized
for a specific electronic state, see for example \citep{SF}. 
To get around the problems of having a single orthonormal set of spin-orbitals, the wave function representations
of the two states: $\gamma' P'J'M'$ and $\gamma PJM$, were transformed in such way that the orbital sets became biorthonormal \citep{biotra}.
Standard methods were then used to evaluate the matrix elements of the transformed CSFs. 
\\
\\

%%%%%%%%%%%%%%%%%%%%%%%%%%%%%%%%%%%%%%%%%%%%%%%%%%%%%%
% Table: configuration
%%%%%%%%%%%%%%%%%%%%%%%%%%%%%%%%%%%%%%%%%%%%%%%%%%%%%%
\begin{deluxetable*}{c c c c c c c c c c c }[ht!!]
  \tablewidth{0pt}
\tablecaption{\label{summary} Summary of active space construction}
\tablehead{
Ion &  & Ground & \multicolumn{2}{c}{MR set} & Active space &  \multicolumn{2}{c}{Number of levels} &&  \multicolumn{2}{c}{N$_{CSFs}$} \\ 
\cline{4-5} \cline{7-8} \cline{10-11}
&&conf.& Even & Odd & & Even & Odd && Even & Odd 
}
\startdata
& \multicolumn{8}{c}{\textbf{Strategies A, B.1}}\\
\cline{4-11}
Nd II && $4f^4 6s$ & $4f^4 6s$, $4f^4 5d$   & $4f^3 5d^2$, $4f^4 6p$ & $\{8s,8p,7d,6f,5g\}$ & 3~890 & 2~998 && 24~568 & 23~966 \\
&&& $4f^3 5d 6p$, $4f^3 6s 6p$ & $4f^3 5d 6s$& \\\\ [-0.5cm]
& \multicolumn{8}{c}{\textbf{Additional configuration in Strategy B.2}}\\
\cline{4-11}
      && &$4f^4 6d$, $4f^3 5d 7p$  &$4f^4 7p$, $4f^4 5f$             & $\{10s,10p,9d,8f,7g\}$      & 1~039 & 1~013 &&468~652 & 468~029 \\
   &&& $4f^4 7s$\\\\ [-0.5cm]
%\cline{4-11}
& \multicolumn{8}{c}{\textbf{Strategy C}}\\
\cline{4-11}
   &&& $4f^4 6s$, $4f^4 5d$   & $4f^3 5d^2$, $4f^4 6p$ & $\{8s,8p,7d,6f,5g\}$ & 3~270 & 2~813 && 188~357 & 113~900\\
   &&& $4f^3 5d 6p$, $4f^3 6s 6p$ & $4f^3 5d 6s$                              \\\\ [-0.5cm]
\hline	
& \multicolumn{8}{c}{\textbf{Strategies A, B}}\\
\cline{4-11}
Nd III &  & $4f^4$ & $4f^4$, $4f^3 6p$   & $4f^3 5d$, $4f^3 6s$ & $\{9s,9p,8d,7f,7g,7h\}$ & 1~020 & 468 && 400~440 & 259~948 \\
&&& $4f^2 5d^2$, $4f^2 5d 6s$ \\ \\[-0.5cm]
& \multicolumn{8}{c}{\textbf{Strategy C}}\\	
\cline{4-11}		
 &  && $4f^4$, $4f^3 6p$   & $4f^3 5d$, $4f^3 6s$ & $\{10s,10p,9d,8f,7g,7h\}$ & 747 & 706 && 844~637 & 559~294 \\
&&& $4f^2 5d^2$, $4f^2 5d 6s$ & $4f^3 6d$, $4f^3 7s$ \\ 
&&& $4f^3 5f$, $4f^3 7p$ \\ \\ [-0.5cm]
& \multicolumn{8}{c}{\textbf{Strategy C with 5p,5s}}\\
\cline{4-11}			
 &  && $4f^4$, $4f^3 6p$   & $4f^3 5d$, $4f^3 6s$ & $\{10s,10p,9d,8f,7g,7h\}$ & 747 & 706 && 900~904 & 586~850 \\
&&& $4f^2 5d^2$, $4f^2 5d 6s$ & $4f^3 6d$, $4f^3 7s$ \\ 
&&& $4f^3 5f$, $4f^3 7p$ \\ 
\hline
& \multicolumn{8}{c}{\textbf{Strategy A}}\\
\cline{4-11}
Nd IV &  & $4f^3$ & $4f^2 5d$, $4f^2 6s$ & $4f^3$, $4f^2 6p$  & $\{9s,9p,8d,7f,7g,7h,7i\}$ & 131 & 110 && 33~825 & 26~590 \\
& \multicolumn{8}{c}{\textbf{Strategy B}}\\	
\cline{4-11}
&&& $4f^2 5d$, $4f^2 6s$ & $4f^3$, $4f^2 6p$  & $\{9s,9p,8d,7f,7g,7h\}$ & 1~068 & 465 && 1~445~481 & 587~774 \\
&&& $5p^5 4f^3 5d$ & $5p^5 4f^4$ \\ \\ [-0.5cm]
& \multicolumn{8}{c}{\textbf{Strategy B with 5s}}\\	
\cline{4-11}		
&&& $4f^2 5d$, $4f^2 6s$ & $4f^3$, $4f^2 6p$  & $\{9s,9p,8d,7f,7g,7h\}$ & 1~068 & 465 && 1~474~463 & 603~827 \\
&&& $5p^5 4f^3 5d$ & $5p^5 4f^4$ \\
\enddata 
\end{deluxetable*}
%%%%%%%%%%%%%%%%%%%%%%%%%%%%%%%%%%%%%%%%%%%%%%%%%%%%%%

%%%%%%%%%%%%%%%%%%%%%%%%%%%%%%%%%%%%%%%%%%%%%%%%%%%%%%
% Schemes
%%%%%%%%%%%%%%%%%%%%%%%%%%%%%%%%%%%%%%%%%%%%%%%%%%%%%%
\section{Schemes of the calculations} 
\label{sec:scheme}

\subsection{Active space construction}
\label{Active_space}

Summary of the MCDHF and RCI calculations for each ion is given in Table \ref{summary}.
The description, which explains in what way these calculations were done is given below.
As a starting point DHF calculations were performed in the EOL
scheme for the states of the ground configuration.
The wave functions from these calculations were taken as the initial ones to calculate
even and odd states of multireference (MR) configurations. The set of orbitals belonging to these MR configurations are referred to as 0 layer (0L).

Unless stated otherwise, the inactive core of each ion used in present calculations is [Xe].
The CSF expansions for states of each parity were obtained by allowing single (S) and
double (SD) substitutions from the MR configurations up to
active orbital sets (see Table \ref{summary}). 
The configuration space was increased step by step with increasing the number of layers (L). 
The orbitals of previous layers 
were held fixed and only the orbitals of the newest layer were allowed to vary.
For example, the scheme used to increase the active spaces of the CSF’s for Nd~III ion (in Strategy A) is presented below: \\
AS$_{0L}$ = \{6s, 6p, 5d\}, \\
AS$_{1L}$ = AS$_{0L}$ + \{7s, 7p, 6d, 5f, 5g\}, \\
AS$_{2L}$ = AS$_{1L}$ + \{8s, 8p, 7d, 6f, 6g, 6h\}, \\
AS$_{3L}$ = AS$_{2L}$ + \{9s, 9p, 8d, 7f, 7g, 7h\}.

The MCDHF calculations were followed by RCI calculations, including the Breit interaction and leading QED effects.
The number of CSFs in the final even and odd state expansions distributed
over the different $J$ symmetries is presented in Table \ref{summary}.

\subsection{Strategies for Nd II ion}
Four strategies were tested for Nd II ion. 
All of them were computed in the active space described in the Table \ref{summary}. 
  For the \textbf{Strategy A} a starting point DHF calculations were performed in the EOL
scheme for the states of the ground configuration $4f^4 6s$.
The wave functions from these calculations were taken as the initial ones to calculate
even and odd states of MR configurations. The set of orbitals belonging to these MR configurations are referred to as 0 layer (0L).
The active space were generated as is presented in the Table \ref{summary}.

For \textbf{Strategy B.1} the starting point was computation of the wave functions for the core $4f^46s$.
%Even if $4f$ and $6s$ shells can be treated like valence here they arbitrary named core shells. 
Wave functions were computed in the neutral system of Nd I - ground state $4f^46s^2$.
Then AS$_{0L}$ was computed: the core shells were frozen and only $5d$ and $6p$ shells 
of the configurations of MR listed in Table \ref{summary} were computed. Even and odd states were computed together.
Later, wave functions were optimized separately for states of different parities in the AS$_{1L}$.
AS$_{1L}$ and the next active space were generated by SD substitutions from shells $4f, 5d, 6p, 6s$.  
%For Nd~II just MCDHF calculations were performed, Breit and QEd effects are unincluded.

In the \textbf{Strategy B.2} the configurations of the Rydberg states listed in Table \ref{summary}
were added to the multireference list; therefore, the first active set included subshells bigger by one principal quantum number.
Then the first active space of the \textbf{Strategy B.2} was  
AS$_{1L}$ = AS$_{0L}$ + \{8s, 8p, 7d, 6f, 5g\}.

In \textbf{Strategy C} computation were performed for each configuration separately. For configurations $4f^4 6s$, $4f^4 6p$ and $4f^4 5d$ 
SD substitutions were allowed from $4f^4nl$ (where $l=s,p,d$) shells in to the $AS_{0L,1L}$ and S to the $AS_{2L}$.
For configurations $4f^3 5d6s$, $4f^3 5d6p$, $4f^3 6s6p$, and $4f^3 5d^2$ only S substitutions were allowed. 
Radial wavefunctions up to $4f$ 
orbital was taken from ground configuration for these configurations. 
The Breit interaction and leading QED effects are included in RCI computations.

\subsection{Strategies for Nd III ion}
After AS$_{0L}$ the even and odd states were calculated separately in \textbf{Strategy A}.
For the Nd~III ion calculations the \textbf{Strategy B} was also applied. 
\textbf{Strategy B} differs from \textbf{Strategy A} in the fact that virtual orbitals for odd parity 
were taken from even parity states instead of varying them in layer 1, and higher layers.

In \textbf{Strategy C} as compared to \textbf{Strategy A} additional configurations: $4f^3 6d$, $4f^3 7s$ (odd parity) and $4f^3 5f$, $4f^3 7p$ (even parity) were added to the MR set. 
In {\textbf{Strategy C with $5p,5s$}} just RCI calculations were performed. The wavefunctions were taken from \textbf{Strategy C} and configurations with S substitutions from $5p$ and $5s$ shells to $\{6s,6p,5d,4f\}$ shells were added additionally in the active space.

\subsection{Strategies for Nd IV ion}
In \textbf{Strategy B} as compared to \textbf{Strategy A}
additional configurations: $4p^5 4f^4$ (odd parity) and $4p^5 4f^3 5d$ (even parity) were added to the MR set.
 The AS for even and odd parities were constructed in such way:
SD substitutions were allowed from the $4f, 5d, 6s, 6p$ shells up to
active orbital sets and S substitution from $5p$ shell to $\{6s,6p,5d,4f\}$ shells. 
In \textbf{{Strategy B with 5s} } just RCI calculations were performed. The wave functions were 
taken from \textbf{Strategy B} and configurations with S substitutions from $5s$ shells 
to $\{6s,6p,5d,4f\}$ shells were added  additionally in the active space.

%%%%%%%%%%%%%%%%%%%%%%%%%%%%%%%%%%%%%%%%%%%%%%%%%%%%%%
% Evaluation or Results
%%%%%%%%%%%%%%%%%%%%%%%%%%%%%%%%%%%%%%%%%%%%%%%%%%%%%%
%\section{Evaluation of data}
%\label{sec:evaluation}
\section{Results}
\label{sec:results}

%%%%%%%%%%%%%%%%%%%%%%%%%%%%%%%%%%%%%%%%%%%%%%%%%%%%%%
% Subsection: Nd II
%%%%%%%%%%%%%%%%%%%%%%%%%%%%%%%%%%%%%%%%%%%%%%%%%%%%%%
%\subsection{Energy Levels for Nd II ion}
\subsection{Nd II}
\label{Nd_II_energies}

A part of all computed excitation energies for Nd II 
are listed in Table \ref{NII_NIST}.
These data were compared with NIST database by evaluating 
relative difference $\Delta E/E =  (E_{NIST} - E) / E_{NIST} $. 
Energy levels computed with Breit interaction and QED effects are presented in columns marked by *. Levels with changed notations are given in Table \ref{NII_NIST_ex}. 
%
%To provide the $LSJ$
%labeling system used in databases such as the NIST
%ASD \citep{NIST}, the wavefunctions are transformed from
%a $jj$-coupled CSF basis into a $LSJ$-coupled CSF basis using the
%methods developed by \citep{jj2lsj_atoms}. 
%
%
%According angular momentum theory such notations of the energy levels does not exist. 

Note that the energy levels of Nd II are also provided by \cite{WyartNdII_1}.
They interpreted 596 levels of 
odd configurations ($4f^35d6s$, $4f^35d^2$, $4f^36s^2$, $4f^46p$ and $4f^5$) in semi-empirical way following the Racah-Slater parametric method, 
by using the Cowan computer codes. 
In their method, radial parameters obtained in a least-squares fit 
were compared with Hartree-Fock (HFR) {\em ab initio} integrals. 
In such a way, obtained energy levels naturally have very small 
disagreement with NIST values, therefore are not presented in this paper.

%%%%%%%%%%%%%%%%%%%%%%%%%%%%%%%%%%%%%%%%%%%%%%%%%%%%%%
% Figure: Energy levels of Nd II
%%%%%%%%%%%%%%%%%%%%%%%%%%%%%%%%%%%%%%%%%%%%%%%%%%%%%%
\begin{figure}
  %\plotone{Nd1_energies.pdf}
  \includegraphics[width=0.45\textwidth]{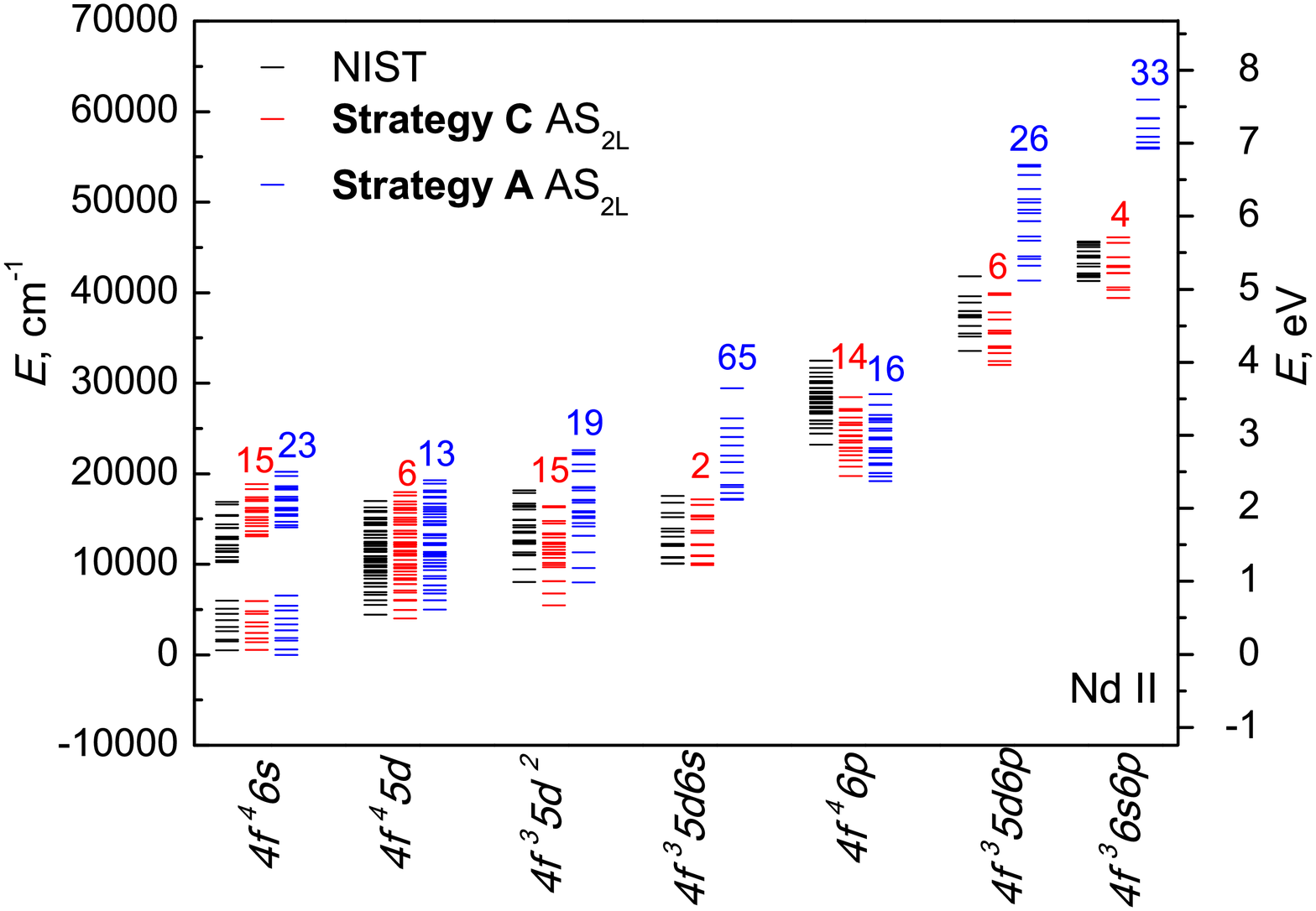}
	\caption{Energy levels for configurations of Nd~II are compared with data of NIST. Black color is representing NIST data, 
	next column of levels in red is our computed energy levels in \textbf{Strategy C} $AS_{2L}$, blue color data are 
	based on \textbf{Startegy A} $AS_{2L}$ \citep[used in][]{tanaka18}. Number on top of red and blue column is averaged disagreement in \% 
	for levels of each configuration comparing with NIST database.
	\label{C_startegy_E_Nd_II}}
\end{figure}
%%%%%%%%%%%%%%%%%%%%%%%%%%%%%%%%%%%%%%%%%%%%%%%%%%%%%%

%\textcolor[rgb]{1,0,0}{Energy levels for each configurations are compared with NIST in the figure \ref{C_startegy_E_Nd_II}. 
%Averaged difference between our computed data and NIST presented values is 10\%. }

 Energy levels for each configurations are compared with NIST in the figure \ref{C_startegy_E_Nd_II}.
Among different strategies, \textbf{Strategy C} $AS_{2L}$ gives the best 
agreement with the NIST database.
Averaged difference between our computed data and NIST presented values is 10 \%. 
This is significant improvement as compared with \textbf{Strategy A} $AS_{2L}$ (blue in Figure \ref{C_startegy_E_Nd_II}),
which was used to compute the opacity of the neutron star mergers in
\cite{tanaka18}.
The averaged difference with the NIST is 22 \% in \textbf{Strategy A} $AS_{2L}$.
For the comparison with the NIST, expression $\overline{\Delta E/E} =\frac{\sum|\Delta E_{i}/E_i|}{N}$ was used,
 where $N$ is the number of compared levels.

\begin{figure}
  \includegraphics[width=0.45\textwidth]{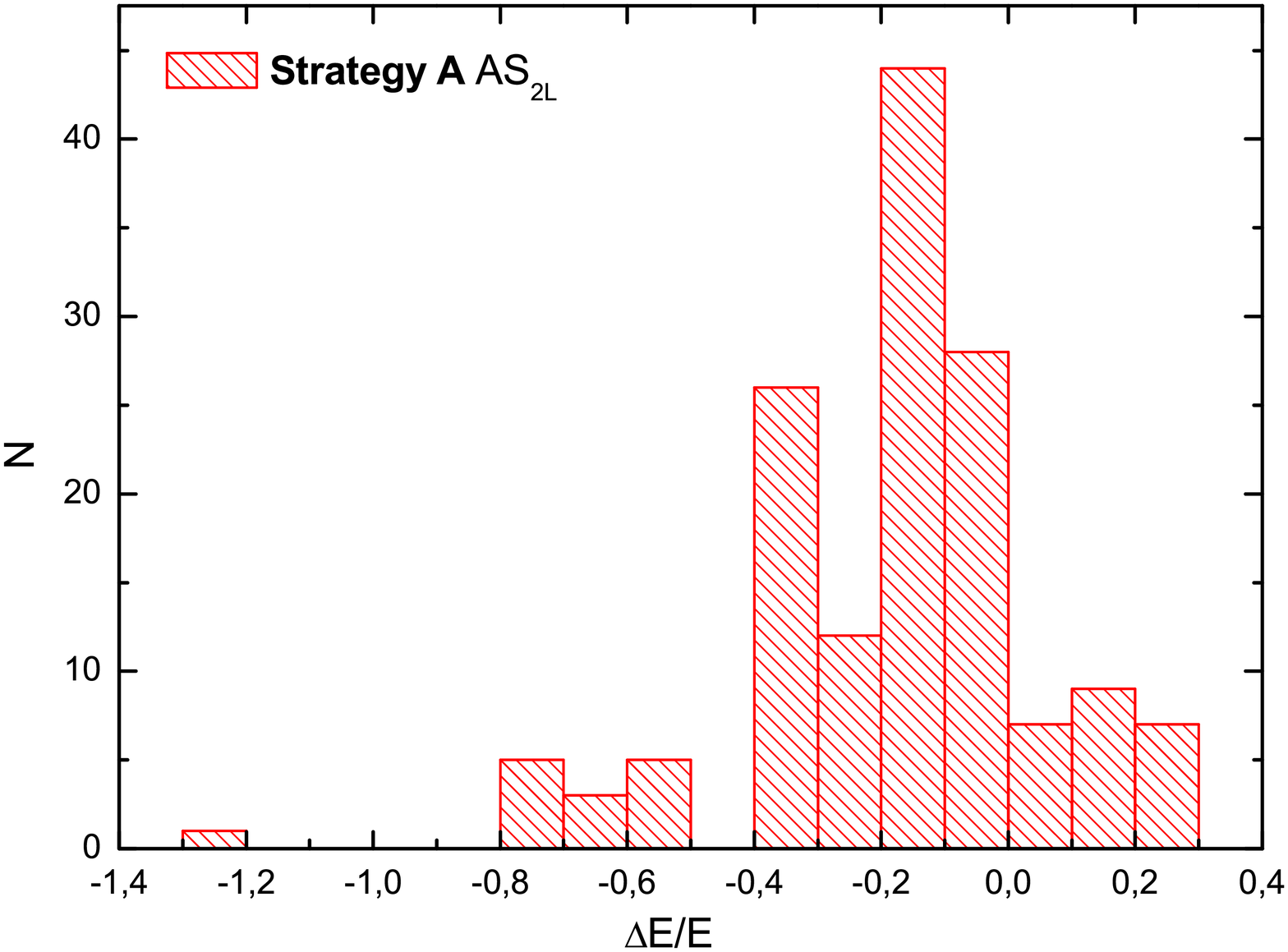}
		\includegraphics[width=0.45\textwidth]{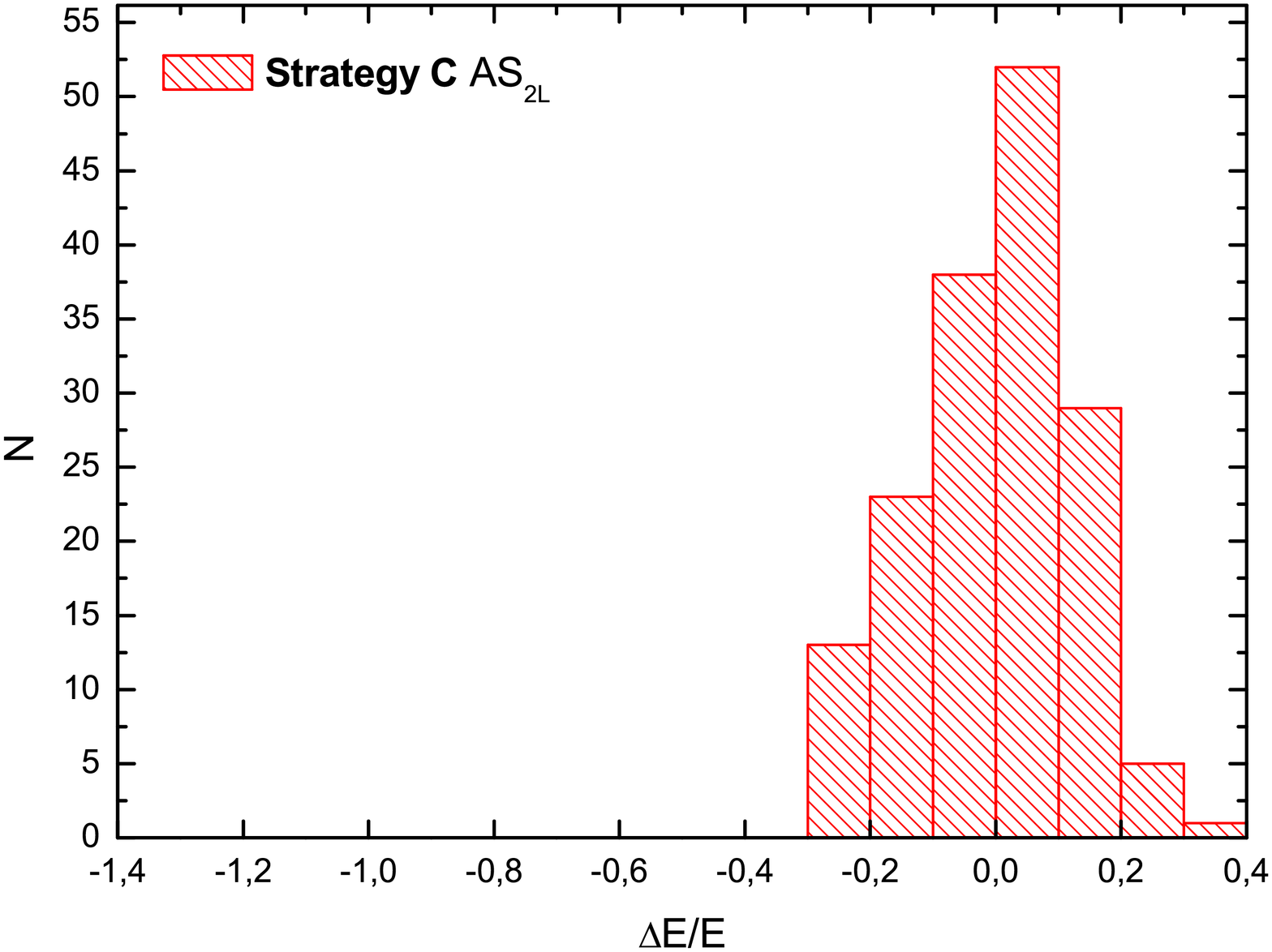}
	\caption{Distribution of the energy levels (N) according the disagreement with NIST database
	for Nd II using \textbf{Strategy A} and \textbf{Strategy C} in $AS_{2L}$.
	\label{statistica_A_C}}
\end{figure}

The figure \ref{statistica_A_C} shows distribution of the energy 
levels number over relative difference comparing with the NIST for \textbf{Strategy A} in active space $AS_2L$.
For the strategies \textbf{A, B.1} and \textbf{B.2} in all active spaces
 the view of the distribution is very similar.
%\begin{figure}
%	\includegraphics[width=0.45\textwidth]{statistica_C_AS2.eps}
%	\caption{Distribution of the energy levels (N) according the disagreement with NIST database for Nd II using \textbf{Strategy C} in $AS_2$.
%	\label{statistica_C}}
%\end{figure}
In case of \textbf{Strategy C} in $AS_2$ (see figure \ref{statistica_A_C}) normal distribution with smaller $\Delta E/E$ range  is observed.
%Energy levels data of Nd II (and consequently transitions data) computed in \textbf{Strategy A} $AS_{2L}$ were used to compute 
%opacity of the neutron star mergers in \cite{tanaka18}.

%\subsection{Results of Nd II E1 type transitions}

%Averaged $dT$ values over number of transitions with the value $log(A)$ in the interval $n$ --- $(n+1)$ where n=-8...+8
%are presented in Fig~\ref{1.pic}.
%These averaged $dT$ values were computed applying \textbf{Strategy B.1} at layer 2 
%\textcolor[rgb]{0,0,1}{and applying \textbf{Strategy B.2} at layer 3-4}.% and represented in Graph \ref{1.pic}.

%\begin{figure}
%	\plotone{B1_2_B2_3_B2_4.pdf}
%	\caption{Averaged $dT$ values at the intervals of the transition probabilities \label{1.pic}}
%\end{figure}

Energy data computed in \textbf{Strategy C} at layer 2 are given in machine readable Table \ref{Levels_NdII}.
This includes number, label, $J$ and $P$ values, and energy value. 
Transitions data obtained from \textbf{Strategy C} at layer 2 are given in machine readable Table \ref{Transition_NdII}. 
This includes identification of upper and lower levels in $LSJ$ coupling, 
transition energy, 
wavelength, 
line strength,
weighted oscillator strength, and transitions probabilities in length form.

%%%%%%%%%%%%%%%%%%%%%%%%%%%%%%%%%%%%%%%%%%%%%%%%%%%%%%
% Table: Energy levels of Nd II
%%%%%%%%%%%%%%%%%%%%%%%%%%%%%%%%%%%%%%%%%%%%%%%%%%%%%%
\begin{deluxetable*}{llcr r@{\hskip 0.05pt} rrr@{\hskip 0.05pt}r@{\hskip 0.05pt}r@{\hskip 0.05pt}rrr@{\hskip 0.05pt}r@{\hskip 0.05pt}rrr@{\hskip 0.05pt}r} 
\tablewidth{0pt}                                                                                                                             
\tablecaption{\label{NII_NIST} Computations of energy values (in cm$^{-1}$) of Nd II by increase of the active space 
performed applying \textbf{Strategy A}, \textbf{Strategy B.1}, \textbf{Strategy B.2}, and \textbf{Strategy C} and their comparison                             
with NIST data base (in \%). With the '*' energy levels were computed with Breit interaction and QED effects are presented.}
\tablehead{&&
&&\multicolumn{2}{c}{\textbf{Strategy A}} 
&&\multicolumn{4
}{c}{\textbf{Strategy B.1}} 
&&\multicolumn{3}{c}{\textbf{Strategy B.2}} 
&&\multicolumn{2}{c}{\textbf{Strategy C}} \\
 \cline{5-6} \cline{8-11} \cline{13-15}\cline{17-18}
Config.                 & Term    & $J$  &   NIST     &AS$_{1L}$/  &AS$_{2L}$
                                                      &&AS$_{1L}$/  &AS$_{2L}$ /&AS$_{2L}$*/     &AS$_{3L}$
																											&&AS$_{1-2L}$/& AS$_{2-3L}$/ &AS$_{3-4L}$                  
																											&& AS$_{1L}$*/& AS$_{2L}$*}
\startdata                                                                                                                                   
$4f^4(^5I)6s$           & $^6I $  &  7/2 &     0.000  &        &      &&        &        &        &       &&         &        &        &&        &     \\  
                        & $^6I $  &  9/2 &   513.330  & $-$14 /& $-$9 && $-$16/ & $-$10 /&  $-$2 /& $-$10 &&  $-$9  /&  $-$9 /& $-$9   && $-$6  /&  $-$7 \\  
                        & $^6I $  & 11/2 &  1470.105  & $-$5  /& $-$3 &&  $-$6/ &  $-$3 /&   8   /&  $-$3 &&  $-$0  /&   0   /& $-$3   && 5     /&   4   \\  
$4f^4(^5I)6s         $  & $^4I $  &  9/2 &  1650.205  & $-$13 /& $-$8 && $-$15/ &  $-$9 /&  $-$3 /&  $-$8 &&  $-$8  /&  $-$7 /& $-$8   && $-$8  /&  $-$8 \\  
$4f^4(^5I)6s         $  & $^6I $  & 13/2 &  2585.460  & $-$5  /& $-$3 &&  $-$4/ &  $-$3 /&   9   /&  $-$3 &&   1    /&   1   /& $-$3   && 6     /&   6   \\  
$4f^4(^5I)6s         $  & $^4I $  & 11/2 &  3066.755  & $-$9  /& $-$6 && $-$10/ &  $-$6 /&   2   /&  $-$6 &&  $-$4  /&  $-$4 /& $-$6   && $-$1  /&  $-$2 \\  
$4f^4(^5I)6s         $  & $^6I $  & 15/2 &  3801.930  & $-$5  /& $-$4 &&  $-$5/ &  $-$4 /&   8   /&  $-$4 &&   0    /&   0   /& $-$4   && 6     /&   6   \\  
$4f^4(^5I)5d         $  & $^6L $  & 11/2 &  4437.560  & $-$13 /& $-$7 && $-$38/ &  $-$3 /&   2   /&  $-$3 &&  $-$2  /&  $-$3 /&  1     && 4     /&   9   \\  
$4f^4(^5I)6s         $  & $^4I $  & 13/2 &  4512.495  & $-$9  /& $-$6 &&  $-$9/ &  $-$6 /&   0   /&  $-$6 &&  $-$4  /&  $-$3 /& $-$6   && 0     /&   0   \\  
$4f^4(^5I)6s         $  & $^6I $  & 17/2 &  5085.640  & $-$6  /& $-$5 &&  $-$6/ &  $-$5 /&   7   /&  $-$5 &&  $-$1  /&  $-$1 /& $-$5   && 6     /&   6   \\  
$4f^4(^5I)5d         $  & $^6L $  & 13/2 &  5487.655  & $-$10 /& $-$5 && $-$30/ &  $-$2 /&   1   /&  $-$2 &&  $-$1  /&  $-$1 /&  1     && 6     /&  10   \\  
$4f^4(^5I)6s         $  & $^4I $  & 15/2 &  5985.580  & $-$9  /& $-$7 &&  $-$9/ &  $-$7 /&   4   /&  $-$7 &&  $-$4  /&  $-$4 /& $-$7   && 1     /&   1   \\  
$4f^4(^5I)5d         $  & $^6K $  &  9/2 &  6005.270  & $-$12 /& $-$7 && $-$36/ &  $-$4 /&  $-$3 /&  $-$4 &&  $-$5  /&  $-$5 /& $-$2   && $-$6  /&  $-$1 \\  
$4f^4(^5I)5d         $  & $^6L $  & 15/2 &  6637.430  & $-$8  /& $-$4 && $-$24/ &  $-$1 /&   3   /&  $-$1 &&   0    /&   0   /&  0     && 7     /&  10   \\  
$4f^4(^5I)5d         $  & $^6K $  & 11/2 &  6931.800  & $-$10 /& $-$6 && $-$31/ &  $-$3 /&  $-$1 /&  $-$3 &&  $-$3  /&  $-$3 /& $-$2   && $-$3  /&   1   \\  
$4f^4(^5I)5d         $  & $^6I $  &  7/2 &  7524.735  & $-$24 /& $-$19&& $-$50/ & $-$17 /& $-$16 /& $-$17 && $-$19  /& $-$19 /&$-$15   && $-$16 /&  $-$1 \\  
$4f^4(^5I)5d         $  & $^6L $  & 17/2 &  7868.910  & $-$7  /& $-$3 && $-$20/ &  $-$1 /&   4   /&  $-$1 &&   1    /&   1   /&  0     && 7     /&  10   \\  
$4f^4(^5I)5d         $  & $^6K $  & 13/2 &  7950.075  & $-$9  /& $-$5 && $-$27/ &  $-$3 /&   1   /&  $-$3 &&  $-$3  /&  $-$3 /& $-$2   && $-$2  /&   2   \\  
$4f^3(^4I)5d^2(^3F)  $  & $^6M^o$ & 13/2 &  8009.810  & 0     /& $-$12&&  $-$7/ & $-$11 /&  $-$6 /& $-$14 &&  $-$7  /& $-$10 /&$-$14   && 32    /&  32   \\  
$4f^4(^5I)5d         $  & $^6I $  &  9/2 &  8420.320  & $-$21 /& $-$16&& $-$43/ & $-$14 /& $-$12 /& $-$14 && $-$15  /& $-$16 /&$-$13   && $-$13 /&  $-$9 \\  
$4f^4(^5I)5d         $  & $^6G $  &  3/2 &  8716.445  & $-$20 /& $-$15&& $-$42/ & $-$13 /& $-$12 /& $-$13 && $-$16  /& $-$16 /&$-$13   && $-$14 /& $-$11 \\  
                        & $^6G $  &  5/2 &  8796.365  & $-$23 /& $-$17&& $-$46/ & $-$15 /& $-$14 /& $-$15 && $-$19  /& $-$19 /&$-$15   && $-$17 /& $-$14 \\  
$4f^4(^5I)5d         $  & $^6K $  & 15/2 &  9042.760  & $-$8  /& $-$5 && $-$24/ &  $-$3 /&   2   /&  $-$3 &&  $-$2  /&  $-$2 /& $-$2   && 0     /&   3   \\  
$4f^4(^5I)5d         $  & $^6L $  & 19/2 &  9166.210  & $-$6  /& $-$3 && $-$17/ &  $-$1 /&   5   /&  $-$1 &&   1    /&   1   /& $-$1   && 7     /&  10   \\  
$4f^4(^5I)5d         $  & $^6G $  &  7/2 &  9198.395  & $-$24 /& $-$18&& $-$46/ & $-$16 /& $-$15 /& $-$16 && $-$19  /& $-$19 /&$-$15   && $-$17 /& $-$14 \\  
$4f^4(^5I)5d         $  & $^6I $  & 11/2 &  9357.910  & $-$18 /& $-$14&& $-$39/ & $-$12 /& $-$10 /& $-$12 && $-$13  /& $-$13 /&$-$11   && $-$10 /&  $-$7 \\  
$4f^3(^4I)5d^2(^3F)  $  & $^6M^o$ & 15/2 &  9448.185  & $-$1  /& $-$12&&  $-$9/ & $-$12 /&  $-$6 /& $-$14 &&  $-$8  /& $-$11 /&$-$14   && 29    /&  29   \\  
$4f^4(^5I)5d         $  & $^6H $  &  5/2 &  9674.835  & $-$28 /& $-$22&& $-$55/ & $-$20 /& $-$20 /& $-$30 && $-$24  /& $-$25 /&$-$19   && $-$23 /& $-$19 \\  
$4f^3(^4I)5d(^5L)6s  $  & $^6L^o$ & 11/2 & 10054.195  & $-$71 /& $-$50&& $-$73/ & $-$77 /&   73  /& $-$79 && $-$73  /& $-$75 /&$-$79   && 1     /&   1   \\  
$4f^3(^4I)5d(^5K)6s  $  & $^6K^o$ &  9/2 & 10091.360  & $-$70 /& $-$51&& $-$71/ & $-$75 /& $-$101/& $-$77 && $-$71  /& $-$73 /&$-$77   && $-$1  /&   0   \\  
$4f^4(^5I)5d         $  & $^6K $  & 17/2 & 10194.805  & $-$8  /& $-$5 && $-$21/ &  $-$3 /&   2   /&  $-$3 &&  $-$2  /&  $-$2 /& $-$3   && 1     /&   3   \\  
$4f^4(^5F)6s         $  & $^6F $  &  1/2 & 10256.040  & $-$37 /& $-$37&& $-$38/ & $-$37 /& $-$36 /& $-$35 && $-$37  /& $-$35 /&$-$35   && $-$28 /& $-$28 \\  
$4f^4(^5I)5d         $  & $^6I $  & 13/2 & 10337.100  & $-$17 /& $-$13&& $-$35/ & $-$11 /&  $-$8 /& $-$11 && $-$12  /& $-$12 /&$-$11   && $-$8  /&  $-$6 \\  
$4f^4(^5F)6s         $  & $^6F $  &  3/2 & 10439.225  & $-$37 /& $-$36&& $-$37/ & $-$48 /& $-$36 /& $-$35 && $-$36  /& $-$34 /&$-$35   && $-$28 /& $-$27 \\  
$4f^4(^5I)5d         $  & $^6L $  & 21/2 & 10516.790  & $-$6  /& $-$3 && $-$15/ &  $-$1 /&   5   /&  $-$1 &&   0    /&   0   /& $-$2   && 7     /&   9   \\  
$4f^4(^5I)5d         $  & $^6H $  &  7/2 & 10666.780  & $-$25 /& $-$20&& $-$36/ & $-$19 /& $-$17 /& $-$18 && $-$11  /& $-$11 /& $-$8   && $-$19 /& $-$16 \\  
$4f^3(^4I)5d(5K)6s   $  & $^6K^o$ & 11/2 & 10720.295  & $-$67 /& $-$82&& $-$82/ & $-$82 /& $-$69 /& $-$75 && $-$69  /& $-$80 /&$-$75   && $-$3  /&  $-$2 \\  
$4f^4(^5F)6s         $  & $^6F $  &  5/2 & 10786.775  & $-$36 /& $-$36&& $-$37/ & $-$50 /& $-$35 /& $-$34 && $-$35  /& $-$47 /&$-$34   && $-$27 /& $-$27 \\  
\enddata
\end{deluxetable*}

%%%%%%%%%%%%%%%%%%%%%%%%%%%%%%%%%%%%%%%%%%%%%%%%%%%%%%

%%%%%%%%%%%%%%%%%%%%%%%%%%%%%%%%%%%%%%%%%%%%%%%%%%%%%%
% Table: Change of NIST notations
%%%%%%%%%%%%%%%%%%%%%%%%%%%%%%%%%%%%%%%%%%%%%%%%%%%%%%
\begin{deluxetable}{l l crrrr r r r }                                                                                                       
\tablewidth{0pt}                                                                                                                             
\tablecaption{\label{NII_NIST_ex} NIST recommended energy levels notations changed by the authors for the Nd II.}                                           
\tablehead{NIST label   & Our label}                                                     
\startdata 
$4f^3(^4I*)5d(^5K*)6s$ $^6L*$ & $4f^3(^4I*)5d(^5L*)6s$ $^6L*$\\
$4f^2(^4I*)5d(^5K*)6s$ $^6I*$ & $4f^3(^4I*)5d(^5I*)6s$ $^6I*$\\ 
\enddata
\end{deluxetable}  
%%%%%%%%%%%%%%%%%%%%%%%%%%%%%%%%%%%%%%%%%%%%%%%%%%%%%%

%%%%%%%%%%%%%%%%%%%%%%%%%%%%%%%%%%%%%%%%%%%%%%%%%%%%%%
% Subsection: Nd III
%%%%%%%%%%%%%%%%%%%%%%%%%%%%%%%%%%%%%%%%%%%%%%%%%%%%%%
%\subsection{Energy Levels for Nd III ion}
\subsection{Nd III}
\label{Nd_III_energies}

\begin{figure}
  %\plotone{Nd1_energies.pdf}
  \includegraphics[width=0.45\textwidth]{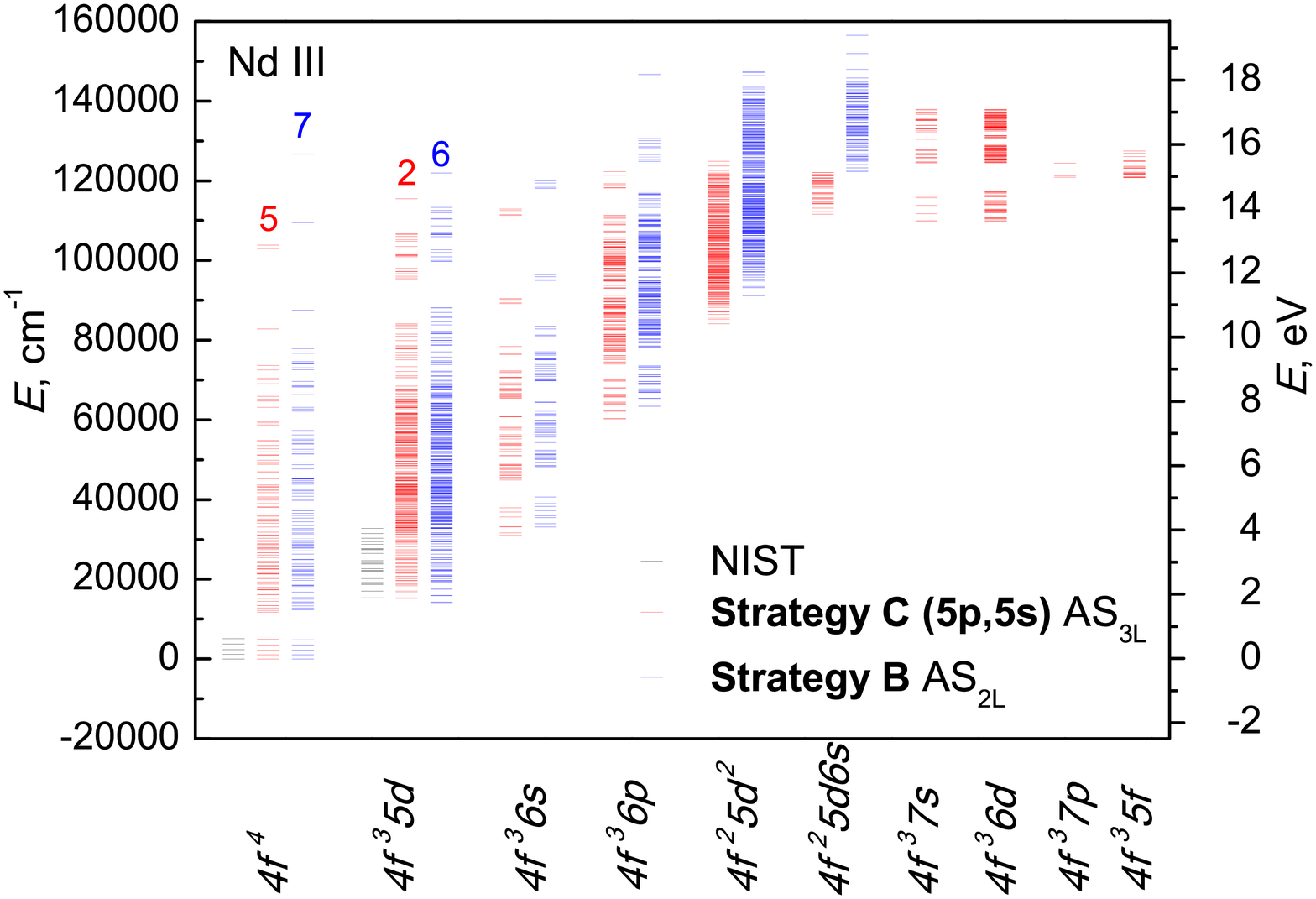}
	\caption{Energy levels for configurations of Nd~III are compared with data of NIST. Black color is representing NIST data, 
	next column of levels in red is our computed energy levels in \textbf{Strategy C (5p,5s)} $AS_{3L}$, blue color data are 
	based on \textbf{Startegy B} $AS_{2L}$. Number on top of red and blue column is averaged disagreement in \% 
	for levels of each configuration comparing with NIST database.
	\label{E_Nd_III}}
\end{figure}

Results of the energy levels for Nd III obtained applying \textbf{Strategies: A, B C, and C with 5p, 5s} are compared with the NIST database and presented in Table \ref{NIII_NIST}.
Among different strategies, \textbf{Strategy C with 5p, 5s} gives the 
best agreement with the NIST database although the number of 
availabe levels is smaller than in the case of Nd II.
All the energy levels and transition data obtained from this strategy
are given the machine readable Tables \ref{Levels_NdIII} and \ref{Transition_NdIII}.
Figure \ref{E_Nd_III} shows the comparison of the energy levels 
with the NIST database.
The averaged difference between our calculations with
\textbf{Strategy C with 5p, 5s} $AS_{3L}$ and the NIST data is {3 \%}.
For comparison, the difference for the case of 
\textbf{Strategy B} $AS_{2L}$, which as used by \cite{tanaka18},
is {5 \%} (blue in Figure \ref{E_Nd_III}).

Results of the energy levels obtained from \textbf{Strategy C with $5p,5s$} 
are also compared with those by \cite{Dzuba}
in Table \ref{NIII_NIST_Z_S}. 
They evaluated energy levels and lifetimes of configurations $4f^4$, $4f^35d$ 
 using relativistic Hartree-Fock and configuration-interaction (RCI) codes as well as a set of
computer codes written by \cite{Cowan}.
Note that \cite{Zhang} also presented low-lying odd energy levels 
(below 33 000 cm$^{-1}$) belonging to the 
configurations: $4f^35d$ and $4f^36s$. 
To compute these energy levels, the HFR, described and coded by \cite{Cowan} 
but modified with the inclusion of core-polarization effects was used. 
It should be mentioned, however, that core-core correlation was not included in energy levels computations.

Disagreement between our data obtained applying \textbf{Strategy C} and \textbf{Strategy C with 5p, 5s} as compared to recommended data by NIST is 
slightly larger than disagreement between data computed by \cite{Dzuba} (Cowan) as compared to recommended data by NIST. 
In this paper we present the lowest 1453 levels of energy spectra and transitions between these states whereas \cite{Dzuba} presented only small part of the spectra (88 levels).
This paper aims at presenting a more complete set of atomic data for astrophysics.
This is clearly reflected in the figure \ref{E_Nd_III} where
energy levels for each configurations at different strategies are presented and compared 
with only a few levels of configuration $4f^4$ and $4f^35d$ available in the NIST.

%%%%%%%%%%%%%%%%%%%%%%%%%%%%%%%%%%%%%%%%%%%%%%%%%%%%%%
% Table: Energy levels of Nd III (strategy A & B & C & C with 5p, 5s & compared with other authors)
%%%%%%%%%%%%%%%%%%%%%%%%%%%%%%%%%%%%%%%%%%%%%%%%%%%%%%
\begin{longrotatetable}
\begin{deluxetable*}{l l c r rrr r rrr r rrr r rrr }
\tabletypesize{\footnotesize}
  %\begin{table}{l l c r r r r}
%\tablewidth{0pt}
\tablecaption{\label{NIII_NIST} Comparison of energy levels with NIST database (in \%) of Nd III by increase of the active space 
performed applying \textbf{Strategies: A, B, C, and C with 5p, 5s}.
States marked by subscript $^{*}$ in term column are without term identification in the NIST database.}
\tablehead{              &&&& \multicolumn{3}{c}{\textbf{Strategy A} }&&\multicolumn{3}{c}{\textbf{Strategy B}} &&\multicolumn{3}{c}{\textbf{Strategy C}} &&\multicolumn{3}{c}{\textbf{Strategy C (5p,5s)}}  \\
 \cline{5-7} \cline{9-11} \cline{13-15} \cline{17-19} 
Config.      & Term & $J$ &   NIST    &AS$_{1L}$/ &AS$_{2L}$/ &AS$_{3L}$ && AS$_{1L}$/ &AS$_{2L}$/ &AS$_{3L}$ && AS$_{1L}$/ &AS$_{2L}$/ &AS$_{3L}$ && AS$_{1L}$/ &AS$_{2L}$/ &AS$_{3L}$  }
\startdata
$4f^4$       &  $^5I$    & 4 &     0.0 &         &         &        &&         &         &        &&      &        &      &&       &       &      \\ 
             &           & 5 &  1137.8 & 8.8/ & 8.5/ & 8.4 && 8.7/ & 8.5/ & 8.4 && 8.8/ & 8.5/ & 8.4 && 6.1/ & 5.9/ & 5.7 \\ 
             &           & 6 &  2387.6 & 7.9/ & 7.7/ & 7.7 && 7.9/ & 7.7/ & 7.7 && 7.9/ & 7.8/ & 7.7 && 5.4/ & 5.3/ & 5.2 \\ 
             &           & 7 &  3714.9 & 7.0/ & 7.0/ & 7.0 && 7.0/ & 7.0/ & 7.0 && 7.1/ & 7.1/ & 7.0 && 4.7/ & 4.7/ & 4.6 \\ 
             &           & 8 &  5093.3 & 6.2/ & 6.2/ & 6.3 && 6.2/ & 6.2/ & 6.3 && 6.3/ & 6.4/ & 6.3 && 4.0/ & 4.1/ & 4.1 \\ 
$4f^3(^4I)5d$&  $^5K^o$  & 5 & 15262.2 & 7.2/ & 11.6/ & 7.4 && 6.8/ & 6.4/ & 6.8 && 7.5/ & 7.0/ & 6.8 && 1.4/ & 1.1/ & 0.9 \\ 
             &           & 6 & 16938.1 & 7.0/ & 11.0/ & 7.1 && 6.7/ & 6.3/ & 6.6 && 7.4/ & 6.8/ & 6.6 && 1.9/ & 1.5/ & 1.3 \\ 
             &           & 7 & 18656.3 & 6.6/ & 10.2/ & 6.7 && 6.4/ & 6.0/ & 6.3 && 7.0/ & 6.4/ & 6.2 && 2.1/ & 1.7/ & 1.5 \\ 
$4f^3(^4I)5d$&  $^5I^o$  & 4 & 18883.7 & $-$2.7/ & 2.1/ & $-$1.2 && $-$3.5/ & $-$2.8/ & $-$1.8 && $-$2.3/ & $-$1.7/ & $-$1.7 && 0.5/ & 0.9/ & 0.9 \\ 
$4f^3(^4I)5d$&  $^5H^o$  & 3 & 19211.0 & $-$5.3/ & $-$0.5/ & $-$3.8 && $-$6.2/ & $-$5.1/ & $-$4.3 && $-$5.0/ & $-$4.2/ & $-$4.2 && $-$3.7/ & $-$2.8/ & $-$2.8 \\ 
             &           & 4 & 20144.3 & $-$3.2/ & 1.3/ & $-$1.8 && $-$3.9/ & $-$3.1/ & $-$2.4 && $-$2.9/ & $-$2.3/ & $-$2.3 && $-$2.3/ & $-$1.6/ & $-$1.6 \\ 
$4f^3(^4I)5d$&  $^5I^o$  & 5 & 20388.9 & $-$1.9/ & 2.5/ & $-$0.6 && $-$2.6/ & $-$2.1/ & $-$1.1 && $-$1.5/ & $-$1.1/ & $-$1.1 && 0.5/ & 0.9/ & 0.9 \\ 
$4f^3(^4I)5d$&  $^5K^o$  & 8 & 20410.9 & 6.1/ & 9.4/ & 6.2 && 5.9/ & 5.5/ & 5.8 && 6.4/ & 5.9/ & 5.7 && 2.0/ & 1.7/ & 1.5 \\ 
$4f^3(^4I)5d$&  $^5H^o$  & 5 & 21886.8 & $-$2.5/ & 1.6/ & $-$1.3 && $-$3.1/ & $-$2.4/ & $-$1.8 && $-$2.2/ & $-$1.7/ & $-$1.7 && $-$1.6/ & $-$1.0/ & $-$1.1 \\ 
$4f^3(^4I)5d$&  $^5I^o$  & 6 & 22047.8 & $-$1.3/ & 2.7/ & $-$0.2 && $-$1.9/ & $-$1.4/ & $-$0.6 && $-$0.9/ & $-$0.5/ & $-$0.6 && 0.6/ & 0.9/ & 0.9 \\ 
$4f^3(^4I)5d$&  $^5K^o$  & 9 & 22197.0 & 5.5/ & 8.6/ & 5.6 && 5.4/ & 5.0/ & 5.2 && 5.8/ & 5.4/ & 5.2 && 1.9/ & 1.6/ & 1.4 \\ 
$4f^3(^4I)5d$&  $^5I^o$  & 7 & 22702.9 & $-$1.0/ & 2.9/ & 0.2 && $-$1.5/ & $-$1.1/ & $-$0.3 && $-$0.6/ & $-$0.2/ & $-$5.8 && 0.6/ & 0.9/ & 0.9 \\ 
$4f^3(^4I)5d$&  $^5H^o$  & 6 & 23819.3 & $-$1.8/ & 1.9/ & $-$0.7 && $-$2.3/ & $-$1.8/ & $-$1.2 && $-$1.5/ & $-$1.1/ & $-$1.1 && $-$1.2/ & $-$0.7/ & $-$0.7 \\ 
$4f^3(^4I)5d$& $^{o*}$& 7 & 24003.2 &       \\ 
$4f^3(^4I)5d$&  $^5I^o$  & 8 & 24686.4 & $-$1.3/ & 2.3/ & $-$0.2 && $-$1.8/ & $-$1.4/ & $-$0.6 && $-$0.9/ & $-$0.6/ & $-$0.6 && 1.4/ & 1.6/ & 1.6 \\ 
$4f^3(^4I)5d$& $^{o*}$& 6 & 26503.2 &       \\ 
$4f^3(^4I)5d$&  $^3K^o$  & 8 & 27391.4 & $-$0.4/ & 3.0/ & 0.8 && $-$0.8/ & $-$0.4/ & 0.4 && $-$0.1/ & 0.4/ & 0.5 && $-$0.9/ & $-$0.4/ & $-$0.3 \\ 
$4f^3(^4F)5d$& $^{o*}$& 3 & 27569.8 &       \\ 
$4f^3(^4F)5d$&  $^5H^o$  & 3 & 27788.2 & $-$10.3/ & $-$6.9/ & $-$8.9 && $-$10.7/ & $-$10.2/ & $-$9.3 && $-$9.9/ & $-$9.4/ & $-$9.2 && $-$6.7/ & $-$6.2/ & $-$6.1 \\ 
             &           & 4 & 28745.3 & $-$10.1/ & $-$6.8/ & $-$8.8 && $-$10.5/ & $-$10.0/ & $-$9.2 && $-$9.7/ & $-$9.2/ & $-$9.1 && $-$6.6/ & $-$6.2/ & $-$6.1 \\ 
$4f^3(^4F)5d$& $^{o*}$& 5 & 29397.3 &       \\ 
$4f^3(^4F)5d$&  $^5H^o$  & 5 & 30232.3 & $-$8.9/ & $-$5.7/ & $-$7.6 && $-$9.3/ & $-$8.8/ & $-$8.0 && $-$8.5/ & $-$8.0/ & $-$7.9 && $-$6.0/ & $-$5.5/ & $-$5.4 \\ 
             &           & 6 & 31394.6 & $-$7.5/ & $-$4.4/ & $-$6.1 && $-$7.8/ & $-$7.3/ & $-$6.5 && $-$7.2/ & $-$6.6/ & $-$6.4 && $-$4.8/ & $-$4.2/ & $-$5.0 \\ 
             &           & 7 & 32832.6 & $-$7.4/ & $-$4.4/ & $-$6.1 && $-$7.8/ & $-$7.3/ & $-$6.5 && $-$7.1/ & $-$6.5/ & $-$6.4 && $-$5.0/ & $-$4.4/ & $-$4.3 \\     
\enddata
\end{deluxetable*}
\end{longrotatetable}

\begin{deluxetable*}{l l c r  r@{\hskip 0.05pt}r @{\hskip 5pt} r @{\hskip 5pt} r@{\hskip 0.05pt}r r@{\hskip 0.05pt}r}
\tabletypesize{\footnotesize}
  %\begin{table}{l l c r r r r}
%\tablewidth{0pt}
\tablecaption{\label{NIII_NIST_Z_S} Comparison of energy levels from present (\textbf{Strategy C with 5p, 5s}) and other theoretical computations with NIST database (in \%) of Nd III.
States marked by subscript $^{*}$ in term column are without term identification in the NIST database.}
\tablehead{              &&&& \multicolumn{2}{c}{Present}  &&\multicolumn{4}{c}{ \cite{Dzuba}} \\
 \cline{5-6} \cline{8-11} 
Config.      & Term & $J$ &   NIST   & \multicolumn{2}{c}{AS$_{3L}$}  && \multicolumn{2}{c}{Cowan} &\multicolumn{2}{c}{RCI}   }
\startdata
$4f^4$       &  $^5I$    & 4 &     0.0 &     0 &     &&      0 &      &     0 &       \\       
             &           & 5 &  1137.8 &  1073/& 5.7 &&   1137/& 0.1 &  1162/& $-$2.1 \\       
             &           & 6 &  2387.6 &  2264/& 5.2 &&   2397/& $-$0.4 &  2471/& $-$3.5 \\    
             &           & 7 &  3714.9 &  3543/& 4.6 &&   3743/& $-$0.8 &  3898/& $-$4.9 \\    
             &           & 8 &  5093.3 &  4885/& 4.1 &&   5148/& $-$1.1 &  5414/& $-$6.3 \\    
$4f^3(^4I)5d$&  $^5K^o$  & 5 & 15262.2 & 15128/& 0.9 &&  14742/& 3.4 & 15357/& $-$0.6 \\       
             &           & 6 & 16938.1 & 16721/& 1.3 &&  16338/& 3.5 & 17380/& $-$2.6 \\       
             &           & 7 & 18656.3 & 18383/& 1.5 &&  18000/& 3.5 & 19485/& $-$4.4 \\       
$4f^3(^4I)5d$&  $^5I^o$  & 4 & 18883.7 & 18714/& 0.9 &&  18467/& 2.2 & 20284/& $-$7.4 \\       
$4f^3(^4I)5d$&  $^5H^o$  & 3 & 19211.0 & 19753/& $-$2.8 &&  19427/& $-$1.1 & 20946/& $-$9.0 \\ 
             &           & 4 & 20144.3 & 20465/& $-$1.6 &&  20189/& $-$0.2 & 21926/& $-$8.8 \\ 
$4f^3(^4I)5d$&  $^5I^o$  & 5 & 20388.9 & 20208/& 0.9 &&  20006/& 1.9 & 21254/& $-$4.2 \\       
$4f^3(^4I)5d$&  $^5K^o$  & 8 & 20410.9 & 20105/& 1.5 &&  19725/& 3.4 & 21666/& $-$6.1 \\       
$4f^3(^4I)5d$&  $^5H^o$  & 5 & 21886.8 & 22119/& $-$1.1 &&  21866/& 0.1 & 22167/& $-$1.3 \\    
$4f^3(^4I)5d$&  $^5I^o$  & 6 & 22047.8 & 21845/& 0.9 &&  21672/& 1.7 & 22664/& $-$2.8 \\       
$4f^3(^4I)5d$&  $^5K^o$  & 9 & 22197.0 & 21882/& 1.4 &&  21503/& 3.1 & 21919/& 1.3 \\          
$4f^3(^4I)5d$&  $^5I^o$  & 7 & 22702.9 & 22499/& 0.9 &&  22244/& 2.0 & 26537/& $-$16.9 \\      
$4f^3(^4I)5d$&  $^5H^o$  & 6 & 23819.3 & 23992/& $-$0.7 &&  23733/& 0.4 & 24076/& $-$1.1 \\    
$4f^3(^4I)5d$& $^{o*}$& 7 & 24003.2 &      &      &&        &      &       &       \\      
$4f^3(^4I)5d$&  $^5I^o$  & 8 & 24686.4 & 24301/& 1.6 &&  24158/& 2.1 & 27396/& $-$11.0 \\      
$4f^3(^4I)5d$& $^{o*}$& 6 & 26503.2 &      &      &&   \\         
$4f^3(^4I)5d$&  $^3K^o$  & 8 & 27391.4 & 27465/& $-$0.3 &&        &      &       &       \\    
$4f^3(^4F)5d$& $^{o*}$& 3 & 27569.8 &      &      &&        &      &       &       \\      
$4f^3(^4F)5d$&  $^5H^o$  & 3 & 27788.2 & 29494/& $-$6.1 &&  28824/& $-$3.7 &       &       \\  
             &           & 4 & 28745.3 & 30506/& $-$6.1 &&  29872/& $-$3.9 &       &       \\  
$4f^3(^4F)5d$& $^{o*}$& 5 & 29397.3 &      &      &&        &      &       &       \\      
$4f^3(^4F)5d$&  $^5H^o$  & 5 & 30232.3 & 31852/& $-$5.4 &&  31117/& $-$2.9 &       &       \\  
             &           & 6 & 31394.6 & 32961/& $-$5.0 &&  32054/& $-$2.1 &       &       \\  
             &           & 7 & 32832.6 & 34234/& $-$4.3 &&  33391/& $-$1.7 &       &       \\      
\enddata
\end{deluxetable*}

%%%%%%%%%%%%%%%%%%%%%%%%%%%%%%%%%%%%%%%%%%%%%%%%%%%%%%

%%%%%%%%%%%%%%%%%%%%%%%%%%%%%%%%%%%%%%%%%%%%%%%%%%%%%%
% Figure: Energy levels of Nd III
%%%%%%%%%%%%%%%%%%%%%%%%%%%%%%%%%%%%%%%%%%%%%%%%%%%%%%

%%%%%%%%%%%%%%%%%%%%%%%%%%%%%%%%%%%%%%%%%%%%%%%%%%%%%%
% Subsection: Nd IV
%%%%%%%%%%%%%%%%%%%%%%%%%%%%%%%%%%%%%%%%%%%%%%%%%%%%%%
%\subsection{Energy Levels for Nd IV ion}
\subsection{Nd IV}

\begin{figure}
  %\plotone{Nd1_energies.pdf}
  \includegraphics[width=0.45\textwidth]{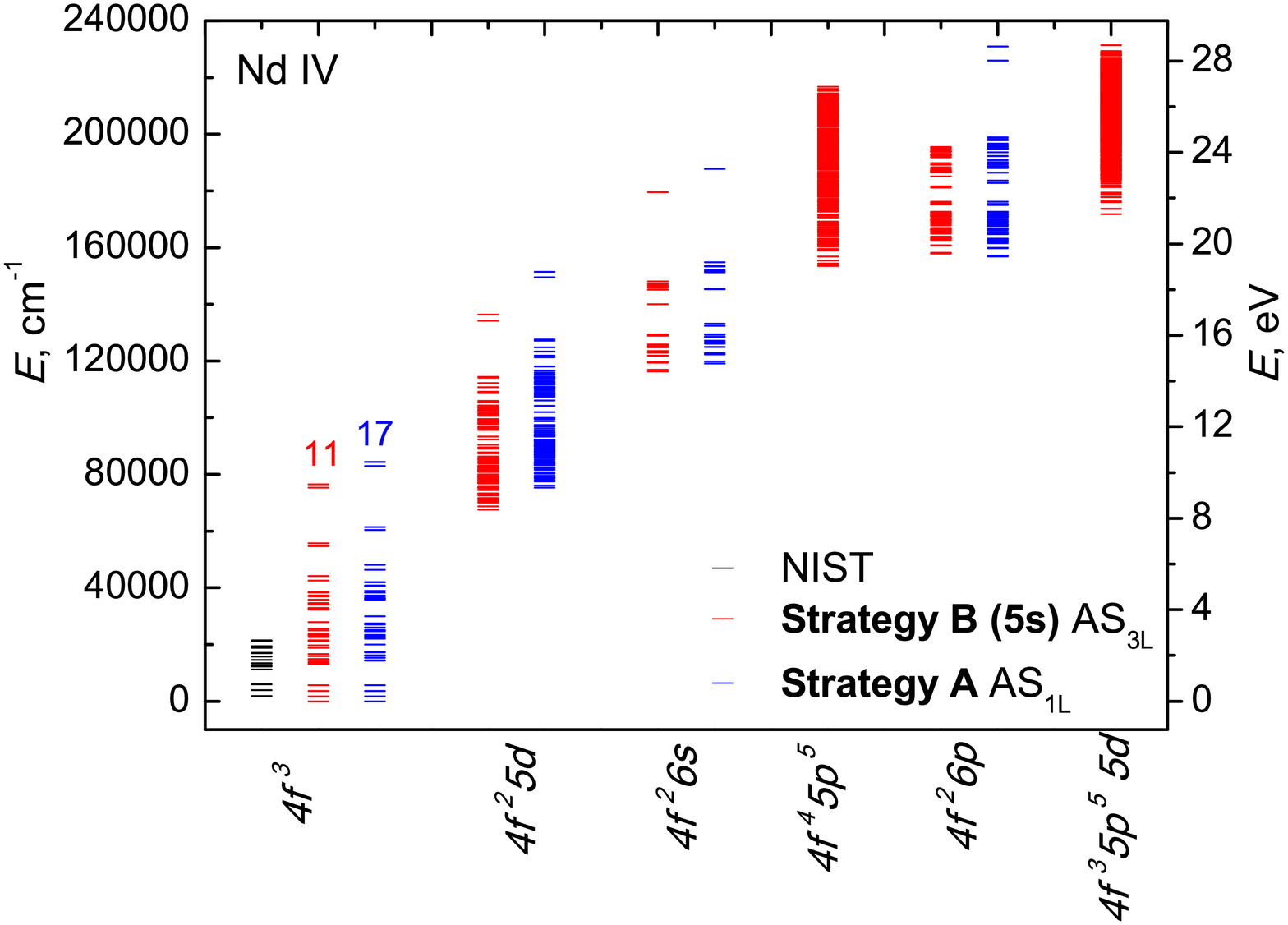}
	\caption{Energy levels for configurations of Nd~IV are compared with data of NIST. Black color is representing NIST data, 
	next column of levels in red is our computed energy levels in \textbf{Strategy B (5s)} $AS_{3L}$, blue color data are 
	based on \textbf{Startegy A} $AS_{1L}$. Number on top of red and blue column is averaged disagreement in \% 
	for levels of each configuration comparing with NIST database.
	\label{E_Nd_IV}}
\end{figure}

Results of the energy levels of \textbf{Strategies A} and \textbf{B} are presented and compared with the NIST database in Table \ref{NIV_NIST}.
The best agreement with the NIST database is obtained for
\textbf{Strategy B with 5s}. 
The energy levels are shown and 
compared with a few levels of configuration $4f^3$ available in the NIST  
in Figure \ref{E_Nd_IV}.
The averaged difference is 11 \% for \textbf{Strategy B with 5s}
while it is 17 \% for \textbf{Strategy A} in active space $AS_{1L}$.

%Energy levels are given in the figure \ref{E_Nd_IV} at different strategies and  compared with a few levels of configuration $4f^3$ available in the NIST.

For Nd IV ion, several experiments and analysis by 
semi-emperical methods have been performed.
The emission spectrum produced by vacuum spark sources was
observed in the vacuum ultraviolet on two normal-incidence spectrographs. 
550 lines have been identified as transitions from 85 (out of 107
possible) levels of $4f^25d$ to 37 (out of 41 possible) levels of $4f^3$.
The method and codes of Cowan were used to predict the spectral ranges of the
strong transitions in the spectra Nd IV in the beginning of paper series 
\citep{Wyart_NIV_2006}. 

Later \cite{Wyart_NIV_2007} used the same experiment to observe 
and classify 1426 lines. 
In total, 41 levels of $4f^3$ configuration were reported.
For deriving their energy levels with
the diagonalization code RCG, the input Hartree-Fock radial integrals including relativistic
corrections, treated as parameters (HFR parameters), were scaled according to earlier results
on the neighbouring ions spectra. Altogether 111 odd parity and 121 even parity of configurations $4f^3$,
 $4f^26p$, $5p^54f^4$, $4f^25d$, $4f^26s$, and $5p^54f^35d$ levels were established. Their
optimized values were calculated with the ELCALC code \citep{ELCALC}.

Then \cite{Wyart_NIV_2008} performed a parametric fit of levels 
energies for $4f^3$ 
configuration, previously obtained in the experiment \citep{Wyart_NIV_2007}.
\cite{Dzuba} did computation in the same way as for Nd III (see subsection \ref{Nd_III_energies}).
This included only 72 levels of configurations $4f^3$ and $4f^25d$.
In Table \ref{NIV_NIST_Wyart},
the energy levels obtained applying \textbf{Strategy B with 5s} are 
compared with the experimental values by \citep{Wyart_NIV_2007}
and semi-empirical values by \cite{Dzuba}.

\begin{figure}
  %\plotone{Nd1_energies.pdf}
  \includegraphics[width=0.45\textwidth]{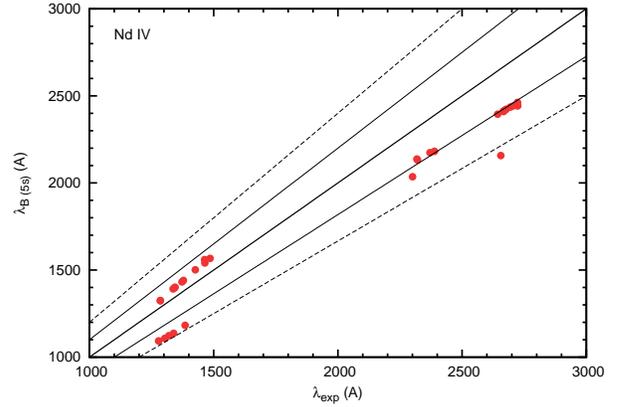}
  \caption{Comparison of transition wavelengths for Nd IV
between our results from \textbf{Strategy B with 5s} 
and experimental data by \cite{Wyart_NIV_2007}.
    The thick line corresponds to the perfect agreement while
    thin solid and dashed lines correspond to 10\% and 20 \% deviation.
	\label{lambda_NdIV}}
\end{figure}

%\textcolor[rgb]{0,0,1}{\subsection{Results of Nd IV E1 type transitions}
In addition to the energy levels, transition data can also be 
compared with experimental data and semi-emperical calculations
(Table \ref{NIV_Wyart_t_3L}).
Our results on the transition wavelengths show good 
agreement with the experimental data by \cite{Wyart_NIV_2007}.
As shown in Figure \ref{lambda_NdIV},
the agreement in the wavelength is within 20 \% for the most transitions.

\begin{figure}
  %\plotone{Nd1_energies.pdf}
  \includegraphics[width=0.45\textwidth]{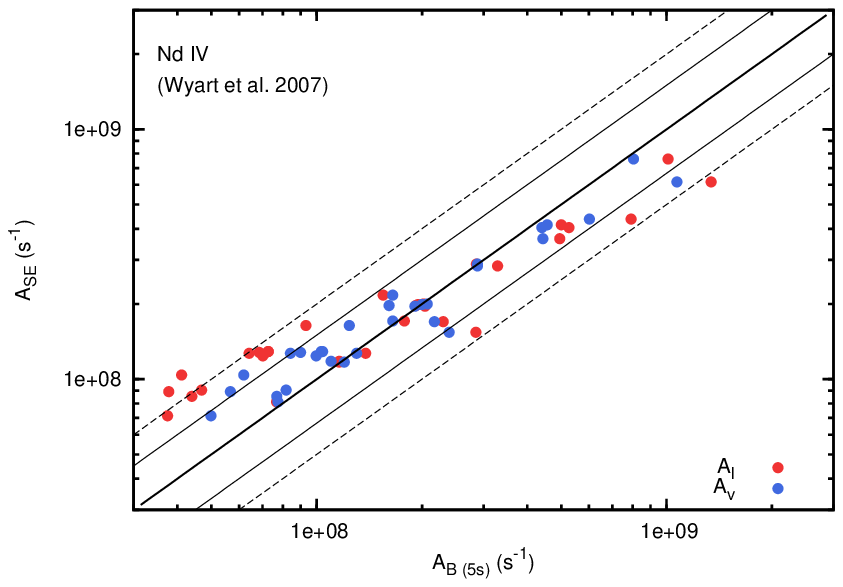}
  \includegraphics[width=0.45\textwidth]{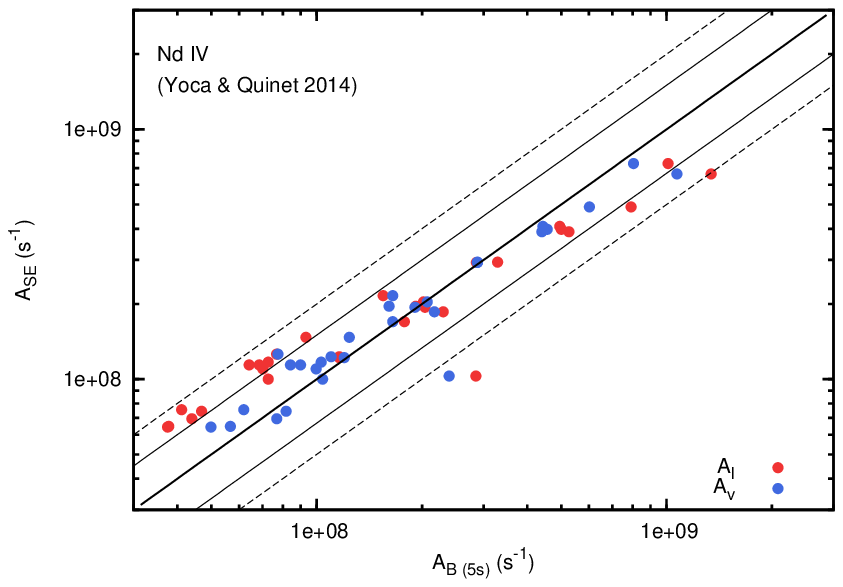}    
  \caption{Comparison of transition probability for Nd IV.
The top and bottom panels show comparison between 
our results from \textbf{Strategy B with 5s} and
semi-emperical results by \cite{Wyart_NIV_2007} and by \cite{Yoca}, respectively.
    The thick line corresponds to the perfect agreement while
    thin solid and dashed lines correspond 
    to the deviation by a factor of 1.5 and 2.0, respectively.
    Red and blue points show the
    values calculated with the length (Babushkin) and velocity (Coulomb) forms,
    respectively.    
	\label{tran_NdIV}}
\end{figure}

We also confirmed a nice agreement in the transition probabilities.
Table \ref{NIV_Wyart_t_3L} and Figure \ref{tran_NdIV}
show transition probabilities for strongest transitions
computed by \cite{Wyart_NIV_2007}.
Our and their results agree within a factor of 2.
Note that semi-emperical calculations have uncertainties.
Using the the same HFR method combined with parametric least-squares fits
to the same experimental data with \cite{Wyart_NIV_2007},
\cite{Yoca} have computed and presented transition probabilities 
(only with $log\,gf \geq -1.0$), 
oscillator strengths and radiative lifetimes in bigger multiconfiguration expansions than \cite{Wyart_NIV_2007}. 
Their results are systematically different, and 
those by \cite{Yoca} in fact show a slightly better agreement with ours 
as shown in the bottom panel of Figure \ref{tran_NdIV}.

%This author for fitting input have used energies of \cite{Wyart_NIV_2007} experiment. 
%Some of these transitions probabilities arepresented in the Table.
%}

%Also compared our wavelength with observed in the \cite{Wyart_NIV_2007} experiment.

%%%%%%%%%%%%%%%%%%%%%%%%%%%%%%%%%%%%%%%%%%%%%%%%%%%%%%
% Figure: Energy levels of Nd IV
%%%%%%%%%%%%%%%%%%%%%%%%%%%%%%%%%%%%%%%%%%%%%%%%%%%%%%
%%%%%%%%%%%%%%%%%%%%%%%%%%%%%%%%%%%%%%%%%%%%%%%%%%%%%%

%%%%%%%%%%%%%%%%%%%%%%%%%%%%%%%%%%%%%%%%%%%%%%%%%%%%%%
% Table: Energy levels of Nd IV (strategy A & B & B with 5s) with NIST
%%%%%%%%%%%%%%%%%%%%%%%%%%%%%%%%%%%%%%%%%%%%%%%%%%%%%%
\begin{deluxetable*}{l l c r rrr r rrr r rrr}
\tabletypesize{\footnotesize}
  %\begin{table}{l l c r r r r}
%\tablewidth{0pt}
\tablecaption{\label{NIV_NIST} Comparison of energy levels with NIST database (in \%) of Nd IV by increase of the active space 
performed applying \textbf{Strategies: A, B, and B with 5s}. State marked by subscript $^{*}$ in term column is without term identification in the NIST database.}
\tablehead{              &&&& \multicolumn{3}{c}{\textbf{Strategy A} }&&\multicolumn{3}{c}{\textbf{Strategy B}} &&\multicolumn{3}{c}{\textbf{Strategy B (5s)}}  \\
 \cline{5-7} \cline{9-11} \cline{13-15}
Config.      & Term & $J$ &   NIST    &AS$_{1L}$/ &AS$_{2L}$/ &AS$_{3L}$ && AS$_{1L}$/ &AS$_{2L}$/ &AS$_{3L}$ && AS$_{1L}$/ &AS$_{2L}$/ &AS$_{3L}$ }
\startdata
$4f^3$   & $^4I^o $  &  9/2 &       0  &       &       &       &&       &       &       &&       &        &   \\
         &           & 11/2 &  [ 1880] & 6.8/ & 6.5/ & 6.4 && 6.9/ & 6.8/ & 6.7   && 7.2  / & 7.1   / & 7.0    \\
         &           & 13/2 &  [ 3860] & 5.6/ & 5.4/ & 5.3 && 5.8/ & 5.8/ & 5.7   && 6.1  / & 6.1   / & 6.0    \\
         &           & 15/2 &  [ 5910] & 4.7/ & 4.6/ & 4.6 && 4.9/ & 5.0/ & 5.0   && 5.2  / & 5.3   / & 5.3    \\
$4f^3$   & $^4F^o $  &  3/2 &  [11290] & $-$27.8/ & $-$27.0/ & $-$26.6 && $-$20.3/ & $-$19.2/ & $-$18.6 && $-$17.6/ & $-$16.4 / & $-$15.8  \\
         &           &  5/2 &  [12320] & $-$24.8/ & $-$24.2/ & $-$23.8 && $-$17.9/ & $-$16.9/ & $-$16.4 && $-$15.3/ & $-$14.4 / & $-$13.8  \\
$4f^3$   & $^2H2^o$  &  9/2 &  [12470] & $-$14.7/ & $-$12.5/ & $-$11.3 && $-$12.8/ & $-$10.4/ & $-$9.4  && $-$12.0/ & $-$9.5  / & $-$8.5   \\
$4f^3$   & $^4F^o $  &  7/2 &  [13280] & $-$22.3/ & $-$21.6/ & $-$21.2 && $-$16.1/ & $-$15.2/ & $-$14.6 && $-$13.7/ & $-$12.8 / & $-$12.3  \\
$4f^3$   & $^4S^o $  &  3/2 &  [13370] & $-$20.2/ & $-$16.8/ & $-$16.5 && $-$15.6/ & $-$12.1/ & $-$11.6 && $-$13.3/ & $-$9.8  / & $-$9.3   \\
$4f^3$   & $^4F^o $  &  9/2 &  [14570] & $-$18.3/ & $-$17.6/ & $-$17.1 && $-$13.2/ & $-$12.1/ & $-$11.6 && $-$11.3/ & $-$10.2 / & $-$9.7   \\
$4f^3$   & $^2H2^o$  & 11/2 &  [15800] & $-$9.6/ & $-$7.8/ & $-$6.8 && $-$8.4/ & $-$6.3/ & $-$5.5  && $-$7.8 / & $-$5.7  / & $-$4.9   \\
$4f^3$   & $^4G^o $  &  5/2 &  [16980] & $-$29.6/ & $-$28.1/ & $-$27.8 && $-$21.7/ & $-$20.2/ & $-$19.6 && $-$18.6/ & $-$17.1 / & $-$16.5  \\
$4f^3$   & $^{o*} $  &  7/2 &  [17100] &    \\
$4f^3$   & $^4G^o $  &  7/2 &  [18890] & $-$23.6/ & $-$22.3/ & $-$22.0 && $-$16.9/ & $-$15.5/ & $-$14.9 && $-$14.3/ & $-$12.9 / & $-$12.3  \\
         &           &  9/2 &  [19290] & $-$16.6/ & $-$27.0/ & $-$26.7 && $-$22.1/ & $-$20.6/ & $-$20.0 && $-$19.7/ & $-$18.3 / & $-$17.7  \\
$4f^3$   & $^2K^o $  & 13/2 &  [19440] & $-$17.7/ & $-$15.0/ & $-$14.2 && $-$15.1/ & $-$12.2/ & $-$11.4 && $-$13.9/ & $-$11.1 / & $-$10.3  \\
$4f^3$   & $^4G^o $  & 11/2 &  [21280] & $-$22.2/ & $-$21.0/ & $-$20.8 && $-$15.9/ & $-$14.6/ & $-$14.2 && $-$13.4/ & $-$12.2 / & $-$11.7  \\
$4f^3$   & $^2K^o $  & 15/2 &  [21430] & $-$16.1/ & $-$13.7/ & $-$12.9 && $-$13.6/ & $-$11.0/ & $-$10.3 && $-$12.5/ & $-$9.9  / & $-$9.2   \\
\enddata
\end{deluxetable*}

%%%%%%%%%%%%%%%%%%%%%%%%%%%%%%%%%%%%%%%%%%%%%%%%%%%%%%

%%%%%%%%%%%%%%%%%%%%%%%%%%%%%%%%%%%%%%%%%%%%%%%%%%%%%%
% Table: Energy levels of Nd IV (B with 5s) with experiment
%%%%%%%%%%%%%%%%%%%%%%%%%%%%%%%%%%%%%%%%%%%%%%%%%%%%%%
\begin{deluxetable*}{l l crrr r@{\hskip 0.05pt}r@{\hskip 0.05pt}r @{\hskip 5pt} r @{\hskip 5pt} r@{\hskip 0.05pt}r@{\hskip 0.05pt}r r@{\hskip 0.05pt}r@{\hskip 0.05pt}r}
\tabletypesize{\footnotesize}
  %\begin{table}{l l c r r r r}
%\tablewidth{0pt}
\tablecaption{\label{NIV_NIST_Wyart} Comparison of energy levels from present (\textbf{Strategy B with 5s}) and other theoretical computations with NIST database (in \%) and with the experiment \cite{Wyart_NIV_2007} (in \%) of Nd IV. State marked by subscript $^{*}$ in term column is without term identification in the NIST database.}
\tablehead{              &&&&&& \multicolumn{3}{c}{Present}  &&\multicolumn{6}{c}{ \cite{Dzuba}} \\
 \cline{7-9} \cline{11-16} 
Config.      & Term & $J$ &   NIST &   Exp.  && \multicolumn{3}{c}{AS$_{3L}$}  && \multicolumn{3}{c}{Cowan} &\multicolumn{3}{c}{RCI}   }
\startdata
$4f^3$   & $^4I^o $  &  9/2 &       0  &     0    &&     0  &        &      &&     0  &        &      &     0  &       &   \\                    
         &           & 11/2 &  [ 1880] &  1897.11 &&  1749/ &   7.0/ &  7.8 &&  1879/ &   0.1/ &   1.0&  1945/ &  $-$3.5/ &  $-$2.5  \\                    
         &           & 13/2 &  [ 3860] &  3907.43 &&  3627/ &   6.0/ &  7.2 &&  3890/ &  $-$0.8/ &   0.4&  4049/ &  $-$4.9/ &  $-$3.6  \\                    
         &           & 15/2 &  [ 5910] &  5988.51 &&  5596/ &   5.3/ &  6.6 &&  5989/ &  $-$1.3/ &   0.0&  6267/ &  $-$6.0/ &  $-$4.7  \\                    
$4f^3$   & $^4F^o $  &  3/2 &  [11290] & 11698.49 && 13076/ & $-$15.8/ &$-$11.8 && 13294/ & $-$17.8/ & $-$13.6& 12490/ & $-$10.6/ &  $-$6.8  \\                    
         &           &  5/2 &  [12320] & 12747.94 && 14022/ & $-$13.8/ &$-$10.0 && 14333/ & $-$16.3/ & $-$12.4& 13545/ &  $-$9.9/ &  $-$6.3  \\                    
$4f^3$   & $^2H2^o$  &  9/2 &  [12470] & 12800.29 && 13536/ &  $-$8.5/ & $-$5.8 && 13272/ &  $-$6.4/ &  $-$3.7& 14522/ & $-$16.5/ & $-$13.5  \\                    
$4f^3$   & $^4F^o $  &  7/2 &  [13280] & 13719.82 && 14911/ & $-$12.3/ & $-$8.7 && 15249/ & $-$14.8/ & $-$11.1& 14622/ & $-$10.1/ &  $-$6.6  \\                    
$4f^3$   & $^4S^o $  &  3/2 &  [13370] & 13792.49 && 14617/ &  $-$9.3/ & $-$6.0 && 15153/ & $-$13.3/ &  $-$9.9& 14452/ &  $-$8.1/ &  $-$4.8  \\                    
$4f^3$   & $^4F^o $  &  9/2 &  [14570] & 14994.87 && 15979/ &  $-$9.7/ & $-$6.6 && 16334/ & $-$12.1/ &  $-$8.9& 16183/ & $-$11.1/ &  $-$7.9  \\                    
$4f^3$   & $^2H2^o$  & 11/2 &  [15800] & 16161.53 && 16581/ &  $-$4.9/ & $-$2.6 && 16456/ &  $-$4.2/ &  $-$1.8& 18142/ & $-$14.8/ & $-$12.3  \\                    
$4f^3$   & $^4G^o $  &  5/2 &  [16980] & 17707.17 && 19780/ & $-$16.5/ &$-$11.7 &&        &        &      &        &        &        \\                    
$4f^3$   & $^{o*} $  &  7/2 &  [17100] & 17655.11 &&   \\                    
$4f^3$   & $^4G^o $  &  7/2 &  [18890] & 19540.80 && 21218/ & $-$12.3/ & $-$8.6 &&                        &       \\                           
         &           &  9/2 &  [19290] & 19969.79 && 22709/ & $-$17.7/ &$-$13.7 &&                        &       \\                           
$4f^3$   & $^2K^o $  & 13/2 &  [19440] & 20005.22 && 21445/ & $-$10.3/ & $-$7.2 &&                        &       \\                           
$4f^3$   & $^4G^o $  & 11/2 &  [21280] & 22047.39 && 23768/ & $-$11.7/ & $-$7.8 &&                        &       \\                           
$4f^3$   & $^2K^o $  & 15/2 &  [21430] & 22043.77 && 23398/ &  $-$9.2/ & $-$6.1 &&                        &       \\
\enddata
\end{deluxetable*}

%%%%%%%%%%%%%%%%%%%%%%%%%%%%%%%%%%%%%%%%%%%%%%%%%%%%%%

%%%%%%%%%%%%%%%%%%%%%%%%%%%%%%%%%%%%%%%%%%%%%%%%%%%%%%
% Figure: transition data for Nd IV (wavelength)
%%%%%%%%%%%%%%%%%%%%%%%%%%%%%%%%%%%%%%%%%%%%%%%%%%%%%%
%%%%%%%%%%%%%%%%%%%%%%%%%%%%%%%%%%%%%%%%%%%%%%%%%%%%%%

%%%%%%%%%%%%%%%%%%%%%%%%%%%%%%%%%%%%%%%%%%%%%%%%%%%%%%
% Figure: transition data for Nd IV (transition prob.)
%%%%%%%%%%%%%%%%%%%%%%%%%%%%%%%%%%%%%%%%%%%%%%%%%%%%%%
%%%%%%%%%%%%%%%%%%%%%%%%%%%%%%%%%%%%%%%%%%%%%%%%%%%%%%

%%%%%%%%%%%%%%%%%%%%%%%%%%%%%%%%%%%%%%%%%%%%%%%%%%%%%%
% Table: Transition for Nd IV (AS_3)
%%%%%%%%%%%%%%%%%%%%%%%%%%%%%%%%%%%%%%%%%%%%%%%%%%%%%%
\begin{deluxetable*}{l l l l r r r l r r r r r}                                                                                                                
%\begin{table}{l l c r r r r}                                                                                                                                  
\tablewidth{0pt}                                                                                                                                               
\tablecaption{\label{NIV_Wyart_t_3L} Computed transitions data of Nd IV in the \textbf{Strategies B with 5s} at \textbf{$AS_{3L}$}                                            
and compared with the experimental wavelength $\lambda_{\rm exp}$ (in \AA) and computed transitions probabilities $A_{SE}$ (in s$^{-1}$) by \cite{Wyart_NIV_2007}.}
\tablehead{&&&& \multicolumn{4}{c}{\textbf{Strategies B with 5s}} && \multicolumn{2}{c}{\cite{Wyart_NIV_2007}}&& \cite{Yoca}\\                                 
 \cline{5-8} \cline{10-11} \cline{12-12}                                                                                                                       
\multicolumn{2}{c}{Upper} & \multicolumn{2}{c}{Lower} & $A_{\rm l}$ & $A_{\rm v}$ & $dT$ & $\lambda$ &&$A_{SE}$  & $\lambda_{\rm exp}$ && $A_{SE}$                             
}                                                                                                                                                              
\startdata                                                                                                                                                     
$4f^2(^3H)6p~^4I  $ & 4.5 & $4f^2(^3H)5d~^2H$ & 4.5 & 5.26E+8 & 4.40E+8 & 0.16 & 1107.72($_{15.0}$)  && 4.04E+8 & 1303.32  && 3.89E+8\\                        
$4f^2(^1I)5d~^2K  $ & 7.5 & $4f^3~^2L       $ & 8.5 & 9.32E+7 & 1.24E+8 & 0.25 & 1540.16($_{-5.1}$)  && 1.64E+8 & 1464.73  && 1.47E+8\\                        
$4f^2(^3H)6p~^4I  $ & 5.5 & $4f^2(^3H)6s~^4H$ & 5.5 & 2.04E+8 & 1.91E+8 & 0.06 & 2412.27($_{~9.5}$)  && 1.96E+8 & 2666.70  && 1.94E+8\\                        
$4f^2(^3H)6p~^4I  $ & 4.5 & $4f^2(^3H)6s~^4H$ & 3.5 & 1.92E+8 & 1.61E+8 & 0.16 & 2410.31($_{~9.6}$)  && 1.97E+8 & 2666.70  && 1.96E+8\\                        
$4f^2(^1G)5d~^2I  $ & 6.5 & $4f^3~^2K       $ & 7.5 & 6.85E+7 & 8.99E+7 & 0.24 & 1559.66($_{-6.6}$)  && 1.28E+8 & 1463.34  && 1.14E+8\\                        
$4f^2(^3F)5d~^4H  $ & 5.5 & $4f^3~^4I       $ & 6.5 & 3.78E+7 & 5.67E+7 & 0.33 & 1324.01($_{-3.0}$)  && 8.92E+7 & 1285.61  && 6.48E+7\\                        
$4f^2(^3F)5d~^4H  $ & 4.5 & $4f^3~^4I       $ & 5.5 & 4.11E+7 & 6.19E+7 & 0.34 & 1325.17($_{-3.1}$)  && 1.04E+8 & 1285.38  && 7.55E+7\\                        
$4f^2(^3H)5d~^4H  $ & 3.5 & $4f^3~^4I       $ & 4.5 & 7.27E+7 & 1.03E+8 & 0.29 & 1401.4 ($_{-4.2}$)  && 1.29E+8 & 1344.74  && 1.17E+8\\                        
$4f^2(^3H)6p~^4I  $ & 4.5 & $4f^2(^3H)5d~^4K$ & 5.5 & 1.01E+9 & 8.05E+8 & 0.20 & 1123.17($_{14.9}$)  && 7.61E+8 & 1319.25  && 7.29E+8\\                        
$4f^2(^3H)6p~^4G  $ & 5.5 & $4f^2(^3H)6s~^4H$ & 6.5 & 1.55E+8 & 1.65E+8 & 0.06 & 2157.57($_{18.8}$)  && 2.17E+8 & 2656.02  && 2.16E+8\\                        
$4f^2(^1I)6p~^2I  $ & 5.5 & $4f^2(^1I)6s~^2I$ & 5.5 & 2.30E+8 & 2.17E+8 & 0.06 & 2443.54($_{10.3}$)  && 1.70E+8 & 2723.51  && 1.86E+8\\                        
$4f^2(^3H)6p~^4H  $ & 4.5 & $4f^2(^3H)6s~^4H$ & 5.5 & 2.02E+8 & 2.07E+8 & 0.02 & 2416.08($_{~9.5}$)  && 2.00E+8 & 2670.03  && 2.04E+8\\                        
$4f^2(^3H)6p~^4H  $ & 3.5 & $4f^2(^3H)6s~^4H$ & 4.5 & 1.95E+8 & 2.00E+8 & 0.02 & 2423.25($_{~9.5}$)  && 1.99E+8 & 2678.00  &&        \\                        
$4f^2(^3H)5d~^4H  $ & 5.5 & $4f^3~^4I       $ & 6.5 & 7.27E+7 & 1.04E+8 & 0.30 & 1391.9 ($_{-4.1}$)  && 1.29E+8 & 1336.98  && 1.00E+8\\                        
$4f^2(^3H)5d~^4I  $ & 7.5 & $4f^3~^4I       $ & 7.5 & 4.69E+7 & 8.18E+7 & 0.43 & 1432.37($_{-4.4}$)  && 9.05E+7 & 1372.20  && 7.44E+7\\                        
$4f^2(^1I)6p~^2I  $ & 5.5 & $4f^2(^1I)6s~^2I$ & 6.5 & 1.16E+8 & 1.20E+8 & 0.03 & 2443.59($_{10.3}$)  && 1.17E+8 & 2723.51  && 1.22E+8\\                        
$4f^2(^3H)6p~^4H  $ & 4.5 & $4f^2(^3H)6s~^2H$ & 4.5 & 1.38E+8 & 1.30E+8 & 0.05 & 2436.05($_{~9.6}$)  && 1.27E+8 & 2694.71  &&        \\                        
$4f^2(^3H)5d~^4H  $ & 4.5 & $4f^3~^4I       $ & 5.5 & 7.02E+7 & 9.97E+7 & 0.30 & 1397.63($_{-4.1}$)  && 1.24E+8 & 1342.01  && 1.10E+8\\                        
$4f^2(^3H)6p~^2I  $ & 6.5 & $4f^2(^3H)6s~^2H$ & 5.5 & 3.29E+8 & 2.88E+8 & 0.13 & 2174   ($_{~8.3}$)  && 2.84E+8 & 2370.51  && 2.94E+8\\                        
$4f^2(^3F)6p~^4D  $ & 3.5 & $4f^2(^3F)6s~^4F$ & 4.5 & 2.86E+8 & 2.88E+8 & 0.01 & 2394.67($_{~9.4}$)  && 2.89E+8 & 2643.03  && 2.93E+8\\                        
$4f^2(^1I)6p~^2K  $ & 6.5 & $4f^2(^1I)6s~^2I$ & 6.5 & 7.68E+7 & 7.75E+7 & 0.01 & 2181.33($_{~8.7}$)  && 8.11E+7 & 2388.79  && 1.26E+8\\                        
$4f^2(^3H)6p~^4I  $ & 7.5 & $4f^2(^3H)5d~^4K$ & 8.5 & 1.34E+9 & 1.07E+9 & 0.20 & 1093.02($_{14.5}$)  && 6.16E+8 & 1278.41  && 6.63E+8\\                        
$4f^2(^3H)6p~^2I  $ & 6.5 & $4f^2(^3H)5d~^2K$ & 7.5 & 7.92E+8 & 6.02E+8 & 0.24 & 1182.67($_{14.6}$)  && 4.37E+8 & 1385.21  && 4.89E+8\\                        
$4f^2(^3H)6p~^4I  $ & 6.5 & $4f^2(^3H)6s~^4H$ & 5.5 & 4.95E+8 & 4.43E+8 & 0.10 & 2132.31($_{~8.1}$)  && 3.65E+8 & 2320.43  && 4.09E+8\\                        
$4f^2(^3F)6p~^4G  $ & 5.5 & $4f^2(^3F)6s~^4F$ & 4.5 & 5.00E+8 & 4.56E+8 & 0.09 & 2137.32($_{~7.8}$)  && 4.15E+8 & 2318.07  && 3.98E+8\\                        
$4f^2(^3H)6p~^2I  $ & 5.5 & $4f^2(^3H)6s~^4H$ & 4.5 & 1.90E+7 & 1.56E+7 & 0.18 & 2035.66($_{11.5}$)  && 3.96E+8 & 2300.68  && 4.16E+8\\                        
$4f^2(^3H)6p~^4H  $ & 5.5 & $4f^2(^3H)5d~^4H$ & 5.5 & 2.85E+8 & 2.39E+8 & 0.16 & 1136.61($_{15.1}$)  && 1.54E+8 & 1338.62  && 1.03E+8\\                        
$4f^2(^1I)5d~^2K  $ & 6.5 & $4f^3~^2L       $ & 7.5 & 6.41E+7 & 8.42E+7 & 0.24 & 1501.86($_{-5.3}$)  && 1.27E+8 & 1426.05  && 1.14E+8\\                        
$4f^2(^3H)6p~^4I  $ & 4.5 & $4f^2(^3H)6s~^4H$ & 4.5 & 1.78E+8 & 1.65E+8 & 0.07 & 2443.44($_{~9.8}$)  && 1.71E+8 & 2708.43  && 1.70E+8\\                        
$4f^2(^3H)6p~^4H  $ & 5.5 & $4f^2(^3H)6s~^2H$ & 5.5 & 1.16E+8 & 1.10E+8 & 0.06 & 2462.26($_{~9.6}$)  && 1.18E+8 & 2723.34  && 1.23E+8\\                        
$4f^2(^3H)5d~^4I  $ & 6.5 & $4f^3~^4I       $ & 6.5 & 4.40E+7 & 7.69E+7 & 0.43 & 1440.32($_{-4.5}$)  && 8.54E+7 & 1378.09  && 6.95E+7\\                        
$4f^2(^1I)5d~^2I  $ & 6.5 & $4f^3~^2K       $ & 7.5 & 3.75E+7 & 4.99E+7 & 0.25 & 1566.59($_{-5.4}$)  && 7.14E+7 & 1485.64  && 6.43E+7\\                        
\enddata                                                                                                                                                                                      
\end{deluxetable*} 
%%%%%%%%%%%%%%%%%%%%%%%%%%%%%%%%%%%%%%%%%%%%%%%%%%%%%%

%%%%%%%%%%%%%%%%%%%%%%%%%%%%%%%%%%%%%%%%%%%%%%%%%%%%%%
% Section: Opacities
%%%%%%%%%%%%%%%%%%%%%%%%%%%%%%%%%%%%%%%%%%%%%%%%%%%%%%
\section{Impact to the Opacities}
\label{sec:opacity}

We calculate bound-bound opacities using our results
to study the impact of the accuracy in the atomic calculations.
By following previous works on NS mergers
\citep{kasen13,barnes13,tanaka13,tanaka14,tanaka18},
we use the formalism of expansion opacity 
\citep{karp77,eastman93,kasen06}:
\begin{equation}
\kappa_{\rm exp}^{\rm bb} (\lambda) = \frac{1}{\rho ct} 
\sum_l \frac{\lambda_l}{\Delta \lambda} (1 - e^{- \tau_l}).
\end{equation}
Here, $\rho$ and $t$ represent density and time after the merger.
The summation is taken over all the transitions in a wavelength bin
($\Delta \lambda$), and $\lambda_l$ and $\tau_l$ are the transition
wavelength and the Sobolev optical depth for each transition.
The Sobolev optical depth $\tau_l$ is expressed as
\begin{equation}
  \tau_l = \frac{\pi e^2}{m_e c} f_l n_{l} t \lambda_l
   = \frac{\pi e^2}{m_e c} \left( \frac{n \lambda_l t}{g_0} \right) g_l f_l {\rm e}^{-E_l/kT},
\end{equation}
where $g_l$, $E_l$, and $f_l$ are the statistical weight and
the energy of the lower level of the transition and the oscillator strength
of the transition, respectively.
For the oscillator strength, we use results computed with the length (Babushkin)
form.
For the number density in the lower level of the transition ($n_l$),
the Boltzmann distribution is assumed, i.e.,
$n_l = (g_l/g_0) n \exp(-E_l/kT)$, where
$g_0$ is the statistical weight for the ground level.
The number density of each ion $n$ is calculated under the assumption
of local thermodynamic equilibrium by using the Saha equation.
In this paper, pure Nd gas is assumed.
We use all the calculated transitions to evaluate the opacity
  without any selection based on the transition strengths,
  which was applied in full radiative transfer simulations \citep{tanaka17}.

\begin{figure*}
  \begin{tabular}{cc}
  %\plotone{Nd1_energies.pdf}
  \includegraphics[width=0.45\textwidth]{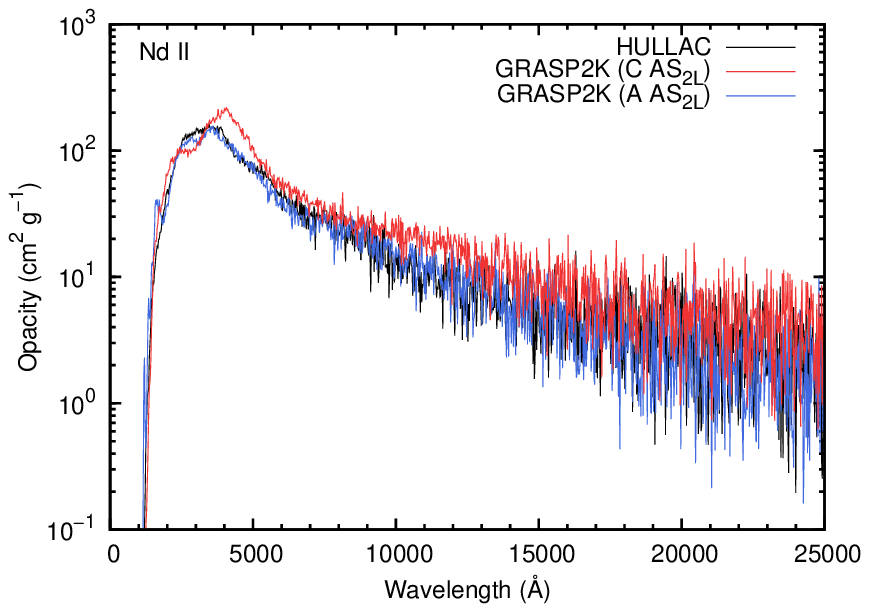}&
  \includegraphics[width=0.45\textwidth]{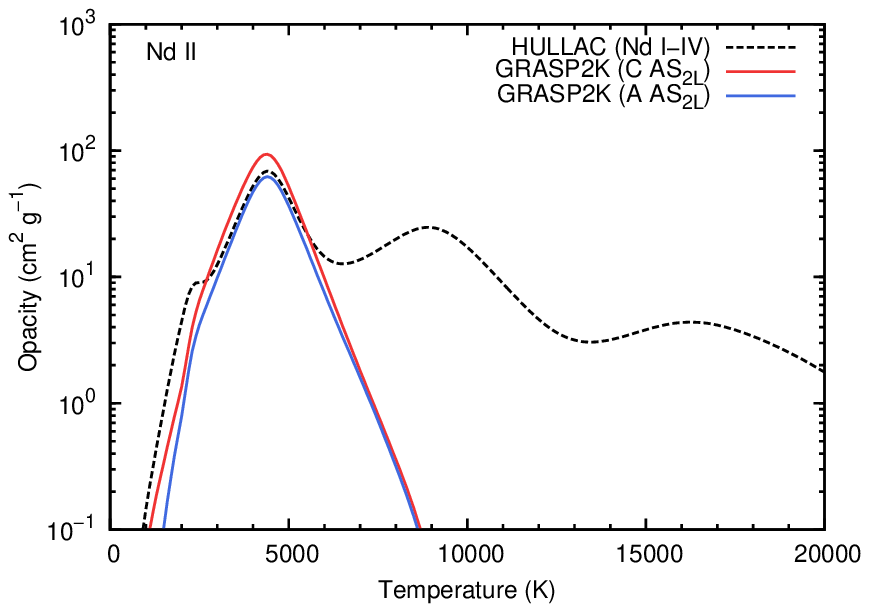}  \\
  \includegraphics[width=0.45\textwidth]{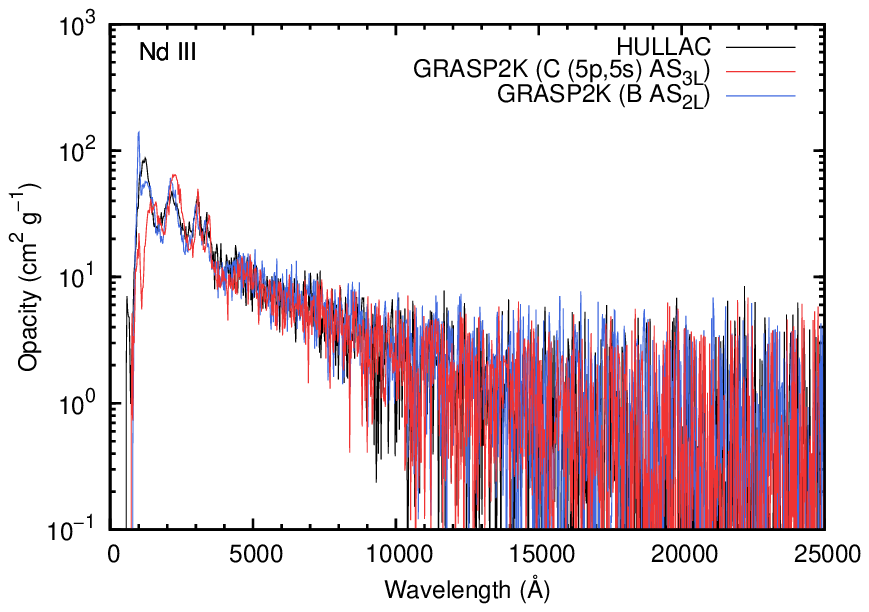}&
  \includegraphics[width=0.45\textwidth]{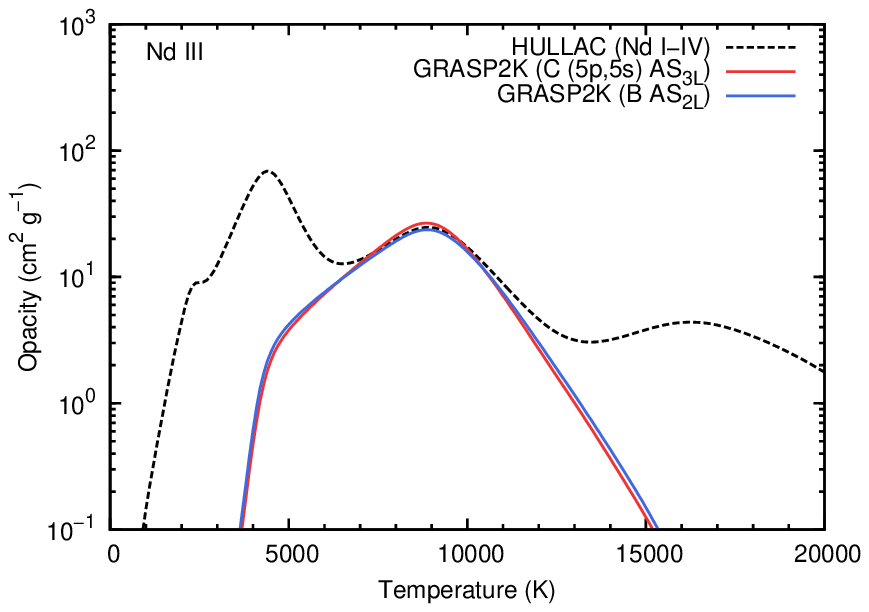}  \\
  \includegraphics[width=0.45\textwidth]{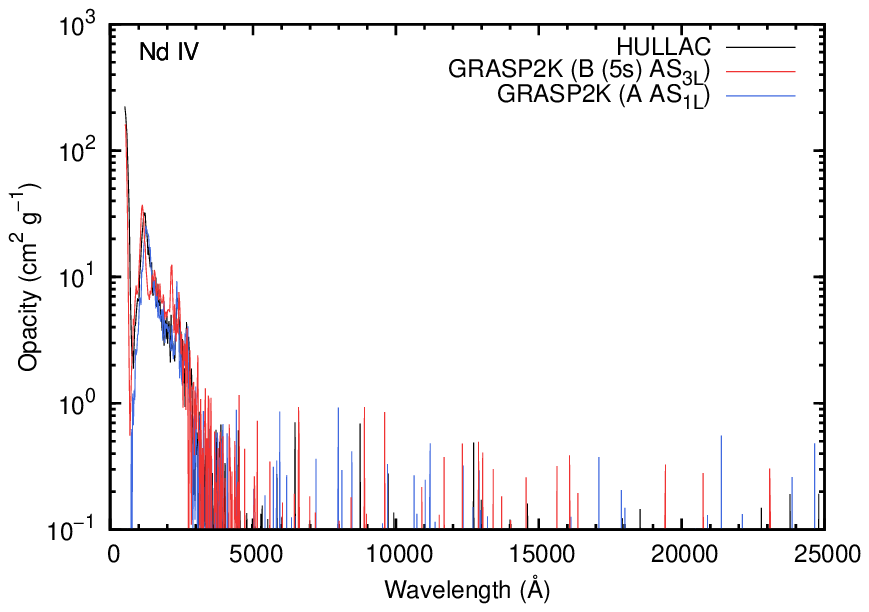}&
  \includegraphics[width=0.45\textwidth]{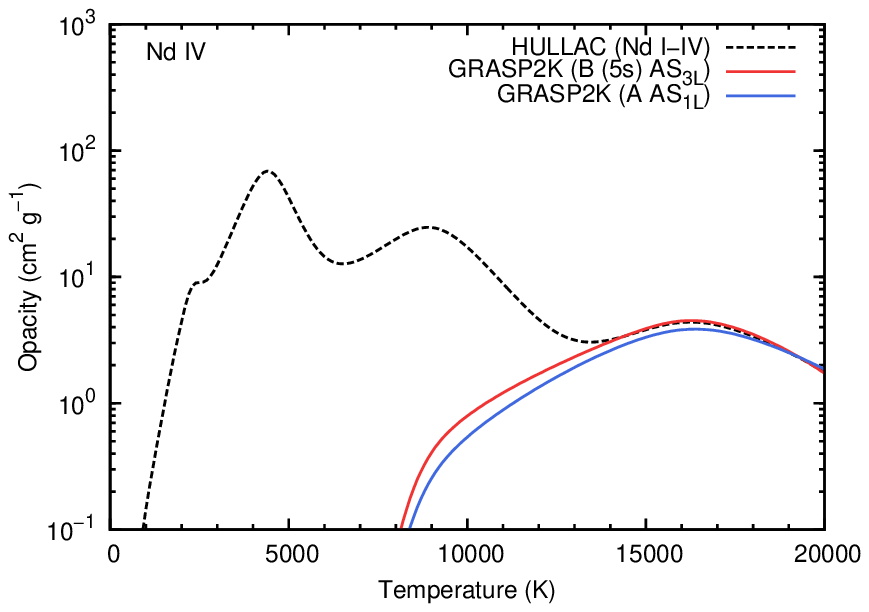}  
  \end{tabular}
  \caption{Opacities for Nd II (top), Nd III (middle), and Nd IV (bottom) ions.
    The left panels show the expansion opacities calculated with
    $T = 5000$ K, 10000 K, and 15000 K for Nd II, Nd III, and Nd IV, respectively.
    The density and time are assumed to be 
    $\rho = 1 \times 10^{-13} \ {\rm g \ cm^{-3}}$ and $t = 1$ day after the merger,
    respectively.
    The right panels show Planck mean opacities for various temperatures.
    The dashed curve shows the Planck mean opacities calculated with
    atomic data for Nd I-IV calculated with the HULLAC code \citep{tanaka18}.
    \label{fig:opacity}}
\end{figure*}

%\subsection{Nd II}
We find that overall properties of opacities are not dramatically
affected by the accuracies of the atomic calculations.
Left panels in figure \ref{fig:opacity} shows the expansion opacities
calculated by using transition data of Nd II, Nd III, and Nd IV.
The temperature is assumed to be 5000 K, 10000 K, and 15000 K for
Nd II, Nd III, and Nd IV, respectively.
The density is $1 \times 10^{-13} \ {\rm g \ cm^{-3}}$
and time after the merger is set to be 1 day.
Overall opacity values and wavelength dependence are quite similar
for different atomic calculations.
The red lines show the best results in this paper while 
blue lines show the previous results used by \cite{tanaka18}.
%Although some features show deviation (see below),

The behaviors of the opacity are also similar for different temperature.
Right panels show the Planck mean opacities calculated for
different temperatures by keeping the density and time to be the same.
The Planck mean opacity from different atomic calculations
agree with each other within a factor of 1.5.
Since the timescale of the kilonova emission scales as $\kappa^{0.5}$
\citep{rosswog15,tanaka16,fernandez16,metzger17},
this level of differences does not significantly 
affect the timescale of kilonova (smaller than $\sim$ 20\%)
compared with those expected from differences in temperature and abundances.

With a close look, however, the wavelength dependent opacities
show some differences.
The most notable difference is the feature around 4000 \AA\
in the case of Nd II.
The new GRASP2K calculations with a better accuracy show
a bump while the old GRASP2K calculations and HULLAC calculations
do not, making a difference in the opacity by a factor of 2.
Interestingly, \citet{kasen13} also showed that this part of the opacity
  is affected by the optimization in the atomic calculations:
  the peak is located near {5000 \AA} in their opt1 case
  while the peak is weaker in their opt2 and opt3 cases.
  The expansion opacities presented by \citet{fontes17}
  also show a peak around 4000 \AA, which is close to our new results.

\begin{figure}
  %\plotone{Nd1_energies.pdf}
  \includegraphics[width=0.45\textwidth]{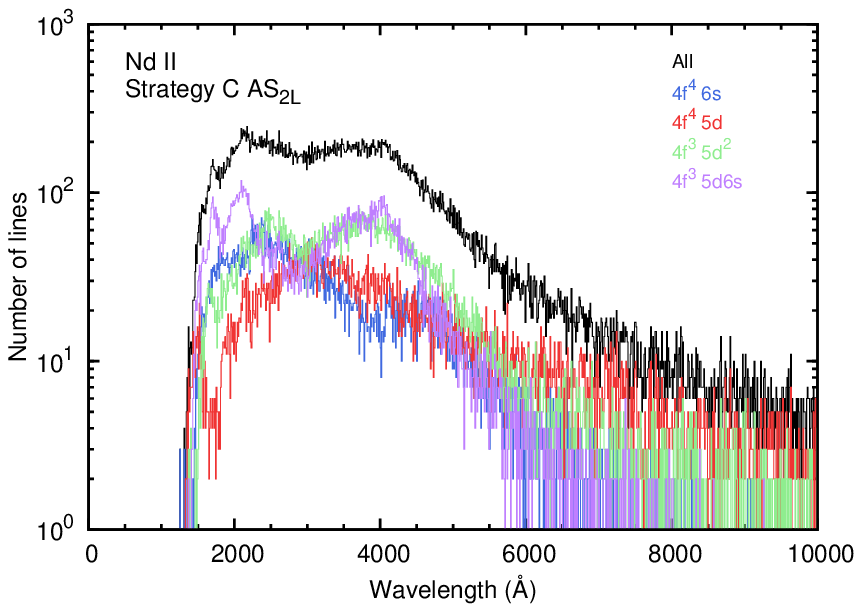}
  \includegraphics[width=0.45\textwidth]{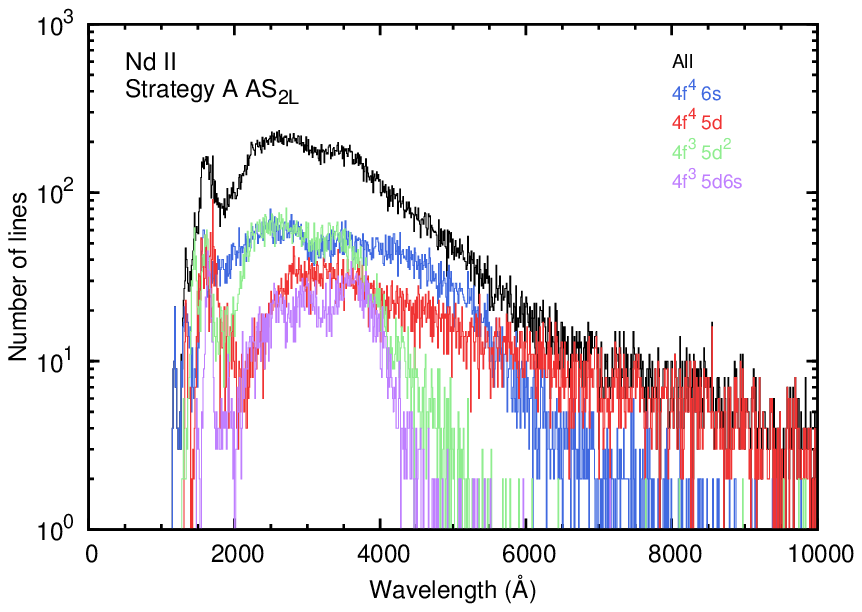}
  \includegraphics[width=0.45\textwidth]{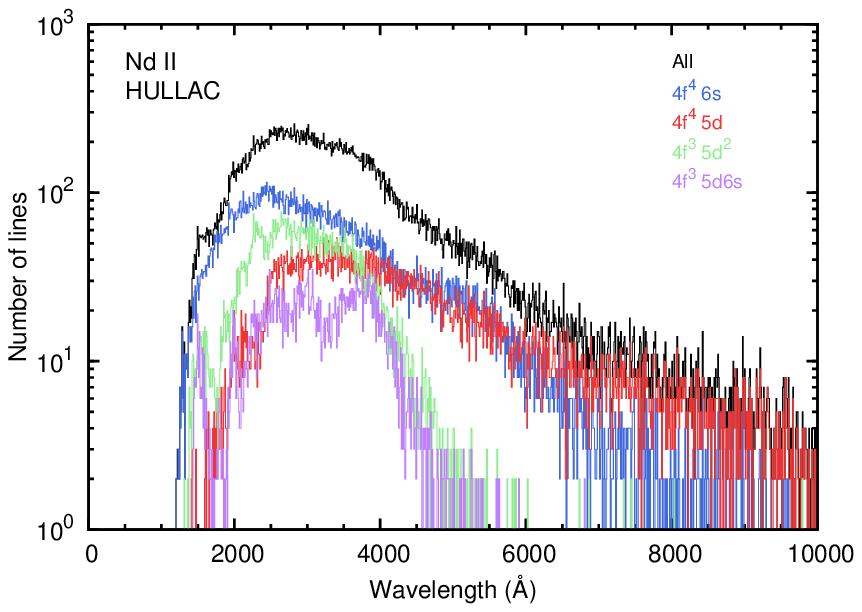} 
  \caption{Number of strong transitions of Nd II as a function of wavelengths.
    The strong transitions are selected by the criterion of
    $g_lf_l \exp(-E_l/kT) > 10^{-5}$ with $T = 5000$ K.
    The black lines show total number of strong transitions
    while color lines show transitions from each lower-level configuration.
    \label{fig:numline}}
\end{figure}

We find that the difference between our new and previous opacities
is caused by the lower energy levels of
$4f^35d^2$ and $4f^35d6s$ configurations in our new calculations
(Figure \ref{C_startegy_E_Nd_II}).
Figure \ref{fig:numline} shows the number of strong transitions
which fulfill $g_lf_l \exp(-E_l/kT) > 10^{-5}$ at $T = 5000$ K.
The numbers of transition are separated according to
the lower level configuration.
The number of strong transition
from the levels of $4f^35d^2$ and $4f^35d6s$ configuration
is enhanced around 4500 \AA\ in our new calculations (\textbf{Strategy C} $AS_{2L}$).
Since the energy of these configurations
are overestimated in our previous calculations (\textbf{Strategy A} $AS_{2L}$, Figure \ref{C_startegy_E_Nd_II}),
the bump structure in the new calculations seems more realistic.
This demonstrates the importance of accurate calculations
for lower energy levels to predict spectra of kilonova.

\begin{figure*}
 \includegraphics[width=1\textwidth]{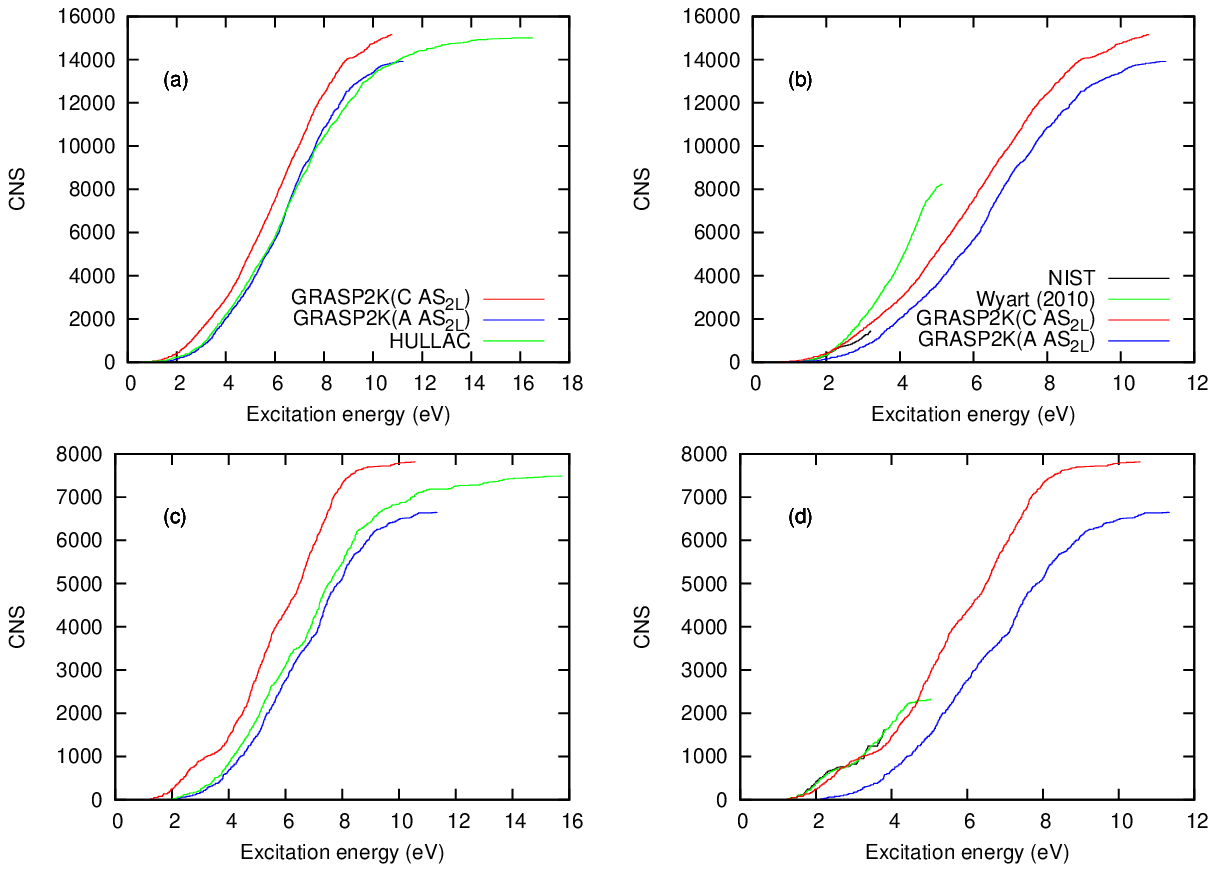}
 \caption{Cumulative number of states for $4f^35d^2$ (a,b) and $4f^35d6s$ (c,d) configurations
 as a function of excitation energy. (a,c): Results with the present and previous GRASP2K and HULLAC calculations. (b,d): Results with the present GRASP2K calculation, NIST, and the semi-empirical method by Wyart (2010).}
% * <kato.daiji@nifs.ac.jp> 2018-08-07T06:46:36.851Z:
%
% ^.
 \label{fig:CNS}
\end{figure*}

%{\bf MT: Please describe energy distributions here (Daiji)}
Figure \ref{fig:CNS} shows the cumulative number of states (CNS) for $4f^35d^2$ and $4f^35d6s$ configurations as a function of excitation energies. It is noted that the number of states takes the statistical-weight (degeneracy) of each level, {\it i.~e.}~$2J+1$, into account. The CNS for the whole energy levels obtained by calculations with GRASP2K and HULLAC is compared in the figure.
The CNS of the new calculations gets rising at a lower energy and has larger values than those of the previous calculations with GRASP2K and HULLAC, indicating that the larger number of states falls into the lower energy region with the new calculations. Since the Boltzmann distribution is assumed for the number density in the lower levels of transitions, it is predicted that the number of strong transitions from the levels of $4f^35d^2$ and $4f^35d6s$ configurations becomes larger with the new calculations as depicted in Fig.~\ref{fig:numline}. This is more remarkable for $4f^35d6s$ configuration as in the CNS.
The CNS of the new calculations is compared also with those of NIST and the semi-empirical results by Wyart (2010).
Overall agreement is good at low energies convincing accuracy of our new calculations. Only exception is the semi-empirical results for $4f^35d^2$ configuration which overshoot significantly at high energies. Reasons of this discrepancy are yet to be investigated.

%\subsection{Nd III}

Another notable difference is a feature around {1000~\AA}:
the opacities in our new calculations are suppressed.
This is due to the inclusion of highly excited energy levels
in the previous calculations (both GRASP2K and HULLAC).
Therefore, the opacities in the ultraviolet wavelengths
depends on the choice of the configurations included in the calculations.
However, if configurations with sufficiently high energy ($E \sim 10-15$ eV)
are included, such a difference appears only in the far ultraviolet
wavelength, and thus, does not affect observable features.

%\subsection{Nd IV}
%Opacities at 15,000 K \\
%Overall similarity \\
%Figure (Nd IV opacity at 15000 K) \\
%Figure (Planck mean) \\

%%%%%%%%%%%%%%%%%%%%%%%%%%%%%%%%%%%%%%%%%%%%%%%%%%%%%%
% Figure: Opacities
%%%%%%%%%%%%%%%%%%%%%%%%%%%%%%%%%%%%%%%%%%%%%%%%%%%%%%
%%%%%%%%%%%%%%%%%%%%%%%%%%%%%%%%%%%%%%%%%%%%%%%%%%%%%%

%%%%%%%%%%%%%%%%%%%%%%%%%%%%%%%%%%%%%%%%%%%%%%%%%%%%%%
% Figure: Opacities
%%%%%%%%%%%%%%%%%%%%%%%%%%%%%%%%%%%%%%%%%%%%%%%%%%%%%%

%%%%%%%%%%%%%%%%%%%%%%%%%%%%%%%%%%%%%%%%%%%%%%%%%%%%%%

%%%%%%%%%%%%%%%%%%%%%%%%%%%%%%%%%%%%%%%%%%%%%%%%%%%%%%
% Section: Summary
%%%%%%%%%%%%%%%%%%%%%%%%%%%%%%%%%%%%%%%%%%%%%%%%%%%%%%
\section{Summary}
\label{sec:summary}

We presented extensive atomic calculations of neodymium
and studied impact of accuracies in the calculations
to the astrophysical opacities.
The extended search of electron correlation effect inclusion strategies 
is presented in this work for the three Nd ions (Nd II, III and IV).
In total, 6 000, 1 453, 1 533 levels are 
presented for Nd II, Nd III and Nd IV respectively,
and E1 type transitions between these levels were computed.
Exclusive accuracy is achieved for atomic energy spectra results.
Compared with NIST database,
the averaged relative differences are 10 \%, 3 \%, and 11 \% for
Nd II, Nd III, and Nd IV, respectively.

Using our new results, we calculated expansion opacities
used in radiative transfer simulations for kilonova,
radioactively-powered EM emission from NS merger.
We found that the overall opacities values and their
wavelength dependence are not very sensitive
to the accuracies of the calculations.
The Planck mean opacities from our previous and new atomic calculations
agree within a factor of 1.5.
This confirms the validity of previous studies on kilonova.

However, some wavelength dependent features are 
affected by the accuracy of atomic calculations.
In particular, the low-lying energy levels ($E < 2-3$ eV)
can affect the opacities and even produce a bump in a certain wavelength range.
Our results highlight importance of accurate atomic calculations
for low-lying energy levels 
to accurately predict the spectra of kilonova.

\acknowledgments

This research was funded by a grant (No. S-LJB-18-1) from the Research Council of Lithuania.
This research was also supported by JSPS Bilateral Joint Research Project.
Computations presented in this paper were performed
at the High Performance Computing Center ``HPC Sauletekis'' of the
Faculty of Physics at Vilnius University
and with Cray XC30 and XC50 at Center for Computational Astrophysics,
National Astronomical Observatory of Japan.
DK is grateful to the support by NINS program of Promoting Research by Networking among Institutions (Grant Number 01411702).
MT is supported by the NINS program for cross-disciplinary science study, Inoue Science Research Award from Inoue Foundation for Science, and the Grant-in-Aid for Scientific Research from JSPS (15H02075, 16H02183) and MEXT (17H06363).

%%%%%%%%%%%%%%%%%%%%%%%%%%%%%%%%%%%%%%%begins of the machine readable tables%%%%%%%%%%%%%%%%%%%%%%%%%%%%%%%%%%%%%%%%%%%%%%%%%%%%%%%%%%%%%%į
\begin{deluxetable*}{rlrrr}
\tabletypesize{\footnotesize}
\setlength{\tabcolsep}{4pt}
\tablecaption{
Energy levels (in cm$^{-1}$) relative to the ground state for the lowest states of Nd~II. \label{Levels_NdII}}
\tablehead{ \colhead{No.} & \colhead{label}  & \colhead{$J$} & \colhead{P} & \colhead{$E$} }
\startdata
     1& $4f^4(^5_1I)6s~^6I          $ &  7/2& +    &    0.00  \\
     2& $4f^4(^5_1I)6s~^6I          $ &  9/2& +    &  547.86  \\
     3& $4f^4(^5_1I)6s~^6I          $ & 11/2& +    & 1404.36  \\
     4& $4f^4(^5_1I)6s~^4I          $ &  9/2& +    & 1789.52  \\
     5& $4f^4(^5_1I)6s~^6I          $ & 13/2& +    & 2425.58  \\
     6& $4f^4(^5_1I)6s~^4I          $ & 11/2& +    & 3120.96  \\
     7& $4f^4(^5_1I)6s~^6I          $ & 15/2& +    & 3564.21  \\
     8& $4f^4(^5_1I)5d~^6L          $ & 11/2& +    & 4019.24  \\
     9& $4f^4(^5_1I)6s~^4I          $ & 13/2& +    & 4506.07  \\
    10& $4f^4(^5_1I)6s~^6I          $ & 17/2& +    & 4789.43  \\
    11& $4f^4(^5_1I)5d~^6L          $ & 13/2& +    & 4941.03  \\
    12& $4f^3(^4_1I)5d^2(^3_2F)~^6M $ & 13/2& $-$  & 5477.69  \\
    13& $4f^4(^5_1I)6s~^4I          $ & 15/2& +    & 5940.53  \\
    14& $4f^4(^5_1I)5d~^6L          $ & 15/2& +    & 5965.42  \\
    15& $4f^4(^5_1I)5d~^6K          $ &  9/2& +    & 6065.11  \\
    16& $4f^3(^4_1I)5d^2(^3_2F)~^6M $ & 15/2& $-$  & 6750.71  \\
    17& $4f^4(^5_1I)5d~^6K          $ & 11/2& +    & 6881.31  \\
    18& $4f^4(^5_1I)5d~^6L          $ & 17/2& +    & 7078.84  \\
    19& $4f^4(^5_1I)5d~^6K          $ & 13/2& +    & 7802.57  \\
    20& $4f^3(^4_1I)5d^2(^3_2F)~^6M $ & 17/2& $-$  & 8147.61  \\
\enddata
\tablecomments{ Table~\ref{Levels_NdII} is published in its entirety in the 
machine-readable format. Part of the values are shown here for guidance regarding its form and content.}
\end{deluxetable*}

\begin{deluxetable*}{llrrllll}
\tabletypesize{\footnotesize}
\setlength{\tabcolsep}{4pt}
\tablecaption{Transition energies $\Delta E$ (in cm$^{-1}$), 
transition wavelengths $\lambda$ (in {\AA}), 
line strengths $S$ (in a.u.), 
weighted oscillator strengths $gf$ and 
transition rates $A$ (in s$^{-1}$) for E1 transitions of  Nd II ion. 
All transition data are in length form. 
$dT$ is the relative difference of the transition rates in length and 
velocity form as given by Eq.~(\ref{accuracy}). \label{Transition_NdII} }
\tablehead{ \colhead{upper} & \colhead{lower} &  \colhead{$\Delta E$} &
\colhead{$\lambda$} & \colhead{$S$} & \colhead{$gf$} & \colhead{$A$} & \colhead{$dT$}      }
\startdata
$4f^3 (^4_1I)\,5d^2 (^3_2F)~^6L_{11/2}$~  & $4f^3 (^4_1I)\,5d~^5K\,6p~^6L_{11/2}$  ~&    22745 &  4396.48 & 4.593E$+$00 & 3.173E$-$01 & 9.127E$+$06 &   0.050\\
$4f^3 (^4_1I)\,5d^2 (^3_2F)~^6L_{11/2}$~  & $4f^3 (^4_1I)\,5d~^5L\,6p~^6L_{11/2}$  ~&    23391 &  4275.02 & 1.001E$+$01 & 7.115E$-$01 & 2.164E$+$07 &   0.154\\
$4f^3 (^4_1I)\,5d^2 (^3_2F)~^6L_{11/2}$~  & $4f^3 (^4_1I)\,5d~^5K\,6p~^6K_{11/2}$  ~&    24959 &  4006.51 & 7.628E$-$01 & 5.783E$-$02 & 2.002E$+$06 &   0.130\\
$4f^3 (^4_1I)\,5d^2 (^3_2F)~^6L_{11/2}$~  & $4f^3 (^4_1I)\,5d~^5K\,6p~^6L_{11/2}$  ~&    25373 &  3941.14 & 1.380E$+$00 & 1.063E$-$01 & 3.807E$+$06 &   0.033\\
$4f^3 (^4_1I)\,5d^2 (^3_2F)~^6L_{11/2}$~  & $4f^3 (^4_1I)\,5d~^5K\,6p~^6I_{11/2}$  ~&    26882 &  3719.95 & 2.049E$-$01 & 1.673E$-$02 & 6.722E$+$05 &   0.041\\
$4f^3 (^4_1I)\,5d^2 (^3_2F)~^6L_{11/2}$~  & $4f^3 (^4_1I)\,5d~^5L\,6p~^6K_{11/2}$  ~&    27650 &  3616.62 & 1.213E$-$01 & 1.018E$-$02 & 4.330E$+$05 &   0.069\\
$4f^3 (^4_1I)\,5d^2 (^3_2F)~^6L_{11/2}$~  & $4f^3 (^4_1I)\,5d~^5K\,6p~^6I_{11/2}$  ~&    28842 &  3467.14 & 5.807E$-$02 & 5.088E$-$03 & 2.352E$+$05 &   0.351\\
$4f^3 (^4_1I)\,5d^2 (^3_2F)~^6L_{11/2}$~  & $4f^3 (^4_1I)\,5d~^5I\,6p~^6K_{11/2}$  ~&    29386 &  3402.93 & 3.867E$-$01 & 3.452E$-$02 & 1.657E$+$06 &   0.120\\
$4f^3 (^4_1I)\,5d^2 (^3_2F)~^6L_{11/2}$~  & $4f^3 (^4_1I)\,5d~^3I\,6p~^4K_{11/2}$  ~&    29450 &  3395.57 & 1.159E$-$02 & 1.037E$-$03 & 5.002E$+$04 &   0.124\\
$4f^3 (^4_1I)\,5d^2 (^3_2F)~^6L_{11/2}$~  & $4f^3 (^4_1I)\,5d~^5I\,6p~^6K_{11/2}$  ~&    30233 &  3307.58 & 5.567E$-$02 & 5.112E$-$03 & 2.597E$+$05 &   0.164\\
$4f^3 (^4_1I)\,5d^2 (^3_2F)~^6L_{11/2}$~  & $4f^3 (^4_1I)\,5d~^3I\,6p~^4H_{11/2}$  ~&    30410 &  3288.31 & 3.589E$-$02 & 3.316E$-$03 & 1.704E$+$05 &   0.320\\
$4f^3 (^4_1I)\,5d^2 (^3_2F)~^6L_{11/2}$~  & $4f^3 (^4_1I)\,5d~^5I\,6p~^6I_{11/2}$  ~&    30982 &  3227.58 & 2.441E$-$04 & 2.297E$-$05 & 1.226E$+$03 &   0.999\\
$4f^3 (^4_1I)\,5d^2 (^3_2F)~^6L_{11/2}$~  & $4f^3 (^4_1I)\,5d~^5I\,6p~^6I_{11/2}$  ~&    31155 &  3209.69 & 5.477E$-$02 & 5.184E$-$03 & 2.797E$+$05 &   0.186\\
$4f^3 (^4_1I)\,5d^2 (^3_2F)~^6L_{11/2}$~  & $4f^3 (^4_1I)\,5d~^5I\,6p~^4K_{11/2}$  ~&    31541 &  3170.46 & 3.420E$-$02 & 3.276E$-$03 & 1.812E$+$05 &   0.373\\
$4f^3 (^4_1I)\,5d^2 (^3_2F)~^6L_{11/2}$~  & $4f^3 (^4_1I)\,5d~^5I\,6p~^4H_{11/2}$  ~&    31848 &  3139.89 & 2.205E$-$04 & 2.133E$-$05 & 1.202E$+$03 &   0.541\\
$4f^3 (^4_1I)\,5d^2 (^3_2F)~^6L_{11/2}$~  & $4f^3 (^4_1I)\,5d~^3H\,6p~^4I_{11/2}$  ~&    32432 &  3083.34 & 5.245E$-$03 & 5.167E$-$04 & 3.021E$+$04 &   0.319\\
$4f^3 (^4_1I)\,5d^2 (^3_2F)~^6L_{11/2}$~  & $4f^3 (^4_1I)\,5d~^5I\,6p~^6H_{11/2}$  ~&    33244 &  3008.02 & 2.221E$-$03 & 2.242E$-$04 & 1.377E$+$04 &   0.427\\
$4f^3 (^4_1I)\,5d^2 (^3_2F)~^6L_{11/2}$~  & $4f^3 (^4_1I)\,5d~^5H\,6p~^6H_{11/2}$  ~&    33332 &  3000.04 & 1.985E$-$05 & 2.010E$-$06 & 1.241E$+$02 &   0.938\\
$4f^3 (^4_1I)\,5d^2 (^3_2F)~^6L_{11/2}$~  & $4f^3 (^4_1I)\,5d~^5H\,6p~^6G_{11/2}$  ~&    33384 &  2995.44 & 1.588E$-$04 & 1.611E$-$05 & 9.980E$+$02 &   0.446\\
$4f^3 (^4_1I)\,5d^2 (^3_2F)~^6L_{11/2}$~  & $4f^3 (^4_1I)\,5d~^5I\,6p~^6I_{11/2}$  ~&    33499 &  2985.08 & 8.480E$-$04 & 8.629E$-$05 & 5.383E$+$03 &   0.424\\
\enddata
\tablecomments{ Table~\ref{Transition_NdII} is published in 
its entirety in the machine-readable format. Part of the values are shown here for guidance regarding its form and content.}
\end{deluxetable*}

%%%%%%%%%%%%%%%%%%%%%%%%%%%%%%%%%%%%%%%%%%%%%%%%%%%%%%%%%%%%%%%%%%%%%%%%%%% Nd III energy and transitions data %%%%%%%%%%%%%%%%%%%%%%%%%%%%

\begin{deluxetable*}{rlrrr}
	\tabletypesize{\footnotesize}
	\setlength{\tabcolsep}{4pt}
	\tablecaption{
Energy levels (in cm$^{-1}$) relative to the ground state for the lowest states of Nd~III. \label{Levels_NdIII}}
	\tablehead{ \colhead{No.} & \colhead{label}  & \colhead{$J$} & \colhead{P} & \colhead{$E$} }
	\startdata
   1&  $4f^{4}(^5_1I)~^{5}I   $      &4&  + &        0.00  \\
   2&  $4f^{4}(^5_1I)~^{5}I   $      &5&  + &     1072.58  \\
   3&  $4f^{4}(^5_1I)~^{5}I   $      &6&  + &     2263.93  \\
   4&  $4f^{4}(^5_1I)~^{5}I   $      &7&  + &     3542.92  \\
   5&  $4f^{4}(^5_1I)~^{5}I   $      &8&  + &     4884.66  \\
   6&  $4f^{4}(^5_1F)~^{5}F   $      &1&  + &    11739.04  \\
   7&  $4f^{4}(^5_1F)~^{5}F   $      &2&  + &    12098.57  \\
   8&  $4f^{4}(^5_1F)~^{5}F   $      &3&  + &    12724.33  \\
   9&  $4f^{4}(^5_0S)~^{5}S   $      &2&  + &    13433.63  \\
  10&  $4f^{4}(^5_1F)~^{5}F   $      &4&  + &    13459.96  \\
  11&  $4f^{4}(^5_1F)~^{5}F   $      &5&  + &    14444.26  \\
  12&  $4f^{4}(^3_2K)~^{3}K   $      &6&  + &    15065.97  \\
  13&  $4f^{3}(^4_1I)5d~^{5}K $      &5& $-$&    15128.43  \\
  14&  $4f^{3}(^4_1I)5d~^{5}L $      &6& $-$&    15257.69  \\
  15&  $4f^{4}(^3_4H)~^{3}H   $      &4&  + &    16151.40  \\
  16&  $4f^{4}(^3_2K)~^{3}K   $      &7&  + &    16180.31  \\
  17&  $4f^{3}(^4_1I)5d~^{5}K $      &6& $-$&    16720.95  \\
  18&  $4f^{3}(^4_1I)5d~^{5}L $      &7& $-$&    16985.08  \\
  19&  $4f^{4}(^5_1G)~^{5}G   $      &2&  + &    17295.05  \\
  20&  $4f^{4}(^5_1G)~^{5}G   $      &3&  + &    17368.09  \\
		\enddata
\tablecomments{ Table~\ref{Levels_NdIII} is published in its entirety in the machine-readable format. Part of the values are shown here for guidance regarding its form and content.}
\end{deluxetable*}

\begin{deluxetable*}{llrrllll}
	\tabletypesize{\footnotesize}
	\setlength{\tabcolsep}{4pt}
	\tablecaption{Transition energies $\Delta E$ (in cm$^{-1}$), transition wavelengths $\lambda$ (in {\AA}), line strengths $S$ (in a.u.), weighted oscillator strengths $gf$ and transition rates $A$ (in s$^{-1}$) for E1 transitions of  Nd III ion. All transition data are in length form. $dT$ is the relative difference of the transition rates in length and velocity form as given by Eq.~(\ref{accuracy}). \label{Transition_NdIII} }
	\tablehead{ \colhead{upper} & \colhead{lower} &  \colhead{$\Delta E$} &
	\colhead{$\lambda$} & \colhead{$S$} & \colhead{$gf$} & \colhead{$A$} & \colhead{$dT$}      }
	\startdata
$4f^{4}(^3_2P)~^{3}P_{0}$      &      $4f^{3}(^4_1F)5d~^{5}D_{1}$    &       6521  &  15334.41 & 2.338E$-$03 & 4.632E$-$05 & 4.380E+02 & 0.948   \\
$4f^{4}(^3_2P)~^{3}P_{0}$      &      $4f^{3}(^4_1F)5d~^{3}P_{1}$    &       9156  &  10920.99 & 1.996E$-$02 & 5.553E$-$04 & 1.035E+04 & 0.861   \\
$4f^{4}(^3_2P)~^{3}P_{0}$      &      $4f^{3}(^4_1F)5d~^{5}P_{1}$    &      10206  &   9798.04 & 5.262E$-$03 & 1.631E$-$04 & 3.778E+03 & 0.787   \\
$4f^{4}(^3_2P)~^{3}P_{0}$      &      $4f^{3}(^4_1F)5d~^{5}F_{1}$    &      10697  &   9348.19 & 8.930E$-$03 & 2.901E$-$04 & 7.382E+03 & 0.822   \\
$4f^{4}(^3_2P)~^{3}P_{0}$      &      $4f^{3}(^4_1S)5d~^{5}D_{1}$    &      11516  &   8683.29 & 3.115E$-$05 & 1.090E$-$06 & 3.214E+01 & 0.890   \\
$4f^{4}(^3_2P)~^{3}P_{0}$      &      $4f^{3}(^4_1S)5d~^{3}D_{1}$    &      12113  &   8255.57 & 1.272E$-$02 & 4.681E$-$04 & 1.527E+04 & 0.752   \\
$4f^{4}(^3_2P)~^{3}P_{0}$      &      $4f^{3}(^4_1F)5d~^{3}D_{1}$    &      14250  &   7017.52 & 9.589E$-$02 & 4.150E$-$03 & 1.874E+05 & 0.710   \\
$4f^{4}(^3_2P)~^{3}P_{0}$      &      $4f^{3}(^2_1G)5d~^{3}D_{1}$    &      17425  &   5738.84 & 7.686E$-$02 & 4.068E$-$03 & 2.746E+05 & 0.647   \\
$4f^{4}(^3_2P)~^{3}P_{0}$      &      $4f^{3}(^4_1G)5d~^{5}F_{1}$    &      18400  &   5434.62 & 5.429E$-$04 & 3.034E$-$05 & 2.284E+03 & 0.634   \\
$4f^{4}(^3_2P)~^{3}P_{0}$      &      $4f^{3}(^2_1P)5d~^{3}P_{1}$    &      19753  &   5062.36 & 1.351E$-$02 & 8.110E$-$04 & 7.036E+04 & 0.786   \\
$4f^{4}(^3_2P)~^{3}P_{0}$      &      $4f^{3}(^4_1G)5d~^{5}D_{1}$    &      20499  &   4878.06 & 4.216E$-$03 & 2.625E$-$04 & 2.453E+04 & 0.562   \\
$4f^{4}(^3_2P)~^{3}P_{0}$      &      $4f^{3}(^2_1D)5d~^{3}P_{1}$    &      20588  &   4857.14 & 6.608E$-$02 & 4.132E$-$03 & 3.894E+05 & 0.679   \\
$4f^{4}(^3_2P)~^{3}P_{0}$      &      $4f^{3}(^2_1D)5d~^{3}S_{1}$    &      21850  &   4576.58 & 4.184E$-$02 & 2.777E$-$03 & 2.948E+05 & 0.711   \\
$4f^{4}(^3_2P)~^{3}P_{0}$      &      $4f^{3}(^4_1F)6s~^{5}F_{1}$    &      22166  &   4511.37 & 1.708E$-$05 & 1.150E$-$06 & 1.257E+02 & 0.812   \\
$4f^{4}(^3_2P)~^{3}P_{0}$      &      $4f^{3}(^2_1D)5d~^{3}S_{1}$    &      22582  &   4428.21 & 7.961E$-$03 & 5.461E$-$04 & 6.192E+04 & 0.707   \\
$4f^{4}(^3_2P)~^{3}P_{0}$      &      $4f^{3}(^2_1P)5d~^{3}D_{1}$    &      23964  &   4172.76 & 1.943E$-$02 & 1.415E$-$03 & 1.806E+05 & 0.137   \\
$4f^{4}(^3_2P)~^{3}P_{0}$      &      $4f^{3}(^4_1S)6s~^{3}S_{1}$    &      24808  &   4030.93 & 4.263E$-$04 & 3.213E$-$05 & 4.396E+03 & 0.318   \\
$4f^{4}(^3_2P)~^{3}P_{0}$      &      $4f^{3}(^4_1S)6s~^{3}S_{1}$    &      25324  &   3948.70 & 3.464E$-$03 & 2.665E$-$04 & 3.800E+04 & 0.784   \\
$4f^{4}(^3_2P)~^{3}P_{0}$      &      $4f^{3}(^2_1D)5d~^{1}P_{1}$    &      26223  &   3813.37 & 2.834E$-$03 & 2.257E$-$04 & 3.452E+04 & 0.389   \\
$4f^{4}(^3_2P)~^{3}P_{0}$      &      $4f^{3}(^4_1D)5d~^{5}F_{1}$    &      27992  &   3572.45 & 4.279E$-$09 & 3.638E$-$10 & 6.339E-02 & 0.999   \\
	\enddata
	\tablecomments{ Table~\ref{Transition_NdIII} is published in its entirety in the machine-readable format. Part of the values are shown here for guidance regarding its form and content.}
\end{deluxetable*}

\begin{deluxetable*}{rlrrr}
	\tabletypesize{\footnotesize}
	\setlength{\tabcolsep}{4pt}
	\tablecaption{
Energy levels (in cm$^{-1}$) relative to the ground state for the lowest states of Nd~IV. \label{Levels_NdIV}}
	\tablehead{ \colhead{No.} & \colhead{label}  & \colhead{$J$} & \colhead{P} & \colhead{$E$} }
	\startdata
   1&  $4f^{3}(^4_1I)~^{4}I$ &      9/2  &$-$&       0.00  \\
   2&  $4f^{3}(^4_1I)~^{4}I$ &     11/2  &$-$&    1748.59  \\
   3&  $4f^{3}(^4_1I)~^{4}I$ &     13/2  &$-$&    3627.05  \\
   4&  $4f^{3}(^4_1I)~^{4}I$ &     15/2  &$-$&    5595.78  \\
   5&  $4f^{3}(^4_1F)~^{4}F$ &      3/2  &$-$&   13076.19  \\
   6&  $4f^{3}(^2_2H)~^{2}H$ &      9/2  &$-$&   13536.44  \\
   7&  $4f^{3}(^4_1F)~^{4}F$ &      5/2  &$-$&   14022.10  \\
   8&  $4f^{3}(^4_1S)~^{4}S$ &      3/2  &$-$&   14617.16  \\
   9&  $4f^{3}(^4_1F)~^{4}F$ &      7/2  &$-$&   14911.07  \\
  10&  $4f^{3}(^4_1F)~^{4}F$ &      9/2  &$-$&   15979.23  \\
  11&  $4f^{3}(^2_2H)~^{2}H$ &     11/2  &$-$&   16581.13  \\
  12&  $4f^{3}(^2_1G)~^{2}G$ &      7/2  &$-$&   18780.28  \\
  13&  $4f^{3}(^4_1G)~^{4}G$ &      5/2  &$-$&   19780.48  \\
  14&  $4f^{3}(^4_1G)~^{4}G$ &      9/2  &$-$&   21211.00  \\
  15&  $4f^{3}(^4_1G)~^{4}G$ &      7/2  &$-$&   21217.90  \\
  16&  $4f^{3}(^2_1K)~^{2}K$ &     13/2  &$-$&   21444.78  \\
  17&  $4f^{3}(^4_1G)~^{4}G$ &      9/2  &$-$&   22708.88  \\
  18&  $4f^{3}(^2_1D)~^{2}D$ &      3/2  &$-$&   23160.03  \\
  19&  $4f^{3}(^2_1K)~^{2}K$ &     15/2  &$-$&   23397.52  \\
  20&  $4f^{3}(^4_1G)~^{4}G$ &     11/2  &$-$&   23768.32  \\
			\enddata
\tablecomments{ Table~\ref{Levels_NdIV} is published in its entirety in the machine-readable format. Part of the values are shown here for guidance regarding its form and content.}
\end{deluxetable*}

\begin{deluxetable*}{llrrllll}
	\tabletypesize{\footnotesize}
	\setlength{\tabcolsep}{4pt}
	\tablecaption{Transition energies $\Delta E$ (in cm$^{-1}$), transition wavelengths $\lambda$ (in {\AA}), line strengths $S$ (in a.u.), weighted oscillator strengths $gf$ and transition rates $A$ (in s$^{-1}$) for E1 transitions of  Nd IV ion. All transition data are in length form. $dT$ is the relative difference of the transition rates in length and velocity form as given by Eq.~(\ref{accuracy}). \label{Transition_NdIV} }
	\tablehead{ \colhead{upper} & \colhead{lower} &  \colhead{$\Delta E$} &
	\colhead{$\lambda$} & \colhead{$S$} & \colhead{$gf$} & \colhead{$A$} & \colhead{$dT$}      }
	\startdata
$4f^{3}(^2_1P)5p^{6}~^{2}P_{1/2}  $    & $4f^{2}(^3_1F)5p^{6}5d~^{2}P_{1/2}$  &   52003 &  1922.95 & 3.607E$-$02 & 5.698E$-$03 & 5.140E+06 & 0.422 \\
$4f^{3}(^4_1D)5p^{6}~^{4}D_{1/2}  $    & $4f^{2}(^3_1F)5p^{6}5d~^{2}P_{1/2}$  &   44283 &  2258.20 & 7.141E$-$03 & 9.606E$-$04 & 6.283E+05 & 0.235 \\
$4f^{2}(^3_1F)5p^{6}5d~^{2}P_{1/2}$    & $4f^{4}(^5_1F)5p^{5}~^{6}D_{1/2}  $  &   83619 &  1195.89 & 3.829E$-$07 & 9.725E$-$08 & 2.268E+02 & 0.945 \\
$4f^{2}(^3_1F)5p^{6}5d~^{2}P_{1/2}$    & $4f^{4}(^5_1G)5p^{5}~^{6}F_{1/2}  $  &   85201 &  1173.69 & 2.099E$-$05 & 5.432E$-$06 & 1.315E+04 & 0.116 \\
$4f^{2}(^3_1F)5p^{6}5d~^{2}P_{1/2}$    & $4f^{2}(^3_1F)5p^{6}6p~^{4}D_{1/2}$  &   92307 &  1083.33 & 1.546E$-$05 & 4.336E$-$06 & 1.232E+04 & 0.249 \\
$4f^{2}(^3_1F)5p^{6}5d~^{2}P_{1/2}$    & $4f^{4}(^3_1D)5p^{5}~^{4}P_{1/2}  $  &   96617 &  1035.01 & 2.052E$-$05 & 6.022E$-$06 & 1.875E+04 & 0.739 \\
$4f^{2}(^3_1F)5p^{6}5d~^{2}P_{1/2}$    & $4f^{4}(^3_3F)5p^{5}~^{4}D_{1/2}  $  &   99613 &  1003.88 & 9.090E$-$05 & 2.750E$-$05 & 9.102E+04 & 0.448 \\
$4f^{2}(^3_1F)5p^{6}5d~^{2}P_{1/2}$    & $4f^{4}(^3_4F)5p^{5}~^{4}D_{1/2}  $  &  103718 &   964.15 & 2.294E$-$06 & 7.228E$-$07 & 2.593E+03 & 0.625 \\
$4f^{2}(^3_1F)5p^{6}5d~^{2}P_{1/2}$    & $4f^{4}(^1_3D)5p^{5}~^{2}P_{1/2}  $  &  105216 &   950.42 & 5.732E$-$04 & 1.832E$-$04 & 6.764E+05 & 0.229 \\
$4f^{2}(^3_1F)5p^{6}5d~^{2}P_{1/2}$    & $4f^{2}(^1_1D)5p^{6}6p~^{2}P_{1/2}$  &  108119 &   924.90 & 1.258E$-$02 & 4.132E$-$03 & 1.611E+07 & 0.069 \\
$4f^{2}(^3_1F)5p^{6}5d~^{2}P_{1/2}$    & $4f^{4}(^5_1D)5p^{5}~^{6}D_{1/2}  $  &  109048 &   917.03 & 6.412E$-$04 & 2.124E$-$04 & 8.424E+05 & 0.310 \\
$4f^{2}(^3_1F)5p^{6}5d~^{2}P_{1/2}$    & $4f^{2}(^3_1P)5p^{6}6p~^{4}D_{1/2}$  &  109666 &   911.85 & 4.762E$-$06 & 1.586E$-$06 & 6.363E+03 & 0.182 \\
$4f^{2}(^3_1F)5p^{6}5d~^{2}P_{1/2}$    & $4f^{2}(^3_1P)5p^{6}6p~^{2}S_{1/2}$  &  110081 &   908.41 & 5.629E$-$05 & 1.882E$-$05 & 7.608E+04 & 0.611 \\
$4f^{2}(^3_1F)5p^{6}5d~^{2}P_{1/2}$    & $4f^{4}(^5_1D)5p^{5}~^{6}D_{1/2}  $  &  110468 &   905.24 & 1.101E$-$03 & 3.695E$-$04 & 1.503E+06 & 0.170 \\
$4f^{2}(^3_1F)5p^{6}5d~^{2}P_{1/2}$    & $4f^{4}(^3_1D)5p^{5}~^{4}D_{1/2}  $  &  112421 &   889.51 & 1.612E$-$03 & 5.508E$-$04 & 2.321E+06 & 0.089 \\
$4f^{2}(^3_1F)5p^{6}5d~^{2}P_{1/2}$    & $4f^{4}(^5_0S)5p^{5}~^{4}P_{1/2}  $  &  113266 &   882.87 & 1.875E$-$04 & 6.451E$-$05 & 2.760E+05 & 0.153 \\
$4f^{2}(^3_1F)5p^{6}5d~^{2}P_{1/2}$    & $4f^{2}(^3_1P)5p^{6}6p~^{4}P_{1/2}$  &  115617 &   864.92 & 3.828E$-$03 & 1.344E$-$03 & 5.994E+06 & 0.192 \\
$4f^{2}(^3_1F)5p^{6}5d~^{2}P_{1/2}$    & $4f^{4}(^3_2F)5p^{5}~^{4}D_{1/2}  $  &  116976 &   854.87 & 6.502E$-$05 & 2.310E$-$05 & 1.054E+05 & 0.454 \\
$4f^{2}(^3_1F)5p^{6}5d~^{2}P_{1/2}$    & $4f^{2}(^3_1P)5p^{6}6p~^{2}P_{1/2}$  &  118179 &   846.17 & 3.156E$-$03 & 1.133E$-$03 & 5.277E+06 & 0.251 \\
$4f^{2}(^3_1F)5p^{6}5d~^{2}P_{1/2}$    & $4f^{4}(^3_2P)5p^{5}~^{4}P_{1/2}  $  &  119038 &   840.07 & 5.618E$-$03 & 2.031E$-$03 & 9.600E+06 & 0.113 \\
	\enddata
	\tablecomments{ Table~\ref{Transition_NdIV} is published in its entirety in the machine-readable format. Part of the values are shown here for guidance regarding its form and content.}
\end{deluxetable*}   

\bibliography{reference}

\begin{thebibliography}{}
\expandafter\ifx\csname natexlab\endcsname\relax\def\natexlab#1{#1}\fi
\providecommand{\url}[1]{\href{#1}{#1}}

\bibitem[{{Abbott} {et~al.}(2017{\natexlab{a}}){Abbott}, {Abbott}, {Abbott},
  {Acernese}, {Ackley}, {Adams}, {Adams}, {Addesso}, {Adhikari}, {Adya}, \&
  et~al.}]{abbott17}
{Abbott}, B.~P., {Abbott}, R., {Abbott}, T.~D., {et~al.} 2017{\natexlab{a}},
  Physical Review Letters, 119, 161101

\bibitem[{{Abbott} {et~al.}(2017{\natexlab{b}}){Abbott}, {Abbott}, {Abbott},
  {Acernese}, {Ackley}, {Adams}, {Adams}, {Addesso}, {Adhikari}, {Adya}, \&
  et~al.}]{abbott17MMA}
---. 2017{\natexlab{b}}, \apjl, 848, L12

\bibitem[{{Andreoni} {et~al.}(2017){Andreoni}, {Ackley}, {Cooke}, {Acharyya},
  {Allison}, {Anderson}, {Ashley}, {Baade}, {Bailes}, {Bannister}, {Beardsley},
  {Bessell}, {Bian}, {Bland}, {Boer}, {Booler}, {Brandeker}, {Brown},
  {Buckley}, {Chang}, {Coward}, {Crawford}, {Crisp}, {Crosse}, {Cucchiara},
  {Cup{\'a}k}, {de Gois}, {Deller}, {Devillepoix}, {Dobie}, {Elmer}, {Emrich},
  {Farah}, {Farrell}, {Franzen}, {Gaensler}, {Galloway}, {Gendre}, {Giblin},
  {Goobar}, {Green}, {Hancock}, {Hartig}, {Howell}, {Horsley}, {Hotan},
  {Howie}, {Hu}, {Hu}, {James}, {Johnston}, {Johnston-Hollitt}, {Kaplan},
  {Kasliwal}, {Keane}, {Kenney}, {Klotz}, {Lau}, {Laugier}, {Lenc}, {Li},
  {Liang}, {Lidman}, {Luvaul}, {Lynch}, {Ma}, {Macpherson}, {Mao},
  {McClelland}, {McCully}, {M{\"o}ller}, {Morales}, {Morris}, {Murphy},
  {Noysena}, {Onken}, {Orange}, {Os{\l}owski}, {Pallot}, {Paxman}, {Potter},
  {Pritchard}, {Raja}, {Ridden-Harper}, {Romero-Colmenero}, {Sadler}, {Sansom},
  {Scalzo}, {Schmidt}, {Scott}, {Seghouani}, {Shang}, {Shannon}, {Shao},
  {Shara}, {Sharp}, {Sokolowski}, {Sollerman}, {Staff}, {Steele}, {Sun},
  {Suntzeff}, {Tao}, {Tingay}, {Towner}, {Thierry}, {Trott}, {Tucker},
  {V{\"a}is{\"a}nen}, {Krishnan}, {Walker}, {Wang}, {Wang}, {Wayth}, {Whiting},
  {Williams}, {Williams}, {Wolf}, {Wu}, {Wu}, {Yang}, {Yuan}, {Zhang}, {Zhou},
  \& {Zovaro}}]{andreoni17}
{Andreoni}, I., {Ackley}, K., {Cooke}, J., {et~al.} 2017, \pasa, 34, e069

\bibitem[{{Arcavi} {et~al.}(2017){Arcavi}, {Hosseinzadeh}, {Howell}, {McCully},
  {Poznanski}, {Kasen}, {Barnes}, {Zaltzman}, {Vasylyev}, {Maoz}, \&
  {Valenti}}]{arcavi17}
{Arcavi}, I., {Hosseinzadeh}, G., {Howell}, D.~A., {et~al.} 2017, \nat, 551, 64

\bibitem[{{Barnes} \& {Kasen}(2013)}]{barnes13}
{Barnes}, J., \& {Kasen}, D. 2013, \apj, 775, 18

\bibitem[{{Chornock} {et~al.}(2017){Chornock}, {Berger}, {Kasen},
  {Cowperthwaite}, {Nicholl}, {Villar}, {Alexander}, {Blanchard}, {Eftekhari},
  {Fong}, {Margutti}, {Williams}, {Annis}, {Brout}, {Brown}, {Chen}, {Drout},
  {Farr}, {Foley}, {Frieman}, {Fryer}, {Herner}, {Holz}, {Kessler}, {Matheson},
  {Metzger}, {Quataert}, {Rest}, {Sako}, {Scolnic}, {Smith}, \&
  {Soares-Santos}}]{chornock17}
{Chornock}, R., {Berger}, E., {Kasen}, D., {et~al.} 2017, \apjl, 848, L19

\bibitem[{{Coulter} {et~al.}(2017){Coulter}, {Foley}, {Kilpatrick}, {Drout},
  {Piro}, {Shappee}, {Siebert}, {Simon}, {Ulloa}, {Kasen}, {Madore},
  {Murguia-Berthier}, {Pan}, {Prochaska}, {Ramirez-Ruiz}, {Rest}, \&
  {Rojas-Bravo}}]{coulter17}
{Coulter}, D.~A., {Foley}, R.~J., {Kilpatrick}, C.~D., {et~al.} 2017, Science,
  358, 1556

\bibitem[{Cowan(1981)}]{Cowan}
Cowan, R. 1981, The Theory of Atomic Structure and Spectra (University of
  California Press, Berkeley, CA)

\bibitem[{{Cowperthwaite} {et~al.}(2017){Cowperthwaite}, {Berger}, {Villar},
  {Metzger}, {Nicholl}, {Chornock}, {Blanchard}, {Fong}, {Margutti},
  {Soares-Santos}, {Alexander}, {Allam}, {Annis}, {Brout}, {Brown}, {Butler},
  {Chen}, {Diehl}, {Doctor}, {Drout}, {Eftekhari}, {Farr}, {Finley}, {Foley},
  {Frieman}, {Fryer}, {Garc{\'{\i}}a-Bellido}, {Gill}, {Guillochon}, {Herner},
  {Holz}, {Kasen}, {Kessler}, {Marriner}, {Matheson}, {Neilsen}, {Quataert},
  {Palmese}, {Rest}, {Sako}, {Scolnic}, {Smith}, {Tucker}, {Williams},
  {Balbinot}, {Carlin}, {Cook}, {Durret}, {Li}, {Lopes}, {Louren{\c c}o},
  {Marshall}, {Medina}, {Muir}, {Mu{\~n}oz}, {Sauseda}, {Schlegel}, {Secco},
  {Vivas}, {Wester}, {Zenteno}, {Zhang}, {Abbott}, {Banerji}, {Bechtol},
  {Benoit-L{\'e}vy}, {Bertin}, {Buckley-Geer}, {Burke}, {Capozzi}, {Carnero
  Rosell}, {Carrasco Kind}, {Castander}, {Crocce}, {Cunha}, {D'Andrea}, {da
  Costa}, {Davis}, {DePoy}, {Desai}, {Dietrich}, {Drlica-Wagner}, {Eifler},
  {Evrard}, {Fernandez}, {Flaugher}, {Fosalba}, {Gaztanaga}, {Gerdes},
  {Giannantonio}, {Goldstein}, {Gruen}, {Gruendl}, {Gutierrez}, {Honscheid},
  {Jain}, {James}, {Jeltema}, {Johnson}, {Johnson}, {Kent}, {Krause}, {Kron},
  {Kuehn}, {Nuropatkin}, {Lahav}, {Lima}, {Lin}, {Maia}, {March}, {Martini},
  {McMahon}, {Menanteau}, {Miller}, {Miquel}, {Mohr}, {Neilsen}, {Nichol},
  {Ogando}, {Plazas}, {Roe}, {Romer}, {Roodman}, {Rykoff}, {Sanchez},
  {Scarpine}, {Schindler}, {Schubnell}, {Sevilla-Noarbe}, {Smith}, {Smith},
  {Sobreira}, {Suchyta}, {Swanson}, {Tarle}, {Thomas}, {Thomas}, {Troxel},
  {Vikram}, {Walker}, {Wechsler}, {Weller}, {Yanny}, \&
  {Zuntz}}]{cowperthwaite17}
{Cowperthwaite}, P.~S., {Berger}, E., {Villar}, V.~A., {et~al.} 2017, \apjl,
  848, L17

\bibitem[{{D{\'{\i}}az} {et~al.}(2017){D{\'{\i}}az}, {Macri}, {Garcia Lambas},
  {Mendes de Oliveira}, {Nilo Castell{\'o}n}, {Ribeiro}, {S{\'a}nchez},
  {Schoenell}, {Abramo}, {Akras}, {Alcaniz}, {Artola}, {Beroiz}, {Bonoli},
  {Cabral}, {Camuccio}, {Castillo}, {Chavushyan}, {Coelho}, {Colazo},
  {Costa-Duarte}, {Cuevas Larenas}, {DePoy}, {Dom{\'{\i}}nguez Romero},
  {Dultzin}, {Fern{\'a}ndez}, {Garc{\'{\i}}a}, {Girardini}, {Gon{\c c}alves},
  {Gon{\c c}alves}, {Gurovich}, {Jim{\'e}nez-Teja}, {Kanaan}, {Lares}, {Lopes
  de Oliveira}, {L{\'o}pez-Cruz}, {Marshall}, {Melia}, {Molino}, {Padilla},
  {Pe{\~n}uela}, {Placco}, {Qui{\~n}ones}, {Ram{\'{\i}}rez Rivera}, {Renzi},
  {Riguccini}, {R{\'{\i}}os-L{\'o}pez}, {Rodriguez}, {Sampedro}, {Schneiter},
  {Sodr{\'e}}, {Starck}, {Torres-Flores}, {Tornatore}, \&
  {Zadro{\.z}ny}}]{diaz17}
{D{\'{\i}}az}, M.~C., {Macri}, L.~M., {Garcia Lambas}, D., {et~al.} 2017,
  \apjl, 848, L29

\bibitem[{{Drout} {et~al.}(2017){Drout}, {Piro}, {Shappee}, {Kilpatrick},
  {Simon}, {Contreras}, {Coulter}, {Foley}, {Siebert}, {Morrell}, {Boutsia},
  {Di Mille}, {Holoien}, {Kasen}, {Kollmeier}, {Madore}, {Monson},
  {Murguia-Berthier}, {Pan}, {Prochaska}, {Ramirez-Ruiz}, {Rest}, {Adams},
  {Alatalo}, {Ba{\~n}ados}, {Baughman}, {Beers}, {Bernstein}, {Bitsakis},
  {Campillay}, {Hansen}, {Higgs}, {Ji}, {Maravelias}, {Marshall}, {Bidin},
  {Prieto}, {Rasmussen}, {Rojas-Bravo}, {Strom}, {Ulloa},
  {Vargas-Gonz{\'a}lez}, {Wan}, \& {Whitten}}]{drout17}
{Drout}, M.~R., {Piro}, A.~L., {Shappee}, B.~J., {et~al.} 2017, Science, 358,
  1570

\bibitem[{Dyall {et~al.}(1989)Dyall, Grant, Johnson, Parpia, \& Plummer}]{EOL}
Dyall, K., Grant, I., Johnson, C., Parpia, F., \& Plummer, E. 1989, Computer
  Physics Communications, 55, 425

\bibitem[{Dzuba {et~al.}(2003)Dzuba, Safronova, \& Johnson}]{Dzuba}
Dzuba, V.~A., Safronova, U.~I., \& Johnson, W.~R. 2003, Phys. Rev. A, 68,
  032503

\bibitem[{{Eastman} \& {Pinto}(1993)}]{eastman93}
{Eastman}, R.~G., \& {Pinto}, P.~A. 1993, \apj, 412, 731

\bibitem[{{Evans} {et~al.}(2017){Evans}, {Cenko}, {Kennea}, {Emery}, {Kuin},
  {Korobkin}, {Wollaeger}, {Fryer}, {Madsen}, {Harrison}, {Xu}, {Nakar},
  {Hotokezaka}, {Lien}, {Campana}, {Oates}, {Troja}, {Breeveld}, {Marshall},
  {Barthelmy}, {Beardmore}, {Burrows}, {Cusumano}, {D'A{\`i}}, {D'Avanzo},
  {D'Elia}, {de Pasquale}, {Even}, {Fontes}, {Forster}, {Garcia}, {Giommi},
  {Grefenstette}, {Gronwall}, {Hartmann}, {Heida}, {Hungerford}, {Kasliwal},
  {Krimm}, {Levan}, {Malesani}, {Melandri}, {Miyasaka}, {Nousek}, {O'Brien},
  {Osborne}, {Pagani}, {Page}, {Palmer}, {Perri}, {Pike}, {Racusin}, {Rosswog},
  {Siegel}, {Sakamoto}, {Sbarufatti}, {Tagliaferri}, {Tanvir}, \&
  {Tohuvavohu}}]{evans17}
{Evans}, P.~A., {Cenko}, S.~B., {Kennea}, J.~A., {et~al.} 2017, Science, 358,
  1565

\bibitem[{{Fern{\'a}ndez} \& {Metzger}(2016)}]{fernandez16}
{Fern{\'a}ndez}, R., \& {Metzger}, B.~D. 2016, Annual Review of Nuclear and
  Particle Science, 66, 23

\bibitem[{Fischer {et~al.}(2016)Fischer, Godefroid, Brage, J{\"o}nsson, \&
  Gaigalas}]{topical_rev}
Fischer, C.~F., Godefroid, M., Brage, T., J{\"o}nsson, P., \& Gaigalas, G.
  2016, Journal of Physics B: Atomic, Molecular and Optical Physics, 49, 182004

\bibitem[{{Fontes} {et~al.}(2017){Fontes}, {Fryer}, {Hungerford}, {Wollaeger},
  {Rosswog}, \& {Berger}}]{fontes17}
{Fontes}, C.~J., {Fryer}, C.~L., {Hungerford}, A.~L., {et~al.} 2017,
  arXiv:1702.02990, arXiv:1702.02990

\bibitem[{Fritzsche \& Grant(1994)}]{SF}
Fritzsche, S., \& Grant, I. 1994, Physics Letters A, 186, 152

\bibitem[{Gaigalas {et~al.}(2017)Gaigalas, Fischer, Rynkun, \&
  J{\"o}nsson}]{jj2lsj_atoms}
Gaigalas, G., Fischer, C., Rynkun, P., \& J{\"o}nsson, P. 2017, Atoms, 5, 6

\bibitem[{Gaigalas \& Rudzikas(1996)}]{Gaigalas_1996}
Gaigalas, G., \& Rudzikas, Z. 1996, Journal of Physics B: Atomic, Molecular and
  Optical Physics, 29, 3303

\bibitem[{Gaigalas {et~al.}(1997)Gaigalas, Rudzikas, \&
  Fischer}]{Gaigalas_1997}
Gaigalas, G., Rudzikas, Z., \& Fischer, C.~F. 1997, Journal of Physics B:
  Atomic, Molecular and Optical Physics, 30, 3747

\bibitem[{Grant(1974)}]{gauge}
Grant, I.~P. 1974, Journal of Physics B: Atomic and Molecular Physics, 7, 1458

\bibitem[{Grant(2007)}]{grant}
---. 2007, Relativistic Quantum Theory of Atoms and Molecules (Springer, New
  York)

\bibitem[{{Hotokezaka} {et~al.}(2018){Hotokezaka}, {Beniamini}, \&
  {Piran}}]{hotokezaka17}
{Hotokezaka}, K., {Beniamini}, P., \& {Piran}, T. 2018, ArXiv e-prints,
  arXiv:1801.01141

\bibitem[{J{\"o}nsson {et~al.}(2013)J{\"o}nsson, Gaigalas, Biero\'n, Fischer,
  \& Grant}]{graspV3}
J{\"o}nsson, P., Gaigalas, G., Biero\'n, J., Fischer, C.~F., \& Grant, I. 2013,
  Computer Physics Communications, 184, 2197

\bibitem[{{Karp} {et~al.}(1977){Karp}, {Lasher}, {Chan}, \&
  {Salpeter}}]{karp77}
{Karp}, A.~H., {Lasher}, G., {Chan}, K.~L., \& {Salpeter}, E.~E. 1977, \apj,
  214, 161

\bibitem[{{Kasen} {et~al.}(2013){Kasen}, {Badnell}, \& {Barnes}}]{kasen13}
{Kasen}, D., {Badnell}, N.~R., \& {Barnes}, J. 2013, \apj, 774, 25

\bibitem[{{Kasen} {et~al.}(2015){Kasen}, {Fern{\'a}ndez}, \&
  {Metzger}}]{kasen15}
{Kasen}, D., {Fern{\'a}ndez}, R., \& {Metzger}, B.~D. 2015, \mnras, 450, 1777

\bibitem[{{Kasen} {et~al.}(2017){Kasen}, {Metzger}, {Barnes}, {Quataert}, \&
  {Ramirez-Ruiz}}]{kasen17}
{Kasen}, D., {Metzger}, B., {Barnes}, J., {Quataert}, E., \& {Ramirez-Ruiz}, E.
  2017, \nat, 551, 80

\bibitem[{{Kasen} {et~al.}(2006){Kasen}, {Thomas}, \& {Nugent}}]{kasen06}
{Kasen}, D., {Thomas}, R.~C., \& {Nugent}, P. 2006, \apj, 651, 366

\bibitem[{{Kasliwal} {et~al.}(2017){Kasliwal}, {Nakar}, {Singer}, {Kaplan},
  {Cook}, {Van Sistine}, {Lau}, {Fremling}, {Gottlieb}, {Jencson}, {Adams},
  {Feindt}, {Hotokezaka}, {Ghosh}, {Perley}, {Yu}, {Piran}, {Allison},
  {Anupama}, {Balasubramanian}, {Bannister}, {Bally}, {Barnes}, {Barway},
  {Bellm}, {Bhalerao}, {Bhattacharya}, {Blagorodnova}, {Bloom}, {Brady},
  {Cannella}, {Chatterjee}, {Cenko}, {Cobb}, {Copperwheat}, {Corsi}, {De},
  {Dobie}, {Emery}, {Evans}, {Fox}, {Frail}, {Frohmaier}, {Goobar}, {Hallinan},
  {Harrison}, {Helou}, {Hinderer}, {Ho}, {Horesh}, {Ip}, {Itoh}, {Kasen},
  {Kim}, {Kuin}, {Kupfer}, {Lynch}, {Madsen}, {Mazzali}, {Miller}, {Mooley},
  {Murphy}, {Ngeow}, {Nichols}, {Nissanke}, {Nugent}, {Ofek}, {Qi}, {Quimby},
  {Rosswog}, {Rusu}, {Sadler}, {Schmidt}, {Sollerman}, {Steele}, {Williamson},
  {Xu}, {Yan}, {Yatsu}, {Zhang}, \& {Zhao}}]{kasliwal17}
{Kasliwal}, M.~M., {Nakar}, E., {Singer}, L.~P., {et~al.} 2017, Science, 358,
  1559

\bibitem[{{Kilpatrick} {et~al.}(2017){Kilpatrick}, {Foley}, {Kasen},
  {Murguia-Berthier}, {Ramirez-Ruiz}, {Coulter}, {Drout}, {Piro}, {Shappee},
  {Boutsia}, {Contreras}, {Di Mille}, {Madore}, {Morrell}, {Pan}, {Prochaska},
  {Rest}, {Rojas-Bravo}, {Siebert}, {Simon}, \& {Ulloa}}]{kilpatrick17}
{Kilpatrick}, C.~D., {Foley}, R.~J., {Kasen}, D., {et~al.} 2017, Science, 358,
  1583

\bibitem[{{Kulkarni}(2005)}]{kulkarni05}
{Kulkarni}, S.~R. 2005, arXiv:astro-ph/0510256, arXiv:astro-ph/0510256

\bibitem[{{Li} \& {Paczy{\'n}ski}(1998)}]{li98}
{Li}, L.-X., \& {Paczy{\'n}ski}, B. 1998, \apjl, 507, L59

\bibitem[{{Lipunov} {et~al.}(2017){Lipunov}, {Gorbovskoy}, {Kornilov},
  {.~Tyurina}, {Balanutsa}, {Kuznetsov}, {Vlasenko}, {Kuvshinov}, {Gorbunov},
  {Buckley}, {Krylov}, {Podesta}, {Lopez}, {Podesta}, {Levato}, {Saffe},
  {Mallamachi}, {Potter}, {Budnev}, {Gress}, {Ishmuhametova}, {Vladimirov},
  {Zimnukhov}, {Yurkov}, {Sergienko}, {Gabovich}, {Rebolo}, {Serra-Ricart},
  {Israelyan}, {Chazov}, {Wang}, {Tlatov}, \& {Panchenko}}]{lipunov17}
{Lipunov}, V.~M., {Gorbovskoy}, E., {Kornilov}, V.~G., {et~al.} 2017, \apjl,
  850, L1

\bibitem[{{McCully} {et~al.}(2017){McCully}, {Hiramatsu}, {Howell},
  {Hosseinzadeh}, {Arcavi}, {Kasen}, {Barnes}, {Shara}, {Williams},
  {V{\"a}is{\"a}nen}, {Potter}, {Romero-Colmenero}, {Crawford}, {Buckley},
  {Cooke}, {Andreoni}, {Pritchard}, {Mao}, {Gromadzki}, \&
  {Burke}}]{muccully17}
{McCully}, C., {Hiramatsu}, D., {Howell}, D.~A., {et~al.} 2017, \apjl, 848, L32

\bibitem[{McKenzie {et~al.}(1980)McKenzie, Grant, \& Norrington}]{Breit}
McKenzie, B., Grant, I., \& Norrington, P. 1980, Computer Physics
  Communications, 21, 233

\bibitem[{{Metzger}(2017)}]{metzger17}
{Metzger}, B.~D. 2017, Living Reviews in Relativity, 20, 3

\bibitem[{{Metzger} \& {Fern{\'a}ndez}(2014)}]{metzger14}
{Metzger}, B.~D., \& {Fern{\'a}ndez}, R. 2014, \mnras, 441, 3444

\bibitem[{{Metzger} {et~al.}(2010){Metzger}, {Mart{\'{\i}}nez-Pinedo},
  {Darbha}, {Quataert}, {Arcones}, {Kasen}, {Thomas}, {Nugent}, {Panov}, \&
  {Zinner}}]{metzger10}
{Metzger}, B.~D., {Mart{\'{\i}}nez-Pinedo}, G., {Darbha}, S., {et~al.} 2010,
  \mnras, 406, 2650

\bibitem[{{Nicholl} {et~al.}(2017){Nicholl}, {Berger}, {Kasen}, {Metzger},
  {Elias}, {Brice{\~n}o}, {Alexander}, {Blanchard}, {Chornock},
  {Cowperthwaite}, {Eftekhari}, {Fong}, {Margutti}, {Villar}, {Williams},
  {Brown}, {Annis}, {Bahramian}, {Brout}, {Brown}, {Chen}, {Clemens},
  {Dennihy}, {Dunlap}, {Holz}, {Marchesini}, {Massaro}, {Moskowitz},
  {Pelisoli}, {Rest}, {Ricci}, {Sako}, {Soares-Santos}, \&
  {Strader}}]{nicholl17}
{Nicholl}, M., {Berger}, E., {Kasen}, D., {et~al.} 2017, \apjl, 848, L18

\bibitem[{Olsen {et~al.}(1995)Olsen, Godefroid, J\"onsson, Malmqvist, \&
  Fischer}]{biotra}
Olsen, J., Godefroid, M.~R., J\"onsson, P., Malmqvist, P.~A., \& Fischer, C.~F.
  1995, Phys. Rev. E, 52, 4499

\bibitem[{{Perego} {et~al.}(2017){Perego}, {Radice}, \& {Bernuzzi}}]{perego17}
{Perego}, A., {Radice}, D., \& {Bernuzzi}, S. 2017, \apjl, 850, L37

\bibitem[{{Pian} {et~al.}(2017){Pian}, {D'Avanzo}, {Benetti}, {Branchesi},
  {Brocato}, {Campana}, {Cappellaro}, {Covino}, {D'Elia}, {Fynbo}, {Getman},
  {Ghirlanda}, {Ghisellini}, {Grado}, {Greco}, {Hjorth}, {Kouveliotou},
  {Levan}, {Limatola}, {Malesani}, {Mazzali}, {Melandri}, {M{\o}ller},
  {Nicastro}, {Palazzi}, {Piranomonte}, {Rossi}, {Salafia}, {Selsing},
  {Stratta}, {Tanaka}, {Tanvir}, {Tomasella}, {Watson}, {Yang}, {Amati},
  {Antonelli}, {Ascenzi}, {Bernardini}, {Bo{\"e}r}, {Bufano}, {Bulgarelli},
  {Capaccioli}, {Casella}, {Castro-Tirado}, {Chassande-Mottin}, {Ciolfi},
  {Copperwheat}, {Dadina}, {De Cesare}, {di Paola}, {Fan}, {Gendre},
  {Giuffrida}, {Giunta}, {Hunt}, {Israel}, {Jin}, {Kasliwal}, {Klose}, {Lisi},
  {Longo}, {Maiorano}, {Mapelli}, {Masetti}, {Nava}, {Patricelli}, {Perley},
  {Pescalli}, {Piran}, {Possenti}, {Pulone}, {Razzano}, {Salvaterra},
  {Schipani}, {Spera}, {Stamerra}, {Stella}, {Tagliaferri}, {Testa}, {Troja},
  {Turatto}, {Vergani}, \& {Vergani}}]{pian17}
{Pian}, E., {D'Avanzo}, P., {Benetti}, S., {et~al.} 2017, \nat, 551, 67

\bibitem[{Radziemski {et~al.}(1970)Radziemski, Fisher, \& Steinhaus}]{ELCALC}
Radziemski, L.~J., Fisher, K., \& Steinhaus, D. 1970

\bibitem[{{Rosswog}(2015)}]{rosswog15}
{Rosswog}, S. 2015, International Journal of Modern Physics D, 24, 1530012

\bibitem[{{Rosswog} {et~al.}(2017){Rosswog}, {Sollerman}, {Feindt}, {Goobar},
  {Korobkin}, {Fremling}, \& {Kasliwal}}]{rosswog17b}
{Rosswog}, S., {Sollerman}, J., {Feindt}, U., {et~al.} 2017, arXiv:1710.05445,
  arXiv:1710.05445

\bibitem[{{Shappee} {et~al.}(2017){Shappee}, {Simon}, {Drout}, {Piro},
  {Morrell}, {Prieto}, {Kasen}, {Holoien}, {Kollmeier}, {Kelson}, {Coulter},
  {Foley}, {Kilpatrick}, {Siebert}, {Madore}, {Murguia-Berthier}, {Pan},
  {Prochaska}, {Ramirez-Ruiz}, {Rest}, {Adams}, {Alatalo}, {Ba{\~n}ados},
  {Baughman}, {Bernstein}, {Bitsakis}, {Boutsia}, {Bravo}, {Di Mille}, {Higgs},
  {Ji}, {Maravelias}, {Marshall}, {Placco}, {Prieto}, \& {Wan}}]{shappee17}
{Shappee}, B.~J., {Simon}, J.~D., {Drout}, M.~R., {et~al.} 2017, Science, 358,
  1574

\bibitem[{{Shibata} {et~al.}(2017){Shibata}, {Fujibayashi}, {Hotokezaka},
  {Kiuchi}, {Kyutoku}, {Sekiguchi}, \& {Tanaka}}]{shibata17}
{Shibata}, M., {Fujibayashi}, S., {Hotokezaka}, K., {et~al.} 2017, \prd, 96,
  123012

\bibitem[{{Siebert} {et~al.}(2017){Siebert}, {Foley}, {Drout}, {Kilpatrick},
  {Shappee}, {Coulter}, {Kasen}, {Madore}, {Murguia-Berthier}, {Pan}, {Piro},
  {Prochaska}, {Ramirez-Ruiz}, {Rest}, {Contreras}, {Morrell}, {Rojas-Bravo},
  \& {Simon}}]{siebert17}
{Siebert}, M.~R., {Foley}, R.~J., {Drout}, M.~R., {et~al.} 2017, \apjl, 848,
  L26

\bibitem[{{Smartt} {et~al.}(2017){Smartt}, {Chen}, {Jerkstrand}, {Coughlin},
  {Kankare}, {Sim}, {Fraser}, {Inserra}, {Maguire}, {Chambers}, {Huber},
  {Kr{\"u}hler}, {Leloudas}, {Magee}, {Shingles}, {Smith}, {Young}, {Tonry},
  {Kotak}, {Gal-Yam}, {Lyman}, {Homan}, {Agliozzo}, {Anderson}, {Angus},
  {Ashall}, {Barbarino}, {Bauer}, {Berton}, {Botticella}, {Bulla}, {Bulger},
  {Cannizzaro}, {Cano}, {Cartier}, {Cikota}, {Clark}, {De Cia}, {Della Valle},
  {Denneau}, {Dennefeld}, {Dessart}, {Dimitriadis}, {Elias-Rosa}, {Firth},
  {Flewelling}, {Fl{\"o}rs}, {Franckowiak}, {Frohmaier}, {Galbany},
  {Gonz{\'a}lez-Gait{\'a}n}, {Greiner}, {Gromadzki}, {Guelbenzu},
  {Guti{\'e}rrez}, {Hamanowicz}, {Hanlon}, {Harmanen}, {Heintz}, {Heinze},
  {Hernandez}, {Hodgkin}, {Hook}, {Izzo}, {James}, {Jonker}, {Kerzendorf},
  {Klose}, {Kostrzewa-Rutkowska}, {Kowalski}, {Kromer}, {Kuncarayakti},
  {Lawrence}, {Lowe}, {Magnier}, {Manulis}, {Martin-Carrillo}, {Mattila},
  {McBrien}, {M{\"u}ller}, {Nordin}, {O'Neill}, {Onori}, {Palmerio},
  {Pastorello}, {Patat}, {Pignata}, {Podsiadlowski}, {Pumo}, {Prentice}, {Rau},
  {Razza}, {Rest}, {Reynolds}, {Roy}, {Ruiter}, {Rybicki}, {Salmon}, {Schady},
  {Schultz}, {Schweyer}, {Seitenzahl}, {Smith}, {Sollerman}, {Stalder},
  {Stubbs}, {Sullivan}, {Szegedi}, {Taddia}, {Taubenberger}, {Terreran}, {van
  Soelen}, {Vos}, {Wainscoat}, {Walton}, {Waters}, {Weiland}, {Willman},
  {Wiseman}, {Wright}, {Wyrzykowski}, \& {Yaron}}]{smartt17}
{Smartt}, S.~J., {Chen}, T.-W., {Jerkstrand}, A., {et~al.} 2017, \nat, 551, 75

\bibitem[{{Soares-Santos} {et~al.}(2017){Soares-Santos}, {Holz}, {Annis},
  {Chornock}, {Herner}, {Berger}, {Brout}, {Chen}, {Kessler}, {Sako}, {Allam},
  {Tucker}, {Butler}, {Palmese}, {Doctor}, {Diehl}, {Frieman}, {Yanny}, {Lin},
  {Scolnic}, {Cowperthwaite}, {Neilsen}, {Marriner}, {Kuropatkin}, {Hartley},
  {Paz-Chinch{\'o}n}, {Alexander}, {Balbinot}, {Blanchard}, {Brown}, {Carlin},
  {Conselice}, {Cook}, {Drlica-Wagner}, {Drout}, {Durret}, {Eftekhari}, {Farr},
  {Finley}, {Foley}, {Fong}, {Fryer}, {Garc{\'{\i}}a-Bellido}, {Gill},
  {Gruendl}, {Hanna}, {Kasen}, {Li}, {Lopes}, {Louren{\c c}o}, {Margutti},
  {Marshall}, {Matheson}, {Medina}, {Metzger}, {Mu{\~n}oz}, {Muir}, {Nicholl},
  {Quataert}, {Rest}, {Sauseda}, {Schlegel}, {Secco}, {Sobreira}, {Stebbins},
  {Villar}, {Vivas}, {Walker}, {Wester}, {Williams}, {Zenteno}, {Zhang},
  {Abbott}, {Abdalla}, {Banerji}, {Bechtol}, {Benoit-L{\'e}vy}, {Bertin},
  {Brooks}, {Buckley-Geer}, {Burke}, {Carnero Rosell}, {Carrasco Kind},
  {Carretero}, {Castander}, {Crocce}, {Cunha}, {D'Andrea}, {da Costa}, {Davis},
  {Desai}, {Dietrich}, {Doel}, {Eifler}, {Fernandez}, {Flaugher}, {Fosalba},
  {Gaztanaga}, {Gerdes}, {Giannantonio}, {Goldstein}, {Gruen}, {Gschwend},
  {Gutierrez}, {Honscheid}, {Jain}, {James}, {Jeltema}, {Johnson}, {Johnson},
  {Kent}, {Krause}, {Kron}, {Kuehn}, {Kuhlmann}, {Lahav}, {Lima}, {Maia},
  {March}, {McMahon}, {Menanteau}, {Miquel}, {Mohr}, {Nichol}, {Nord},
  {Ogando}, {Petravick}, {Plazas}, {Romer}, {Roodman}, {Rykoff}, {Sanchez},
  {Scarpine}, {Schubnell}, {Sevilla-Noarbe}, {Smith}, {Smith}, {Suchyta},
  {Swanson}, {Tarle}, {Thomas}, {Thomas}, {Troxel}, {Vikram}, {Wechsler},
  {Weller}, {Dark Energy Survey}, \& {Dark Energy Camera GW-EM
  Collaboration}}]{soares-santos17}
{Soares-Santos}, M., {Holz}, D.~E., {Annis}, J., {et~al.} 2017, \apjl, 848, L16

\bibitem[{{Tanaka}(2016)}]{tanaka16}
{Tanaka}, M. 2016, Advances in Astronomy, 2016, 634197

\bibitem[{{Tanaka} \& {Hotokezaka}(2013)}]{tanaka13}
{Tanaka}, M., \& {Hotokezaka}, K. 2013, \apj, 775, 113

\bibitem[{{Tanaka} {et~al.}(2014){Tanaka}, {Hotokezaka}, {Kyutoku}, {Wanajo},
  {Kiuchi}, {Sekiguchi}, \& {Shibata}}]{tanaka14}
{Tanaka}, M., {Hotokezaka}, K., {Kyutoku}, K., {et~al.} 2014, \apj, 780, 31

\bibitem[{{Tanaka} {et~al.}(2017){Tanaka}, {Utsumi}, {Mazzali}, {Tominaga},
  {Yoshida}, {Sekiguchi}, {Morokuma}, {Motohara}, {Ohta}, {Kawabata}, {Abe},
  {Aoki}, {Asakura}, {Baar}, {Barway}, {Bond}, {Doi}, {Fujiyoshi}, {Furusawa},
  {Honda}, {Itoh}, {Kawabata}, {Kawai}, {Kim}, {Lee}, {Miyazaki}, {Morihana},
  {Nagashima}, {Nagayama}, {Nakaoka}, {Nakata}, {Ohsawa}, {Ohshima}, {Okita},
  {Saito}, {Sumi}, {Tajitsu}, {Takahashi}, {Takayama}, {Tamura}, {Tanaka},
  {Terai}, {Tristram}, {Yasuda}, \& {Zenko}}]{tanaka17}
{Tanaka}, M., {Utsumi}, Y., {Mazzali}, P.~A., {et~al.} 2017, \pasj, 69, 102

\bibitem[{{Tanaka} {et~al.}(2018){Tanaka}, {Kato}, {Gaigalas}, {Rynkun},
  {Rad{\v z}i{\= u}t{\.e}}, {Wanajo}, {Sekiguchi}, {Nakamura}, {Tanuma},
  {Murakami}, \& {Sakaue}}]{tanaka18}
{Tanaka}, M., {Kato}, D., {Gaigalas}, G., {et~al.} 2018, \apj, 852, 109

\bibitem[{{Tanvir} {et~al.}(2017){Tanvir}, {Levan},
  {Gonz{\'a}lez-Fern{\'a}ndez}, {Korobkin}, {Mandel}, {Rosswog}, {Hjorth},
  {D'Avanzo}, {Fruchter}, {Fryer}, {Kangas}, {Milvang-Jensen}, {Rosetti},
  {Steeghs}, {Wollaeger}, {Cano}, {Copperwheat}, {Covino}, {D'Elia}, {de Ugarte
  Postigo}, {Evans}, {Even}, {Fairhurst}, {Figuera Jaimes}, {Fontes}, {Fujii},
  {Fynbo}, {Gompertz}, {Greiner}, {Hodosan}, {Irwin}, {Jakobsson},
  {J{\o}rgensen}, {Kann}, {Lyman}, {Malesani}, {McMahon}, {Melandri},
  {O'Brien}, {Osborne}, {Palazzi}, {Perley}, {Pian}, {Piranomonte}, {Rabus},
  {Rol}, {Rowlinson}, {Schulze}, {Sutton}, {Th{\"o}ne}, {Ulaczyk}, {Watson},
  {Wiersema}, \& {Wijers}}]{tanvir17}
{Tanvir}, N.~R., {Levan}, A.~J., {Gonz{\'a}lez-Fern{\'a}ndez}, C., {et~al.}
  2017, \apjl, 848, L27

\bibitem[{{Tominaga} {et~al.}(2018){Tominaga}, {Tanaka}, {Morokuma}, {Utsumi},
  {Yamaguchi}, {Yasuda}, {Tanaka}, {Yoshida}, {Fujiyoshi}, {Furusawa},
  {Kawabata}, {Lee}, {Motohara}, {Ohsawa}, {Ohta}, {Terai}, {Abe}, {Aoki},
  {Asakura}, {Barway}, {Bond}, {Fujisawa}, {Honda}, {Ioka}, {Itoh}, {Kawai},
  {Kim}, {Koshimoto}, {Matsubayashi}, {Miyazaki}, {Saito}, {Sekiguchi}, {Sumi},
  \& {Tristram}}]{tominaga18}
{Tominaga}, N., {Tanaka}, M., {Morokuma}, T., {et~al.} 2018, \pasj,
  arXiv:1710.05865

\bibitem[{{Troja} {et~al.}(2017){Troja}, {Piro}, {van Eerten}, {Wollaeger},
  {Im}, {Fox}, {Butler}, {Cenko}, {Sakamoto}, {Fryer}, {Ricci}, {Lien}, {Ryan},
  {Korobkin}, {Lee}, {Burgess}, {Lee}, {Watson}, {Choi}, {Covino}, {D'Avanzo},
  {Fontes}, {Gonz{\'a}lez}, {Khandrika}, {Kim}, {Kim}, {Lee}, {Lee}, {Kutyrev},
  {Lim}, {S{\'a}nchez-Ram{\'{\i}}rez}, {Veilleux}, {Wieringa}, \&
  {Yoon}}]{troja17}
{Troja}, E., {Piro}, L., {van Eerten}, H., {et~al.} 2017, \nat, 551, 71

\bibitem[{{Utsumi} {et~al.}(2017){Utsumi}, {Tanaka}, {Tominaga}, {Yoshida},
  {Barway}, {Nagayama}, {Zenko}, {Aoki}, {Fujiyoshi}, {Furusawa}, {Kawabata},
  {Koshida}, {Lee}, {Morokuma}, {Motohara}, {Nakata}, {Ohsawa}, {Ohta},
  {Okita}, {Tajitsu}, {Tanaka}, {Terai}, {Yasuda}, {Abe}, {Asakura}, {Bond},
  {Miyazaki}, {Sumi}, {Tristram}, {Honda}, {Itoh}, {Itoh}, {Kawabata},
  {Morihana}, {Nagashima}, {Nakaoka}, {Ohshima}, {Takahashi}, {Takayama},
  {Aoki}, {Baar}, {Doi}, {Finet}, {Kanda}, {Kawai}, {Kim}, {Kuroda}, {Liu},
  {Matsubayashi}, {Murata}, {Nagai}, {Saito}, {Saito}, {Sako}, {Sekiguchi},
  {Tamura}, {Tanaka}, {Uemura}, \& {Yamaguchi}}]{utsumi17}
{Utsumi}, Y., {Tanaka}, M., {Tominaga}, N., {et~al.} 2017, \pasj, 69, 101

\bibitem[{{Valenti} {et~al.}(2017){Valenti}, {David}, {Sand}, {Yang},
  {Cappellaro}, {Tartaglia}, {Corsi}, {Jha}, {Reichart}, {Haislip}, \&
  {Kouprianov}}]{valenti17}
{Valenti}, S., {David}, {Sand}, J., {et~al.} 2017, \apjl, 848, L24

\bibitem[{{Wollaeger} {et~al.}(2017){Wollaeger}, {Korobkin}, {Fontes},
  {Rosswog}, {Even}, {Fryer}, {Sollerman}, {Hungerford}, {van Rossum}, \&
  {Wollaber}}]{wollaeger17}
{Wollaeger}, R.~T., {Korobkin}, O., {Fontes}, C.~J., {et~al.} 2017,
  arXiv:1705.07084, arXiv:1705.07084

\bibitem[{Wyart(2010)}]{WyartNdII_1}
Wyart, J.-F. 2010, Physica Scripta, 82, 035302

\bibitem[{Wyart {et~al.}(2006)Wyart, Meftah, Bachelier, Sinzelle,
  Tchang-Brillet, Champion, Spector, \& Sugar}]{Wyart_NIV_2006}
Wyart, J.-F., Meftah, A., Bachelier, A., {et~al.} 2006, Journal of Physics B:
  Atomic, Molecular and Optical Physics, 39, L77

\bibitem[{Wyart {et~al.}(2008)Wyart, Meftah, Sinzelle, Tchang-Brillet, Spector,
  \& Judd}]{Wyart_NIV_2008}
Wyart, J.-F., Meftah, A., Sinzelle, J., {et~al.} 2008, Journal of Physics B:
  Atomic, Molecular and Optical Physics, 41, 085001

\bibitem[{Wyart {et~al.}(2007)Wyart, Meftah, Tchang-Brillet, Champion, Lamrous,
  Spector, \& Sugar}]{Wyart_NIV_2007}
Wyart, J.-F., Meftah, A., Tchang-Brillet, W.-.~L., {et~al.} 2007, Journal of
  Physics B: Atomic, Molecular and Optical Physics, 40, 3957

\bibitem[{Yoca \& Quinet(2014)}]{Yoca}
Yoca, S.~E., \& Quinet, P. 2014, Journal of Physics B: Atomic, Molecular and
  Optical Physics, 47, 035002

\bibitem[{{Zhang, Z. G.} {et~al.}(2002){Zhang, Z. G.}, {Svanberg, S.},
  {Palmeri, P.}, {Quinet, P.}, \& {Bi\'emont, E.}}]{Zhang}
{Zhang, Z. G.}, {Svanberg, S.}, {Palmeri, P.}, {Quinet, P.}, \& {Bi\'emont, E.}
  2002, A\&A, 385, 724

\end{thebibliography}



\end{document}